\documentclass[]{fairmeta}
\DeclareMathOperator*{\argmin}{arg\,min}

\title{Scaling Small Agents Through Strategy Auctions}

\author[\dagger, 2]{Lisa Alazraki}
\author[\dagger, 3]{William F. Shen}
\author[1]{Yoram Bachrach}
\author[1]{Akhil Mathur}

\affiliation[1]{Meta Superintelligence Labs}
\affiliation[2]{Imperial College London}
\affiliation[3]{University of Cambridge}

\contribution[\dagger]{Work done at Meta}

\abstract{Small language models are increasingly viewed as a promising, cost-effective approach to agentic AI, with proponents claiming they are sufficiently capable for agentic workflows. However, while smaller agents can closely match larger ones on simple tasks, it remains unclear how their performance scales with task complexity, when large models become necessary, and how to better leverage small agents for long-horizon workloads. In this work, we empirically show that small agents' performance fails to scale with task complexity on deep search and coding tasks, and we introduce \textit{Strategy Auctions for Workload Efficiency} (\textit{\textsc{sale}}), an agent framework inspired by freelancer marketplaces. In \textsc{sale}, agents bid with short strategic plans, which are scored by a systematic cost–value mechanism and refined via a shared auction memory, enabling per-task routing and continual self-improvement without training a separate router or running all models to completion. Across deep search and coding tasks of varying complexity, \textsc{sale} reduces reliance on the largest agent by 52\%, lowers overall cost by 35\%, and consistently improves upon the largest agent's pass@1 with only a negligible overhead beyond executing the final trace. In contrast, established routers that rely on task descriptions either underperform the largest agent or fail to reduce cost, often both, underscoring their poor fit for agentic workflows. These results suggest that while small agents may be insufficient for complex workloads, they can be effectively ``scaled up'' through coordinated task allocation and test-time self-improvement. More broadly, they motivate a systems-level view of agentic AI in which performance gains come less from ever-larger individual models and more from market-inspired coordination mechanisms that organize heterogeneous agents into efficient, adaptive ecosystems.}

\date{\today}
\correspondence{ \email{lisa.alazraki20@imperial.ac.uk},
\email{akhilm@meta.com}}

\begin{document}

\maketitle

\section{Introduction}
\label{section:intro}

Recent work on tool-augmented AI agents has led to growing optimism that small language models may be sufficient for many real-world applications. By offloading computation and knowledge to external tools and environments, small agents are argued to need less parametric capacity while still supporting complex, multi-step behavior~\citep{houliston2025provablebenefitsintoollearning}. Combined with advances that narrow the performance gap between small and large language models~\citep{sarahooker2025deathofscaling}, this has led to claims that small, inexpensive agents can replace large ones as the foundation of agentic AI~\citep{belcak2025smalllanguagemodelsfuture}.

Yet much of the current optimism around small agents is framed in terms of model size and agentic capabilities, with comparatively little attention to how these interact with the structure and complexity of the tasks they are meant to solve. In practice, agentic workloads span a wide spectrum: from short, well-specified tasks with simple evaluation criteria to open-ended, long-horizon problems that require extended reasoning, integrating different types of information, and maintaining coherence over many steps~\citep{wang2025odysseybenchevaluatingllmagents}. It is not obvious that the same small agent that performs well on the former regime will also succeed on the latter, especially as demands on reasoning, planning, and context management grow with task complexity.

This perspective raises two central questions for the design of agentic AI systems. First, \emph{how does task complexity mediate the relative effectiveness of small and large agents?} Second, given an increasingly heterogeneous landscape of models with different capabilities and costs, \emph{how should we route tasks across agents to balance accuracy and cost, maximizing the workload handled by small, cheap agents without degrading performance on complex tasks?} Existing routing approaches provide only a partial answer. Non-predictive strategies that generate full outputs from all candidate models are tractable for single-shot QA but become infeasible for agents, whose trajectories can span tens of thousands to millions of tokens. Predictive routers, in turn, require training separate routing models that are costly to fit, do not generalize well to new models, and have been shown to degrade as task difficulty increases~\citep{dhrif2025reasoningawarepromptorchestrationfoundation}. It remains unclear how to design routing mechanisms for agentic systems that incur minimal additional inference cost, apply directly to off-the-shelf agents, remain effective on complex, long-horizon tasks, and ideally also help smaller agents shoulder more of the workload over time, effectively ``scaling them up'' without sacrificing accuracy.

\begin{figure*}[t]
     \centering
     \hspace{0.05cm}
     \begin{subfigure}[b]{0.48\textwidth}
         \centering
         \includegraphics[height=4.7cm]{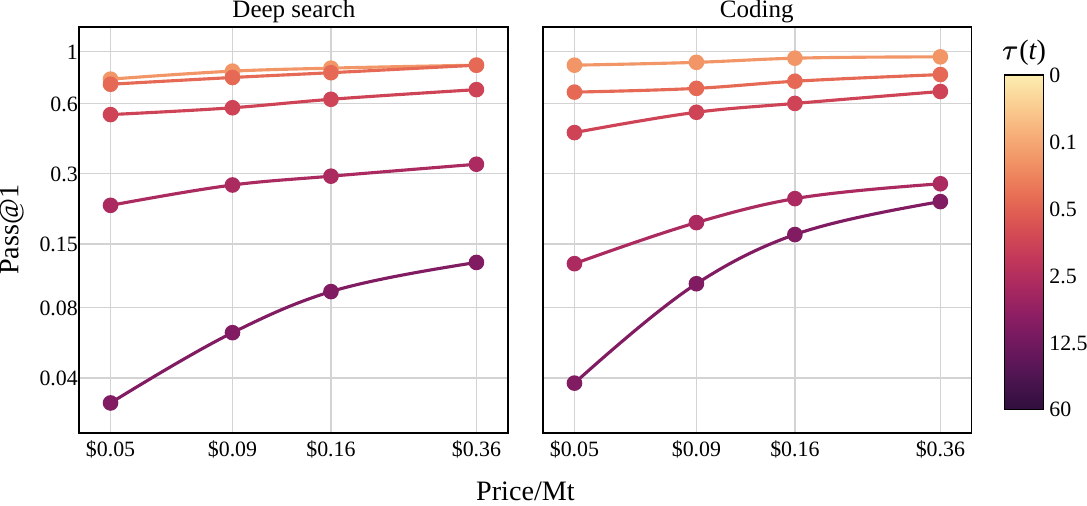}
         \caption{Pass@1 as a function of $\tau(t)$ and agent price.}
         \label{fig:plots:a}
     \end{subfigure}
     \hfill
     \begin{subfigure}[b]{0.48\textwidth}
        \hspace{1.4cm}
         \centering
         \includegraphics[height=4.7cm]{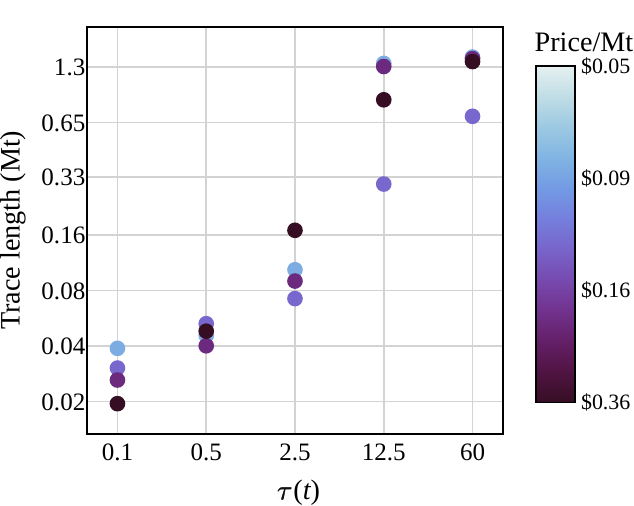}
         \caption{Trace length by $\tau(t)$ for agents of different prices.}
         \label{fig:plots:b}
     \end{subfigure}
     \caption{Pass@1 accuracy on deep search and coding tasks (a) and average trace length in million tokens (b). We show the effective price per million tokens $\pi(a_d)$ for Qwen3 agents from smallest to largest ($d =$ \{4B, 8B, 14B, 32B\}).}
     \label{fig:plots}
\end{figure*}

To study how task complexity shapes the relative usefulness of small and large agents, we empirically evaluate deep search and coding tasks across multiple horizons. We choose these domains as they typify agentic workflows: deep search requires extended reasoning and information integration~\citep{zhang2025websearchagenticdeep}, while coding demands multi-step planning and precise execution~\citep{wang2025aiagenticprogrammingsurvey}. Both domains span short, well-specified tasks as well as open-ended, long-horizon problems, making them ideal for probing how agent capabilities scale with complexity. Following~\cite{kwa2025measuring}, we operationalize \textbfrm{task complexity} via human solution time: the average time expert annotators need to complete each task, from a few seconds to one hour. We apply this annotation protocol primarily to tasks from existing public benchmarks, with a small number of ad hoc tasks. Using off-the-shelf models from the Qwen3 series ranging from 4B to 32B parameters, and the Agent Research Environment (ARE)~\citep{froger2025arescalingagentenvironments}, we find that on the simplest tasks the smallest agent attains $\sim$90\% of the pass@1 performance of the largest agent, but on the most complex tasks this relative performance drops to only $\sim$21\% (ref. Figure~\ref{fig:plots} and Section \ref{sec:performance-complexity}). Thus, while small agents can closely match larger ones on simple tasks, their performance fails to scale with task complexity. This suggests that small agents alone are unlikely to be sufficient for many high-value applications and that model size should be treated as a \emph{per-task} decision rather than a global choice about whether small agents can \emph{replace} large ones.

%%%%%%%%%%%%%%%%%%%%%%%%%%%%%%%%%%%%%%%%%%%%%%%%%%%
%%%%%%%%%%%%%%%%%%%%%%%%%%%%%%%%%%%%%%%%%%%%%%%%%%

%%%%%%%%%%%%%%%%%%%%%%%%%%%%%%%%%%%%%%%%%%%%%%%%%%
\begin{figure*}[!t]
    \centering
    \includegraphics[width=\linewidth]{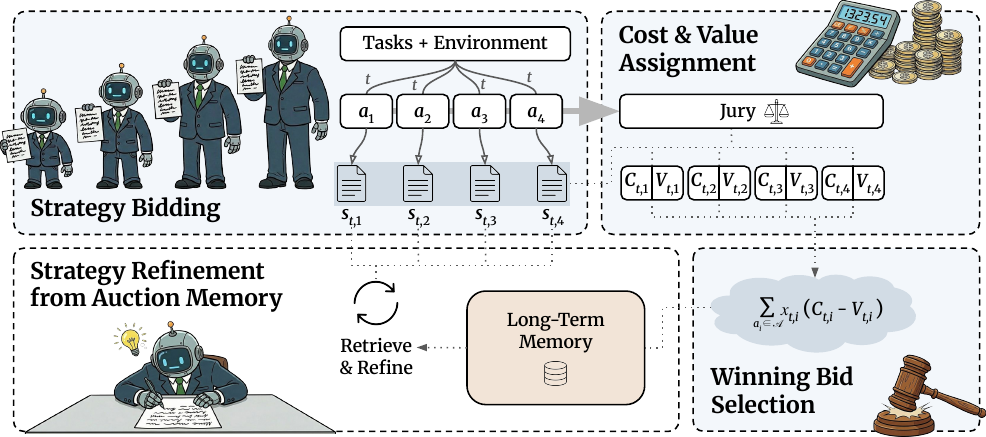}
    \caption{An illustration of the \textsc{sale} pipeline. Given a task $t$, each agent $a_i$ proposes a strategic plan $s_{t,i}$ as its bid. Bids are evaluated by cost $C_{t,i}$ and value $V_{t,i}$, and a provisional winner is selected by minimizing cost-minus-value. Agents cheaper than the provisional winner may then refine their strategies using similar past successes and failures retrieved from the auction memory, after which a final winner is selected and its strategy is executed.}
    \label{fig:main}
\end{figure*}
%%%%%%%%%%%%%%%%%%%%%%%%%%%%%%%%%%%%%%%%%%%%%%%%%%

In response, we develop a routing mechanism that is compute-efficient, applicable to off-the-shelf agents, and preserves performance on complex, long-horizon tasks. Inspired by human freelance marketplaces and virtual agent economies~\citep{tomasev2025virtualagenteconomies}, we introduce \textbfrm{S}trategy \textbfrm{A}uctions for Work\textbfrm{l}oad \textbfrm{E}fficiency (\textsc{sale}), a test-time auction framework that leverages a well-established correlation between plan quality and execution quality~\citep{sun-etal-2024-enhancing-code}. For each task, candidate agents propose strategic solution plans that are scored by predicted value and cost via peer assessment and heuristic predictors. The winning agent is selected based on this cost–value trade-off and its plan is executed, yielding an adaptive allocation of work across the agent pool. Crucially, this process is not static: plan refinement using the outcomes of past auctions can overturn the initial ranking before any strategy is executed; a self-improvement process analogous to how freelancers upskill over time to secure more work. In this way, \textsc{sale} functions not only as a router but also as a mechanism that systematically increases the share of work handled by smaller, cheaper agents where possible, effectively ``scaling up'' small models via market-like coordination.

We find that \textsc{sale} not only matches but even exceeds the largest agent's pass@1 ($+3.5\%$ on deep search and $+2.7\%$ on coding) while offloading much of its workload ($-65\%$ and $-40\%$, respectively) and reducing total spend ($-42\%$ on deep search and $-25\%$ on coding). These gains come with only a negligible increase in inference tokens. In contrast, established routers that rely on task descriptions either underperform the largest agent or fail to meaningfully reduce spend, often both. This underscores their poor fit for agentic workflows, where complex trajectories decouple task inputs from downstream success and strategic planning proves a more reliable routing signal. We also observe that, as the auction memory grows, the smallest agents are selected increasingly often, suggesting that they progressively capitalize on accumulated experience. Overall, \textsc{sale} extends the performance–cost Pareto frontier beyond any single agent: it reduces reliance on large agents and total inference cost while improving accuracy across task complexities.
 
In summary, our contributions are:\vspace{2pt}
\begin{enumerate}
\item We empirically study how task complexity affects the performance gap between small and large agents on deep search and coding tasks, finding that small agents nearly match large ones on simple tasks but diverge sharply as complexity increases. To the best of our knowledge, this is the first such investigation on realistic workloads; prior literature has examined agentic scaling behavior only on synthetic tasks.\vspace{3pt}
\item We develop \textsc{HST-Bench}\footnote{\url{https://github.com/facebookresearch/hst-bench}}, a benchmark that pairs agentic tasks with human solution times as a proxy for task complexity, enabling fine-grained evaluation of agent scaling behavior.\vspace{3pt}
\item We introduce \textsc{sale}, a marketplace-inspired framework in which heterogeneous AI agents bid with solution plans, are selected based on predicted value and cost, and use auction feedback to refine subsequent bids, yielding a unified mechanism that couples per-task model routing with test-time self-improvement.\vspace{3pt}
\item We show that \textsc{sale} achieves a better performance–cost Pareto frontier than any individual agent in the pool or existing routers on deep search and coding tasks. This demonstrates that strategy-based routing with continual agent self-improvement outperforms single-model and conventional routing baselines.\vspace{3pt}
\item More broadly, by providing \textsc{sale} as a marketplace-inspired framework, we illustrate how auction-based coordination can structure competition and collaboration among heterogeneous agents at test time, contributing to emerging discussions about how labor-like market dynamics and adaptive orchestration may shape future ecosystems of interacting AI agents.
\end{enumerate}

\section{Related Work}

\para{Agent Performance Under Task Complexity.} 
Scaling AI agents to handle increasingly long and difficult tasks has become a central focus of recent work~\citep{chan2025mlebench, chen2025mlrbench, froger2025arescalingagentenvironments, wang2025odysseybenchevaluatingllmagents}. \cite{kwa2025measuring} address this by tying capability to task duration, defining a 50\%-success time horizon in terms of human solution time and studying how it scales on research and software-engineering tasks. \cite{sinha2025illusiondiminishingreturnsmeasuring} instead examine how performance degrades as tasks are extended, using synthetic, controlled multi-step tasks with explicit plans to argue that small per-step accuracy gains can yield much longer executable sequences and that many long-horizon failures reflect compounding execution errors as models condition on their own past outputs. We build on both perspectives by analyzing these scaling phenomena on real-world deep search and coding workloads and by shifting from isolated model behavior to system-level performance in a marketplace that allocates tasks across heterogeneous agents.

\para{Multi-Agent Routing.} Routing has emerged as a key strategy for harnessing the diversity of heterogeneous AI systems. There are two main approaches to routing: non-predictive routing, which selects outputs after running multiple models, and predictive routing, which chooses a model in advance based on input features or learned decision policies~\citep{hu2024routerbench}. Non-predictive methods~\citep{chen2024frugalgpt} can be prohibitively expensive in agentic settings, where trajectories involve extended tool use and long interaction histories~\citep{tsiourvas2025causal}. Predictive approaches~\citep{hu2024routerbench, stripelis-etal-2024-tensoropera, somerstep2025carrotcostawarerate} mitigate this cost by learning separate routing models, but these are themselves costly to fit, tightly coupled to specific model sets, and have been shown to degrade as task difficulty increases~\citep{dhrif2025reasoningawarepromptorchestrationfoundation}. Moreover, existing routers are typically static: once trained, their routing policies do not incorporate test-time feedback, and thus do not improve with experience. In contrast, our framework, \textsc{sale}, implements a lightweight, strategy-based, partially predictive routing mechanism in which agents bid with short plans rather than full solutions, leveraging empirical evidence that plan quality correlates with downstream task success~\citep{sun-etal-2024-enhancing-code, kang2025distillingllmagentsmall, xiong-etal-2025-mpo}. Auction feedback and shared memory refine future bids, progressively shifting more work onto smaller agents. Thus, \textsc{sale} couples routing with continual adaptation, turning agent selection from a purely passive assignment into a mechanism that actively improves small agents' effective capabilities under compute constraints.

\para{Memory-Driven Adaptation.} Memory systems help agents improve by reusing past behavior. Existing work typically uses memory to improve an agent's reasoning, either by extracting reusable routines from successful trajectories to guide future actions~\citep{cao2025remembermerefineme, wang2025agent}, or by maintaining structured records of past interactions that provide richer context and user-specific knowledge~\citep{salama2025meminsightautonomousmemoryaugmentation, wang2025mirixmultiagentmemoryllmbased, xu2025amem}. In contrast, \textsc{sale} differs both in what is stored and in how that information is used: rather than logging answers, execution traces, or user histories, we treat bidding strategies and their auction outcomes (wins and losses) as the primary memory signal. This makes memory an explicit mechanism for reallocating work and upgrading the effective capabilities of smaller agents, as feedback from past auctions is used to refine future bids and adjust the division of labor in the marketplace.

\para{Agent Systems as Virtual Economies.} Prior work has argued that as autonomous agents become economically significant, they should be coordinated through explicit market mechanisms, including auction-based interaction~\citep{duetting2024mechanism, zhu2024evidence, jiang2025harborexploringpersonadynamics,yang2025agentexchangeshapingfuture}, virtual sandbox economies with controlled links to human markets~\citep{tomasev2025virtualagenteconomies}, and settings in which assistant and service agents transact directly on behalf of users and firms~\citep{rothschild2025agenticeconomy}. Building on this perspective, \textsc{sale} treats agents as freelancers in a job marketplace, where each job is awarded on the basis of both price and quality through a centralised scoring rule, a setting formally studied as an auction~\citep{3f72338c-0e9d-3b84-9736-5fcbd092d6eb, 9c70fefc-5b5a-37ce-9a1f-c911e138ac22}. In \textsc{sale}, bids over strategic plans allocate work and learning opportunities, illustrating how labor-like dynamics can shape future agent ecosystems.

\section{Experimental Setup}
To evaluate agentic performance across task complexities and model scales, we run all experiments within the Agent Research Environment (ARE) framework~\citep{froger2025arescalingagentenvironments}. ARE provides a standardized platform for benchmarking agent behavior, enabling consistent measurement and comparison across domains.

\subsection{Data}\label{sec:datasets}
We evaluate agentic performance on two domains: deep search and coding, as they broadly represent agentic workflows requiring extended reasoning and multi-step planning. For deep search, we sample from SimpleQA~\citep{wei2024measuringshortformfactualitylarge}, PopQA~\citep{mallen-etal-2023-trust}, HotpotQA \citep{yang-etal-2018-hotpotqa}, GAIA \citep{mialon2024gaia}, and an expert-validated portion of Humanity's Last Exam~\citep{phan2025humanitysexam, white2025hlewrong}. Coding tasks are drawn from MBPP~\citep{austin2021programsynthesislargelanguage} and LeetCode~\citep{xia2025leetcodedatasettemporaldatasetrobust}, supplemented with custom multiple-choice questions to cover lower-complexity cases. We select these datasets because they span a wide range of task horizons and require genuinely agentic capabilities: deep search demands dynamic tool use, iterative retrieval, and cross-source synthesis, while coding involves iterative debugging and test-driven refinement. These benchmarks have been widely adopted for evaluating agentic AI systems, ensuring both breadth and comparability with prior work~\citep{Coignion_2024,  labruna2024retrieveteachingllmsutilize, liu2024bolaa, amini2025openagentspecificationagent, gan-etal-2025-master, zoom2025hle, xie2025profileawaremaneuveringdynamicmultiagent}.

\citet{kwa2025measuring} validate human solution time as the primary metric for agentic task complexity, showing that it naturally integrates reasoning, planning and execution into a single scale. We adopt this measure and define the \emph{task complexity} of $t \in \mathcal{D}$ as
$\tau(t)$, the average time (in minutes) required by expert annotators to solve $t$. Human solution times are annotated by three expert annotators, yielding reliably reproducible estimates (Krippendorff's $\alpha = 0.86$; details in Appendix~\ref{appendix:data-annotation}). 
To enable fine-grained analysis of complexity effects, we group tasks into five non-overlapping bins according to $\tau(t)$, corresponding to average human solution times of up to 6 seconds ($0 < \tau(t) \le 0.1$), 30 seconds ($0.1 < \tau(t) \le 0.5$), 2.5 minutes ($0.5 < \tau(t) \le 2.5$), 12.5 minutes ($2.5 < \tau(t) \le 12.5$), and 60 minutes ($12.5 < \tau(t) \le 60$). As we show in Section~\ref{sec:performance-complexity}, pass@1 degrades monotonically with $\tau(t)$ across all four agent sizes, indicating that HST tracks LLM-perceived difficulty closely on our task distribution. Bin boundaries follow a geometric progression (5× between adjacent bins), yielding equal spacing on a log scale. This is appropriate given that human solution times span nearly three orders of magnitude, and produces approximately balanced sample sizes across bins. A breakdown of the data composition for each time bin is provided in Appendix~\ref{appendix:data}. In total, the resulting human-timed dataset \textsc{HST-Bench}, contains 753 tasks.

\subsection{Models}
For all experiments, we utilize the Qwen3 family of language models~\citep{yang2025qwen3technicalreport}, chosen for their open-weight availability and broad range of sizes. Qwen3 provides checkpoints at 4B, 8B, 14B, and 32B parameters, which prior work has treated as a matched set for studying scaling behavior~\citep{sinha2025illusiondiminishingreturnsmeasuring}. To support cost-aware evaluation, we define an effective price per million tokens $\pi(a_d)$ for each agent of size $d$, based on published API rates (see Appendix~\ref{appendix:cost}) and an observed average input-to-output token ratio of $4{:}1$. Under this convention, we obtain $\pi(a_{\text{4B}}) = \$0.05$, $\pi(a_{\text{8B}}) = \$0.09$, $\pi(a_{\text{14B}}) = \$0.16$, and $\pi(a_{\text{32B}}) = \$0.36$. Because model size and $\pi$ are monotonically aligned, we use \textit{smaller/cheaper} and \textit{larger/more expensive} interchangeably when discussing these particular agents. However, in figures we plot `price per million tokens' on the $x$-axis to emphasize the cost dimension explicitly. Note that we run all models with greedy decoding.

\section{Agent Performance vs. Task Complexity}\label{sec:performance-complexity}

We systematically evaluate Qwen3 agents of different sizes and costs on deep search and coding tasks, conditioning performance on task complexity as measured by $\tau(t)$ (Section~\ref{sec:datasets}). This setup enables direct comparison of agent performance as task demands increase. We measure performance via pass@1, scored via LLM-as-a-judge evaluation against ground-truth answers (see Appendix~\ref{appendix:are_setup}).

Across both domains, agents perform very similarly on the simplest tasks, as shown in Figure~\ref{fig:plots:a}. For deep search, the cheapest agent achieves about 87\% of the most expensive agent's pass@1 on tasks with $\tau(t) \le 0.1$; for coding, this relative performance is about 92\%. In this regime, the scaling curves are nearly flat: moving from cheaper/smaller to more expensive/larger agents yields only modest gains. As task complexity increases, the scaling curves gradually become steeper, and by the most complex tasks ($\tau(t)  \le 60$), the separation between agents is sharp. For deep search, the cheapest agent attains only 25\% of the most expensive agent's pass@1 on these tasks; for coding, it reaches just 17\%. In this long-horizon regime, performance is strongly stratified by model size and cost.

One might hope that, although larger agents are more expensive per token, they implicitly ``pay for themselves'' by solving tasks with shorter trajectories, for instance by requiring fewer reasoning steps, tool calls, or revisions. In practice, as shown in Figure~\ref{fig:plots:b}, we observe this pattern only for low-complexity tasks. As $\tau(t)$ increases, total token usage grows across all models, and larger agents do not consistently achieve shorter traces than smaller ones. Indeed, on many long-horizon instances, they incur comparable or greater token counts, and where shorter traces do occur, the savings are too small to offset the higher per-token cost. Thus, increased parametric capacity does not generally yield more token-efficient solutions on complex workloads, and higher per-token costs for larger agents are not naturally offset by reduced test-time compute.

In sum, cheaper agents are effective for tasks with low $\tau(t)$, but their limitations become starkly apparent as task demands intensify. More expensive agents appear indispensable for complex problems, yet deploying them universally squanders resources on tasks that do not require their power. The challenge, therefore, is to build systems that can dynamically allocate tasks to the right agent, achieving a better balance between resource efficiency and capability.

\section{Strategy Auctions}\label{sec:strategy-auctions}

Agentic pipelines commonly include a planning phase in which agents outline their intended approach before acting. These strategic plans encode task-relevant information such as decomposition strategies, tool selection, anticipated challenges, yet they are rarely leveraged beyond the agent that produced them. Our framework, \textsc{sale} (Figure~\ref{fig:main}), exploits this observation by casting strategic plans as bids in an auction framework. Specifically, given an environment $E$, a task $t$, and a heterogeneous group of agents $\mathcal{A} = \{a_i\}_{i=1}^{|\mathcal{A}|}$, each agent $a_i$ generates a strategy $s_{t,i}$ conditioned on $t$ and $E$ (we omit $E$ from the notation for brevity). We interpret $s_{t,i}$ as the ``bid'' of agent $a_i$ for task $t$, which is then used to compute both the cost and value of $a_i$ with respect to $t$, enabling model selection based on strategic intent rather than task description alone. Please see Appendix~\ref{appendix:are_setup} for the prompts used to obtain $s_{t,i}$. 

\textbf{Cost and Value Assignment.} \ \ Let $C_{t, i}$ and $V_{t, i}$ denote the cost and value, respectively, of deploying agent $a_i$ on task $t$. We estimate the \emph{cost} as
% \begingroup
% \setlength{\abovedisplayskip}{0pt}
% \setlength{\belowdisplayskip}{2pt}
\[
C_{t, i} = w_c \cdot \pi(a_i) \cdot |s_{t,i}|,
\]
% \endgroup
where $\pi(a_i)$ is the price per million tokens for agent $a_i$, $|s_{t,i}|$ is the length of $s_{t,i}$ in tokens, and $w_c$ is a tuned weight. We use strategy length as a cost signal motivated by two prior works. First, \citet{goebel2025llmreasoningmodelsreplaceclassical} show that plan (or strategy) length is correlated with final trace length, hence serving as a proxy for total inference cost. Second, execution reliability degrades with plan length: prior work finds that success rates decline as plans grow longer \citep{xiong2025deliberate}. Because failed executions nonetheless consume compute, longer plans entail higher expected cost, both through greater token usage and increased risk of wasted computation. We also show thorough ablations in Appendix~\ref{appendix:ablations} to validate this design choice.

We estimate the \emph{value} of agent $a_i$ for task $t$ as
\begingroup
\setlength{\abovedisplayskip}{5pt}
\setlength{\belowdisplayskip}{2pt}
\[
V_{t, i} = w_h \cdot H(s_{t,i}) + \sum_{a_j \in \mathcal{A}} w_j \cdot \gamma_j(s_{t,i}),
\]
\endgroup
where the weights $w_h$ and $w_j$ are tunable. Here, $H(s_{t,i})$ is the normalized entropy of $s_{t,i}$, computed as the mean per-token entropy from the generation log-probabilities, and each $\gamma_j(s_{t,i})$ is an integer judgment score in $\lbrace 0,1,...,5 \rbrace$ produced by agent $a_j$ in response to a fixed scoring prompt (Appendix~\ref{appendix:prompts}). The jury comprises all agents in $\mathcal{A}$, including $a_i$. Value thus combines two signals: intrinsic quality, captured by entropy, and extrinsic quality, captured by self-and-peer assessment.

The choice of entropy as a proxy for strategy value is motivated by extensive prior literature linking higher-entropy intermediate reasoning to greater informational content and reduced redundancy~\citep{chen2025aresmultimodaladaptivereasoning, cheng2025reasoningexplorationentropyperspective, li2025compressingchainofthoughtllmsstep, wang2025beyond}, and by work suggesting that prioritizing higher-entropy trajectories can be beneficial for planning~\citep{liu2024entropyreinforced} (validated by our ablations in Appendix~\ref{appendix:ablations}). The use of a mixed self-and-peer jury is motivated by work on LLM juries~\citep{badshah-sajjad-2025-reference, verga2024replacingjudgesjuriesevaluating} and evidence that combining self-evaluation with peer assessment yields more reliable judgments~\citep{mousavi2023ncritics}, consistent with recent findings that 0--5 Likert scoring without chain-of-thought yields strong alignment with human judgments on contemporary models including Qwen3~\citep{li2026gradingscaleimpactllmasajudge, wang-etal-2025-improving-llm-judge} (human validation in Appendix~\ref{app:jury_human}; prompts in Appendix~\ref{appendix:are_setup}). Ablations (Appendix~\ref{appendix:ablations}) confirm that excluding self-evaluation or reducing jury size degrades performance. 

The judge prompt is provided in Appendix~\ref{appendix:are_setup}; further details of the cost–value estimation appear in Appendix~\ref{appendix:minmax}.

\textbf{Winning Bid Selection.} \ \ Given the cost and value assignments described above, \textsc{sale} aims to select the agent whose strategy achieves the optimal trade-off between resource efficiency and expected performance. Our goal is thus to learn scoring weights that minimize the worst-case cost-minus-value over a training set of tasks. The cost-minus-value serves as a unified measure of an agent's desirability for a given task: lower costs improve resource efficiency, while higher values reflect stronger expected performance. By minimizing $C - V$, we favor agents that deliver high value at low cost.

Formally, we pose a min--max optimization over both the assignment variables $x$ and the scoring weights $w = (w_c, w_h, \{w_j\}_{a_j \in \mathcal{A}})$. Let $Q$ denote the maximum cost-minus-value over all tasks. The objective is
% \begingroup
% \setlength{\abovedisplayskip}{6pt}
% \setlength{\belowdisplayskip}{4pt}
\[
\min_{w, x, Q} \; Q \quad \text{s.t.} \quad
z_t \le Q \;\; \forall t, \quad
\sum_{a_i \in \mathcal{A}} x_{t,i} = 1 \;\; \forall t, \quad
w \in \mathbb{R}^{2 + |\mathcal{A}|},
\]
% \endgroup
where $z_t$ is the cost-minus-value of the chosen strategy for task $t$, and additional big-$M$ constraints are imposed (see Appendix~\ref{appendix:minmax} for details of the tuning process). The min–max formulation ensures robustness across the training distribution: by optimizing against the worst-case task, we guard against any single task receiving a disproportionately poor assignment.

At inference time, given the learned weights $w$, \textsc{sale} then applies the resulting scoring rule to route new tasks. For each task $t$, we introduce binary assignment variables $x_{t,i} \in \{0,1\}$ indicating whether $a_i$ is selected for task $t$, and define
% \begingroup
% \setlength{\abovedisplayskip}{0pt}
% \setlength{\belowdisplayskip}{2pt}
\[
z_t = \sum_{a_i \in \mathcal{A}} x_{t,i} \left( C_{t,i} - V_{t,i} \right).
\]
% \endgroup
Since exactly one agent is assigned per task, this reduces to $z_t = C_{t,\hat{i}(t)} - V_{t,\hat{i}(t)}$, where $\hat{i}(t) = \arg\max_i x_{t,i}$, thus assigning each task to the agent with the lowest cost-minus-value.

\textbf{Strategy Refinement from Auction Memory.} \ \ After each auction, we store all proposed strategies (both winning and losing bids) in a long-term memory bank $\mathcal{M}$. This enables a self-improvement mechanism in which cost-efficient agents that are not selected in the initial auction round can learn from $\mathcal{M}$, refine their initial strategies, and submit improved bids. Importantly, this refinement is \emph{opportunistic}: we do not use memory for all agents by default. Doing so would require every agent to produce both an initial and a memory-informed bid, increasing latency and token usage. Instead, we first collect baseline strategies without memory. If a cheap agent already wins the auction, no refinement is needed; otherwise, only agents cheaper than the provisional winner output a refined, memory-informed bid, preserving the cost-efficiency goals of \textsc{sale}.

Concretely, for each past task $t'$, we store a record $\mathcal{M}(t') = (t', \{s_{t', i}\}_{a_i \in \mathcal{A}}, y_{t'})$ where $\{s_{t', i}\}_{a_i \in \mathcal{A}}$ are the strategies proposed by all agents for $t'$ and $y_{t'}$ encodes the auction outcome, indicating which strategy won and which ones failed. Let $\mathcal{T}_\mathcal{M}$ denote the set of tasks for which we have stored memory records, i.e.\ $\mathcal{T}_\mathcal{M} = \{t' : \mathcal{M}(t') \in \mathcal{M}\}$. This memory accumulates a diverse set of strategic plans and outcomes, providing a rich resource for agents to learn from past experience. Given a new task $t$, the refinement procedure operates as follows (see Appendix~\ref{appendix:strategy_refinement} for a full algorithmic description):

\begin{enumerate}
    \item \textit{Initial bids.} Each agent $a_i \in \mathcal{A}$ submits an initial strategy $s_{t,i}$, and a provisional winner $\hat{i}(t) = \arg\min_i (C_{t,i} - V_{t,i})$ is selected.
    
    \item \textit{Shared memory retrieval.} For each agent $a_i$ cheaper than the provisional winner (i.e., $\pi(a_i) < \pi(a_{\hat{i}(t)})$), we retrieve a subset $\mathcal{M}_{t,i}$ of relevant past strategy pairs:
    % \begingroup
    % \setlength{\abovedisplayskip}{7pt}
    % \setlength{\belowdisplayskip}{7pt}
    \[
    \mathcal{M}_{t,i} = \Bigl\{
    \bigl(s^{\text{lose}}_{t'},\, s^{\text{win}}_{t'}\bigr)_i
    \;\Big|\;
    t' \in \
    \operatorname*{top\text{-}\mathit{\tilde{k}}}_{t' \in \mathcal{T}_{\mathcal{M}}} \
    \mathrm{sim}\bigl(t, t'\bigr)
    \Bigr\}, 
    \quad \tilde{k} = \text{min}(k, |\mathcal{T}_{\mathcal{M}}|),
    \]
    % \endgroup
    where $\mathrm{sim}$ denotes cosine similarity over text embeddings (see Appendix~\ref{sec:retrieval_hyperparams} for details) and each pair $\bigl(s^{\text{lose}}_{t'},\, s^{\text{win}}_{t'}\bigr)_i$ contains a losing and winning strategy for $t'$, with at least one proposed by $a_i$.
    
    \item \textit{Contrastive prompting.} Retrieved pairs are formatted using a contrastive prompt template (Appendix~\ref{appendix:prompts}) that encourages agents to learn from past auction outcomes.
    
    \item \textit{Reassignment.} Each eligible agent produces a refined bid $s^r_{t,i}$, which is scored to obtain updated cost $C^r_{t,i}$ and value $V^r_{t,i}$. If any refined bid improves upon the provisional winner's cost-value trade-off, the best such bid wins; otherwise, the provisional winner is retained:
    \begingroup
    \setlength{\abovedisplayskip}{6pt}
    \setlength{\belowdisplayskip}{1pt}
    \[
    i^*(t) = 
    \begin{cases}
    \displaystyle\argmin_{i:\, \pi(a_i) < \pi(a_{\hat{i}(t)})} \left( C^r_{t,i} - V^r_{t,i} \right) & \text{if any refined bid satisfies } C^r_{t,i} - V^r_{t,i} < C_{t,\hat{i}(t)} - V_{t,\hat{i}(t)}, \\[8pt]
    \hat{i}(t) & \text{otherwise.}
    \end{cases}
    \]
    \endgroup
    \item \textit{Execution.} After selecting $i^{*}(t)$, we execute agent $a_{i^{*}(t)}$ conditioned on $t$ and its winning strategy.
\end{enumerate}

It is worth noting that both jury scoring and strategy refinement incur only a small additional inference cost, on the order of a few hundred tokens, whereas executing the final agentic trace typically consumes tens of thousands to millions of tokens (see Figure~\ref{fig:plots:b}), depending on task complexity. In both token and wall-clock terms, the auction phase accounts for under 1\% of the total inference cost on average (Appendix~\ref{appendix:overhead}). \emph{Thus, the overhead introduced by the auction mechanism is negligible relative to the overall test-time compute.}

%%%%%%%%%%%%%%%%%%%%%%%%%%%%%%%%%%%%%%%%%%%%%%%%%
\begin{figure}[!t]
    \centering
    \includegraphics[width=\linewidth]{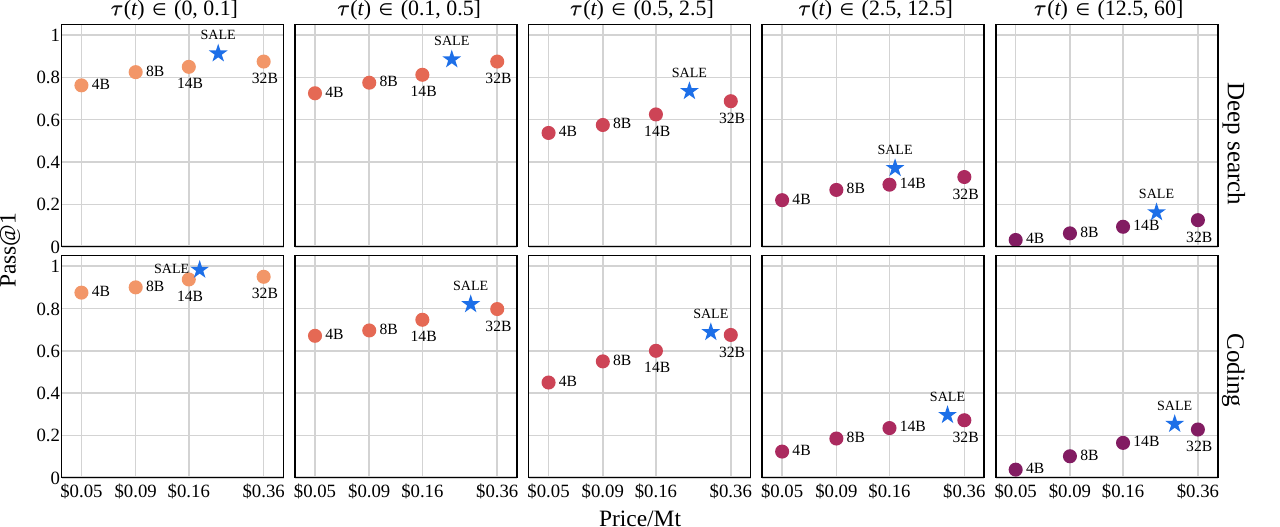}
    \caption{Performance–cost trade-offs for deep search (top row) and coding (bottom row) across task-complexity bins. At a given price per million tokens $\pi$, the \textsc{sale} auction ensemble consistently attains substantially higher pass@1 than would be predicted by the approximate linear scaling trend observed for individual Qwen3 agents, showing that it systematically exceeds the expected performance–cost curve.}
    \label{fig:results}
\end{figure}
%%%%%%%%%%%%%%%%%%%%%%%%%%%%%%%%%%%%%%%%%%%%%%%%%

\section{Results}\label{sec:results}

\begin{table*}[t]
    \centering
    \setlength{\tabcolsep}{1.45pt}%
    \renewcommand{\arraystretch}{1.1}%
    \scalebox{0.75}{%
    \begin{NiceTabular}{lc*{14}{c}}[cell-space-limits=2pt]
    \CodeBefore
      \rectanglecolor{metabg}{8-2}{8-16}
      \rectanglecolor{metabg}{14-2}{14-16}
    \Body
    \toprule
      \Block{2-1}{\small{\textbfrm{Task}}\\\small{\textbfrm{type}}} &
      \Block{2-1}{{$\tau(t)$}} &
      \Block{1-2}{\textbfrm{\footnotesize{Best single agent}}} & &
      \Block{1-2}{\textbfrm{\textsc{wtp}}} & &
      \Block{1-2}{\textbfrm{\textsc{carrot}}} & &
      \Block{1-2}{\footnotesize{\textbfrm{\textsc{TO}-Router}}} & &
      \Block{1-2}{\textbfrm{\footnotesize{FrugalGPT}}} & &
      \Block{1-2}{\textbfrm{\textsc{sale} \footnotesize{w/o memory}}} & &
      \Block{1-2}{\textbfrm{\textsc{sale}}} & \\
      \cmidrule(lr){3-4} \cmidrule(lr){5-6} \cmidrule(lr){7-8} \cmidrule(lr){9-10} \cmidrule(lr){11-12} \cmidrule(lr){13-14} \cmidrule(lr){15-16}
      & & \small{Pass@1($\uparrow$)} & \small{$\$$/Mt($\downarrow$)}
        & \small{Pass@1($\uparrow$)} & \small{$\$$/Mt($\downarrow$)}
        & \small{Pass@1($\uparrow$)} & \small{$\$$/Mt($\downarrow$)}
        & \small{Pass@1($\uparrow$)} & \small{$\$$/Mt($\downarrow$)}
        & \small{Pass@1($\uparrow$)} & \small{$\$$/Mt($\downarrow$)}
        & \small{Pass@1($\uparrow$)} & \small{$\$$/Mt($\downarrow$)}
        & \small{Pass@1($\uparrow$)} & \small{$\$$/Mt($\downarrow$)} \\
    \midrule
    \Block{6-1}{\small{\textit{Deep}}\\\small{\textit{search}}} & $\le0.1\phantom{0}$
      & 87.5 & 0.36
      & 83.8 & 0.32
      & 85.0 & 0.27
      & 86.3 & 0.28
      & 86.3 & 0.47
      & \textbfrm{91.3} & 0.24
      & \textbfrm{91.3}$_{0.0}$ & \textbfrm{0.22}$_{0.01}$ \\
    & $\le0.5\phantom{0}$
      & 87.5 & 0.36
      & 86.3 & 0.33
      & 86.3 & 0.28
      & 86.3 & 0.32
      & 81.3 & 0.48
      & 87.5 & 0.24
      & \textbfrm{88.5}$_{0.5}$ & \textbfrm{0.22}$_{0.01}$ \\
    & $\le2.5\phantom{0}$
      & 68.8 & 0.36
      & 67.5 & 0.31
      & 66.3 & 0.29
      & 67.5 & 0.34
      & 66.3 & 0.53
      & 72.5 & 0.25
      & \textbfrm{73.5}$_{1.2}$ & \textbfrm{0.23}$_{0.01}$ \\
    & $\le12.5$
      & 32.9 & 0.36
      & 34.2 & 0.32
      & 29.3 & 0.29
      & 32.9 & 0.36
      & 30.5 & 0.50
      & 35.4 & 0.19
      & \textbfrm{37.1}$_{1.8}$ & \textbfrm{0.17}$_{0.01}$ \\
    & $\le60\phantom{.}\phantom{0}$
      & 12.5 & 0.36
      & 9.4 & 0.31
      & 9.4 & 0.32
      & 12.5 & 0.36
      & 12.5 & 0.60
      & 15.6 & 0.26
      & \textbfrm{16.3}$_{1.3}$ & \textbfrm{0.23}$_{0.02}$ \\
    & All
      & 63.8 & 0.36
      & 62.4 & 0.32
      & 61.3 & 0.28
      & 63.0 & 0.33
      & 61.0 & 0.51
      & 66.4 & 0.24
      & \textbfrm{67.3}$_{0.5}$ & \textbfrm{0.21}$_{0.00}$ \\
    \midrule
    \Block{6-1}{\small{\textit{Coding}}} & $\le0.1\phantom{0}$
      & 95.0 & 0.36
      & 93.8 & \textbfrm{0.16} 
      & 95.0 & 0.36
      & 95.0 & 0.36
      & 97.5 & 0.39
      & 97.5 & 0.22
      & \textbfrm{98.3}$_{1.0}$ & 0.18$_{0.00}$ \\
    & $\le0.5\phantom{0}$
      & 79.7 & 0.36
      & 76.0 & \textbfrm{0.15}
      & \textbfrm{82.3} & 0.25
      & 79.7 & 0.36
      & 69.6 & 0.61
      & \textbfrm{82.3} & 0.28
      & 82.0$_{0.5}$ & 0.27$_{0.01}$ \\
    & $\le2.5\phantom{0}$
      & 67.5 & 0.36
      & 60.0 & \textbfrm{0.15}
      & 60.0 & 0.26
      & 67.5 & 0.36
      & 56.3 & 0.61
      & 68.8 & 0.31
      & \textbfrm{69.0}$_{0.5}$ & 0.29$_{0.00}$ \\
    & $\le12.5$
      & 27.2 & 0.36
      & 14.8 & \textbfrm{0.05}
      & 27.2 & 0.36
      & 27.2 & 0.36
      & 18.5 & 0.61
      & 27.2 & 0.32
      & \textbfrm{30.4}$_{2.2}$ & 0.30$_{0.02}$ \\
    & $\le60\phantom{.}\phantom{0}$
      & 22.8 & 0.36
      & 6.3 & \textbfrm{0.05}
      & 21.5 & 0.35
      & 22.8 & 0.36
      & 10.1 & 0.61
      & 24.1 & 0.31
      & \textbfrm{26.1}$_{2.4}$ & 0.29$_{0.01}$ \\
    & All
      & 58.4 & 0.36
      & 50.1 & \textbfrm{0.11}
      & 57.1 & 0.31
      & 58.4 & 0.36
      & 50.4 & 0.57
      & 59.9 & 0.27
      & \textbfrm{61.1}$_{0.6}$ & 0.27$_{0.00}$ \\
    \bottomrule
    \end{NiceTabular}
    }
    \caption{
    Deep search and coding performance (pass@1) and price per million tokens ($\$$/Mt) across task-complexity bins. We compare \textsc{sale} with the best single agent, the Willingness-to-Pay router (WTP), the TensorOpera Router (TO-Router), FrugalGPT, and an ablated variant of \textsc{sale} without memory-based self-refinement (\textsc{sale} w/o memory). For \textsc{sale}, we report five runs with distinct randomized test-set orders, with standard deviations shown as subscripts.}
    \label{tab:results}
\end{table*}

We run \textsc{sale} on the full \textsc{HST-Bench} test set, containing tasks from all complexity levels interleaved in random order, and only partition results into complexity bins for analysis. We use greedy decoding in all runs. For the single-model baselines we report results from a single run. For \textsc{sale}, however, task order matters because its auction memory for strategy refinement evolves online. Following established practice for order-sensitive evaluation~\citep{fan2026meiqamemoryenhancedimagequality, 10.5555/3737916.3741314}, we thus report all \textsc{sale} metrics as averages over five independent random permutations of the full test set.

Figure \ref{fig:results} summarizes the performance–cost trade-offs for deep search (top row) and coding (bottom row) across all five task-complexity bins, plotting pass@1 against price per million tokens for individual Qwen3 agents and for the \textsc{sale} ensemble. Detailed numerical results for each bin, along with additional baselines described below, appear in Table~\ref{tab:results}. For deep search, \textsc{sale} exceeds the best single agent's pass@1 on the lowest-complexity tasks while operating at a lower effective price per million tokens (39\% cost reduction, 3.8 pass@1 gain). On medium-complexity tasks, it improves pass@1 by between 1 and 4.7 percentage points over the best single agent, while reducing cost by 36–53\%. On the most complex tasks, it still outperforms the best agent by 3.8 points while lowering cost by 36\%. For coding, \textsc{sale} likewise beats the best single agent on the simplest tasks (50\% cost reduction, 3.3 pass@1 gain), and on medium-complexity tasks it achieves 1.5–3.2 pass@1 improvement over the best single agent while achieving cost reductions of 17–25\%. On the most complex coding tasks, it improves pass@1 by 3.3 points at 19\% lower cost than the best single agent. Across both domains and all complexity levels, the auction ensemble dominates the single-agent Pareto frontier, i.e. no fixed model attains higher pass@1 at equal or lower price per million tokens, indicating that strategy-based routing with self-improvement yields strictly better performance–cost trade-offs than any single model size. All reported improvements are statistically significant even accounting for run-to-run variance (see Appendix~\ref{appendix:tests}), confirming that these gains are robust across random orderings.

\para{Comparison with Existing Routers.}
\textsc{sale} is deliberately lightweight: it leverages agents' existing strategic-planning abilities through a simple, low-dimensional scoring function with few global weights, rather than learning a separate high-capacity routing model. This design makes \textsc{sale} directly applicable to off-the-shelf agents with minimal tuning, while explicitly accounting for agent cost rather than optimizing for performance alone. We compare against four baseline routers. Willingness-to-Pay (WTP)~\citep{hu2024routerbench} uses nearest-neighbor retrieval over pretrained sentence embeddings to predict the model with the best performance–cost trade-off given a task description. \textsc{CARROT}~\citep{somerstep2025carrotcostawarerate} fine-tunes an encoder to jointly estimate per-query cost and accuracy, routing to the model that minimizes a weighted combination of both. TensorOpera Router (TO-Router)~\citep{stripelis-etal-2024-tensoropera} trains a learned task classifier, i.e. a language model encoder fine-tuned on soft performance labels, to predict the best-performing model from the task description alone, without explicitly modeling cost. All three are \emph{predictive} routers that select an agent before execution. FrugalGPT~\citep{chen2024frugalgpt}, by contrast, is a \emph{non-predictive} cascade that executes agent trajectories sequentially until a fine-tuned encoder model accepts a response, potentially running multiple traces per task. For fair comparison, we train all baselines on the same training split used to set \textsc{sale}'s scoring weights (hyperparameters in Appendix~\ref{appendix:baselines}). Unlike the baselines, \textsc{sale} requires no learned predictor, only a fixed scoring form with a small number of tunable weights. 

As shown in Table~\ref{tab:results}, WTP yields modest cost reductions on deep-search tasks (11\% on average) but slightly underperforms the best single agent at most complexity levels. On coding, WTP achieves large savings but increasingly sacrifices pass@1 as complexity rises, dropping to 6.3\% at the highest complexity versus 22.8\% for the best single agent. CARROT strikes a better balance, reducing cost by 22\% on deep search and 14\% on coding while incurring only small accuracy drops overall, though it still underperforms \textsc{sale} on both metrics. TO-Router tends to default to the strongest agent, so both performance and cost remain close to the single-agent baseline. FrugalGPT matches or slightly exceeds the best single agent on low-complexity tasks (e.g., 97.5\% vs.\ 95.0\% on simple coding), but as complexity grows, its pass@1 declines sharply while its average spend increases, rising to 0.61\$/Mt on coding versus the 0.36\$/Mt of the best agent. This exposes a limitation of non-predictive routing in agentic settings: not only does the cascade potentially incur the cost of multiple full traces, but the scoring model also struggles to assess answer reliability when the mapping from task to solution is indirect and mediated by long, complex trajectories. In contrast, \textsc{sale} maintains or improves pass@1 relative to the best single agent across all complexity levels while reducing cost by 36--53\% on deep search and 17--50\% on coding, thereby advancing the performance--cost Pareto frontier more consistently than any alternative router (Figure~\ref{fig:pareto-scatter}). Comparisons to an oracle router are provided in Appendix~\ref{appendix:oracle}.

\para{Ablation.}
To isolate the contribution of self-refinement, we ablate the memory-based stage and evaluate a variant of \textsc{sale} that performs only a single auction: agents bid with strategic plans, and tasks are assigned by minimizing cost-minus-value over these bids. Even without memory, this router either matches or improves the average pass@1 of the best single agent while always reducing effective cost across all task-complexity bins, indicating that strategy-based routing provides a clear benefit. Comparing this ablated variant to the full \textsc{sale} system (Table~\ref{tab:results}) shows that the memory mechanism consistently improves the trade-off: incorporating past auction outcomes either reduces cost at similar accuracy or jointly improves pass@1 and cost, thus pushing the Pareto frontier further outward, and remains effective even under aggressive pruning of memory size and retention window (Appendix \ref{appendix:memory_robustness}). Further ablations (Appendix~\ref{appendix:ablations}) study the effect of removing each term from the cost–value function, finding that all contribute meaningfully to performance, and show that the jury's diversity provides a regularizing effect that no single judge or smaller jury subset can replicate.

\section{Analysis}

%%%%%%%%%%%%%%%%%%%%%%%%%%%%%%%%%%%%%%%%%%%%%%%%%%
\begin{figure}[t]
    \centering
    \begin{minipage}[!b]{0.4975\textwidth}
        \centering
        \includegraphics[width=\linewidth]{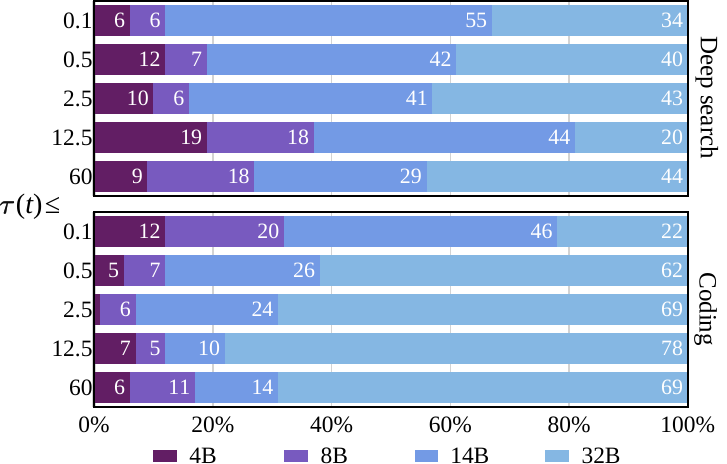}
        \caption{\textsc{sale}'s average workload allocation across the 4B, 8B, 14B, and 32B agents for deep search (top) and coding (bottom) tasks, stratified by task complexity $\tau(t)$. Bar labels indicate the share of all tasks assigned to each agent.}
        \label{fig:alloc}
    \end{minipage}%
    \hfill
    \begin{minipage}[!b]{0.489\textwidth}
        \centering
        \setlength{\tabcolsep}{4.5pt}%
\renewcommand{\arraystretch}{1.1}%
\makebox[\linewidth][c]{%
\scalebox{0.755}{%
\begin{NiceTabular}{lc*{8}{c}}[cell-space-limits=2pt]
\toprule
  \Block{2-1}{\textbfrm{Task}\\ \textbfrm{type}} &
  \Block{2-1}{{$\tau(t)$}} &
  \Block{1-4}{\textbfrm{\textsc{sale} \small{w/o memory}}} & & & &
  \Block{1-4}{\textbfrm{\textsc{sale}}} & & & \\
  \cmidrule(lr){3-6} \cmidrule(lr){7-10}
  & & 4B & 8B & 14B & 32B
    & 4B & 8B & 14B & 32B \\
\midrule
\Block{5-1}{\textit{Deep}\\\textit{search}} & $\le0.1\phantom{0}$
  & 22.0 & 23.9 & 24.1 & 30.0 & 25.6 & 24.2 & 26.3 & 23.9 \\
& $\le0.5\phantom{0}$
  & 21.7 & 23.6 & 24.5 & 30.2 & 24.1 & 25.4 & 25.2 & 24.8 \\
& $\le2.5\phantom{0}$
  & 19.8 & 21.6 & 24.7 & 33.9 & 23.4 & 26.7 & 25.7 & 24.2 \\
& $\le12.5$
  & 10.9 & 23.6 & 29.3 & 36.2 & 20.0 & 22.6 & 27.3 & 30.0 \\
& $\le60\phantom{.}\phantom{0}$
  & 0.0 & 13.9 & 38.9 & 47.2 & 7.1 & 24.2 & 35.6 & 33.1 \\
\midrule
\Block{5-1}{\textit{Coding}} & $\le0.1\phantom{0}$
  & 23.2 & 23.8 & 24.3 & 28.7 & 26.2 & 24.5 & 23.6 & 25.8 \\
& $\le0.5\phantom{0}$
  & 21.4 & 23.2 & 24.0 & 31.4 & 19.1 & 24.8 & 26.4 & 29.7 \\
& $\le2.5\phantom{0}$
  & 16.2 & 20.5 & 28.0 & 35.3 & 25.0 & 25.0 & 25.0 & 25.0 \\
& $\le12.5$
  & 3.8 & 17.4 & 22.0 & 56.8 & 13.2 & 27.3 & 35.8 & 23.7 \\
& $\le60\phantom{.}\phantom{0}$
  & 0.0 & 9.6 & 24.6 & 65.8 & 9.3 & 26.5 & 33.8 & 30.4 \\
\bottomrule
\end{NiceTabular}
}}
        \captionof{table}{Average Shapley values of each agent's marginal contribution to the overall system, with and without memory-based self-refinement, across task types and complexity bins. Values are normalized to sum to 100 and reported as each agent's percentage share of the total contribution.}
        \label{tab:shapley_results}
    \end{minipage}
\end{figure}
%%%%%%%%%%%%%%%%%%%%%%%%%%%%%%%%%%%%%%%%%%%%%%%%%%%

\subsection{Agent Allocation}\label{sec:alloc} 
Figure~\ref{fig:alloc} shows how \textsc{sale} allocates workload across agents of different sizes and task-complexity bins for both deep search and coding tasks. For deep search, across all bins the 32B agent's share ranges from 20\% to 44\%, with the remaining workload routed to smaller agents. The 14B agent handles 29--55\% of tasks, while the 4B and 8B agents together account for approximately 12--37\%. Even in the highest-complexity bin, smaller agents (4B and 8B) still process nearly 30\% of tasks, indicating that \textsc{sale} substantially reduces reliance on the largest model while matching or exceeding its accuracy. For coding tasks, the 32B agent is still used for a substantial proportion of the workload in all but the easiest complexity bin, where its share is only 22\%. The smaller 4B and 8B agents together account for between 7\% and 32\% of coding queries across bins, again demonstrating that a substantial fraction of work can be offloaded from the largest model.

\subsection{Agent Contributions via Shapley Values}

Given the cooperative nature of \textsc{sale}, where agents propose solution strategies and influence one another through jury votes and shared memory, it is important to quantify each agent's overall contribution to the system, both when selected for final inference and through indirect effects on other agents. Formally, we define a cooperative game $(\mathcal{A}, \nu)$ in which the players are agents and the value of a coalition $\mathcal{A}' \subseteq \mathcal{A}$ is the performance achieved by running \textsc{sale} with participation restricted to $\mathcal{A}'$. Following established uses in the analysis of auction outcomes~\citep{VANESSEN2021105315, LINDSAY2018170}, we then quantify each agent's average marginal contribution using Shapley values~\citep{10.5555/3295222.3295230}, where $\phi_i$ denotes agent $i$'s average marginal contribution to ensemble accuracy across all possible coalitions:
% \begingroup
% \setlength{\abovedisplayskip}{8pt}
% \setlength{\belowdisplayskip}{6pt}
\[
\phi_i = \sum_{
\mathcal{A}' \subseteq \mathcal{A} \setminus \{i\}} \frac{|\mathcal{A}'|!\,(|\mathcal{A}|-|
\mathcal{A}'|-1)!}{|\mathcal{A}|!} \left[ \nu(
\mathcal{A}' \cup \{i\}) - \nu(
\mathcal{A}') \right].
\]
% \endgroup
Here, $\mathcal{A}$ is the set of agents, $\nu(\mathcal{A}')$ is the expected utility induced by our cost--value selection mechanism when only agents in $\mathcal{A}'$ participate in all roles (bidding, jury scoring, and memory-based refinement). The weighting is the probability that coalition $\mathcal{A}'$ precedes agent $i$ in a random ordering. Table~\ref{tab:shapley_results} reports these values with and without memory-based self-refinement. We observe that in the without-memory setting, the largest agent has the highest Shapley values across all complexity bins and task domains, even though it is not the most-selected agent in all settings (Figure~\ref{fig:alloc}). This indicates that \textsc{sale} benefits from the largest agent's contribution in jury scoring, and yet saves inference costs by choosing smaller agents for solving the task. Further, we observe that introducing memory consistently lowers the 32B agent's Shapley value across task domains and complexity bins, while the smaller 4B and 8B agents generally gain marginal contribution, especially on higher-complexity coding tasks. 

It is worth noting that, when computing Shapley values, we remove the target agent from all roles in \textsc{sale}, including the candidate pool, the jury, and the memory bank. Thus, an agent's Shapley value captures its total contribution to the system, i.e. its ability to contribute effective strategies directly as well as its indirect impact. This explains, for example, why the 4B model can attain a relatively high Shapley value on coding tasks despite being selected infrequently for final inference: it still improves the ensemble through judgement and memory contributions. Hence the distributions in Figure~\ref{fig:alloc} and Table~\ref{tab:shapley_results} need not correlate.

%%%%%%%%%%%%%%%%%%%%%%%%%%%%%%%%%%%%%%%%%%%%%%%%%%
\begin{figure}[t]
    \centering
    \includegraphics[width=\linewidth]{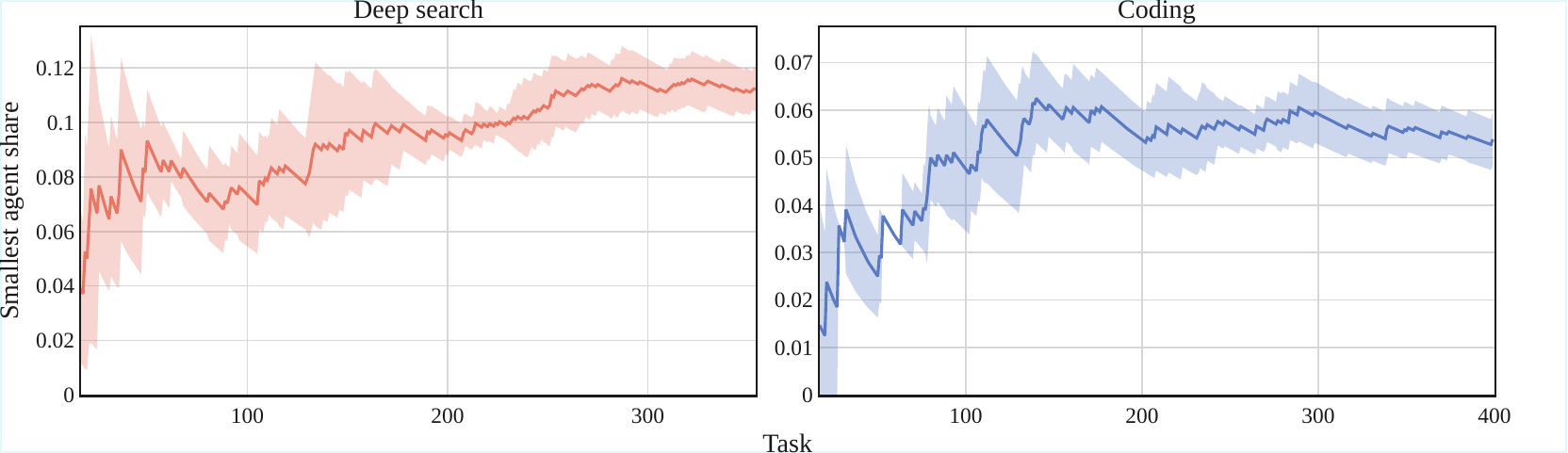}
    \caption{Cumulative share of tasks routed to the smallest agent over time. Solid lines show the mean across 5 runs with randomized task orderings; shading denotes ±1 standard deviation. An upward trend indicates that the local selection rate exceeds the historical average, reflecting increased delegation to the smallest agent as auction history accumulates. Short-term fluctuations reflect the random interleaving of tasks of varying complexity.}
    \label{fig:time}
\end{figure}
%%%%%%%%%%%%%%%%%%%%%%%%%%%%%%%%%%%%%%%%%%%%%%%%%%

\subsection{Smallest Agent Selection Over Time}
Beyond static routing, \textsc{SALE} enables smaller agents to progressively ``scale up'' by learning from auction feedback, effectively expanding their competitiveness over time. We test this by tracking how often the smallest (4B) agent is selected as the final executor over time. Figure~\ref{fig:time} plots the cumulative selection rate of the 4B agent: the running fraction of all tasks processed so far that were ultimately routed to this agent. For deep search, this cumulative share increases from 3.7\% in the early portion of the evaluation to 11.1\% by the final tasks, approximately a threefold increase. For coding, the cumulative share grows from 1.4\% to 5.3\%, nearly a fourfold increase, with most of the rise concentrated in the first $\sim$150 tasks. Overall, the proportion of workload handled by the smallest agent increases over time as a growing memory bank yields increasingly relevant auction feedback that progressively scales up the practical contribution of small agents. This temporal dynamic distinguishes \textsc{SALE} from conventional routers, which treat model selection as a stationary mapping from task features to agents. Similar plots for all agents are shown in Appendix~\ref{appendix:time}.

\subsection{Qualitative Analysis of Refined Strategies}
Consistent with recent observations on reusable behaviors in LLM reasoning~\citep{didolkar2025metacognitivereuseturningrecurring}, we find that memory-guided refinement systematically grounds strategies in auction feedback by reusing recurrent structural elements from past winning bids on similar tasks. For search tasks, refined strategies more frequently (i) explicitly mention tools and their arguments, (ii) impose tighter search-space constraints to specified reputable sources, and (iii) add intermediate cross-reference checks for ambiguity or inconsistencies. For coding tasks, refinement more often (i) specifies function and helper-function names and arguments precisely, (ii) maintains stronger alignment with the final objective (e.g., returning code rather than its runtime output), and (iii) performs systematic testing, including both the use of provided tests and the construction of additional test cases and edge-case checks. Across both domains, refined strategies also adopt a clearer layout and step organization. Table~\ref{tab:refinement_patterns} reports the proportion of selected refined strategies in which each pattern appears, and Appendix~\ref{appendix:examples} shows representative examples illustrating all behaviors.

\subsection{Complementary Failure Modes}
For \textsc{sale} to outperform any single agent (as shown in Table~\ref{tab:results}), there must exist tasks where smaller agents succeed and the largest agent fails, including cases where smaller agents have improved through prior auction feedback. If the largest agent's errors were a strict subset of smaller agents' errors, routing would offer no accuracy benefit and only cost savings. In Appendix~\ref{appendix:failure_modes}, we investigate such complementary failure modes, providing qualitative grounding for the quantitative improvements observed in Section~\ref{sec:results}. 

Across task types and complexity levels, we find that failures of the largest agent often stem less from a lack of underlying capability and more from how that capability is exercised. In particular, the largest agent is more prone to overconfident behavior: it sometimes bypasses available tools in favor of parametric recall, over-engineers straightforward problems, or skips basic verification steps. Smaller agents, by contrast, more often adhere to simpler strategies, lean more heavily on tools, and perform explicit checks. Crucially, these tendencies are visible in the initial strategic plans agents submit before any trajectory is executed, implying that the auction has access to a reliable signal for predicting failure-mode divergence at bid time, without needing to run trajectories to completion. That said, these patterns do not mean that smaller agents are generally more accurate: the largest agent remains superior on aggregate, especially at higher complexities (Section~\ref{sec:performance-complexity}). They do, however, suggest a consistent complementarity in failure modes: some tasks are handled better by simpler, tool-centric strategies than by the largest agent's approach. \textsc{sale} is designed to exploit this by using these early differences in strategy as the main signal for how work should be divided.

\section{Conclusion}
We investigated how task complexity affects the relative performance of small and large language-model agents, and how to allocate work across them efficiently. On deep search and coding tasks spanning multiple horizons, smaller agents perform comparably to larger ones on simple instances but fall substantially behind on more complex ones. This suggests that small agents alone are insufficient for complex workloads, whereas always defaulting to the largest model ignores substantial opportunities for efficiency.

To address this, we proposed \textsc{sale}, a strategy-auction framework where heterogeneous agents bid with short strategic plans, are scored by a cost–value objective, and refine bids using shared auction memory. \textsc{sale} runs entirely at test time on off-the-shelf models, without training a separate router, and introduces only a negligible additional inference overhead beyond executing the final trajectory. Empirically, across both deep search and coding domains, it improves the pass@1 of the strongest single agent while reducing cost and shifting a substantial fraction of workload away from the largest model, adaptively improving smaller agents over time so they can shoulder more of the work and maintaining these gains even on the most complex tasks. Our findings indicate that scaling individual models is only one axis of progress, and that \emph{how} we structure and coordinate agents can be an equally powerful lever. Rather than treating capability as a property of a single, ever-larger model, \textsc{sale} treats it as an emergent property of a system that allocates work, prices compute, and lets agents learn from each other and adapt their strategies over time. This points toward a view of agentic AI where advances are driven not just by stronger models, but by the coordination mechanisms and division of labor that bind them into effective and adaptive systems.

\section*{Limitations}
We study two domains, deep search and coding, that are canonical benchmarks in the agentic AI literature and exercise complementary capabilities: search requires retrieval and multi-step exploration, while coding emphasizes generation and logical reasoning. Together they cover diverse agentic patterns with distinct tool-use profiles and offer objective, automatable evaluation metrics. That said, they do not span all applications of agentic systems; future work can apply \textsc{sale} to additional task families (e.g., data analysis or long-form report writing) to test how broadly our findings generalize.

On the modeling side, we work with Qwen3 models from 4B to 32B parameters. We focus on a single model family for our complexity-scaling analysis because cross-family comparisons would confound scale effects with architecture, tokenizer, and training recipe differences, making it impossible to isolate how model size mediates performance as task complexity grows. Qwen3 is the only contemporary open-weight suite offering a dense, consistently-trained ladder (4B $\rightarrow$ 8B $\rightarrow$ 14B $\rightarrow$ 32B) suitable for this methodology; in contrast, other open-weight families offer narrower size ranges, larger gaps between sizes, or mix architectures across scales, and closed-source models do not disclose parameter counts. Importantly, \textsc{sale} is model-agnostic: the auction mechanism and cost--value objectives do not depend on model-specific properties, so our findings about when to route to larger versus smaller models should transfer qualitatively to other families. Indeed, we provide preliminary empirical support for this in Appendix~\ref{app:cross_family}, where the auction mechanism produces complexity-sensitive allocations on the Gemma 3 family analogous to those observed for Qwen3. As for size, the 4B-32B range already yields a clear task--complexity-dependent performance gap and a roughly 8$\times$ cost spread, but it still sits below the largest frontier models. That said, much of the empirical literature on scaling behavior draws inferences from trends observed across multiple smaller, systematically spaced model sizes; our controlled size ladder is designed to support that style of analysis by isolating scale while holding other factors as constant as possible. Evaluating \textsc{sale} with substantially larger models (e.g., 70B+) would be a useful extension to assess how the cost--value trade-offs behave when agents are even more capable and more expensive. While the controlled ladder suits our scaling analysis, \textsc{sale} is designed to coordinate heterogeneous pools; we provide preliminary evidence of this in Appendix~\ref{app:cross_family}, where Llama 3.1 and Ministral 3 agents are added to a Qwen3 pool as bidders and receive sensible allocations without retraining the MILP weights.

Our auction memory bank grows linearly with the number of tasks. In our experiments, this remained tractable, and our ablations indicate that refinement gains remain robust even under aggressive memory size and retention window pruning. We did not, however, study retrieval at much larger scales, across model-version upgrades, or under task-distribution shift where the memory may contain strategies from earlier distributions. For such settings, approximate nearest-neighbor indexing is recommended to scale retrieval to larger memories, while utility-based retention can prune stale or low-value records. We note that both integrate with our retrieval step without modifying the auction mechanism.

Finally, our cost accounting focuses on language-model tokens and does not explicitly price tool calls. This is appropriate for our setup where token costs dominate, however, the modular cost function can be extended for deployments where tool costs are non-negligible. For instance, the cost term can be augmented with tool-related features extracted directly from the strategy, or per-token price can be replaced with a blended cost-per-step metric incorporating empirical tool latency. Neither extension requires changes to the auction mechanism; we provide a preliminary demonstration of this in Appendix~\ref{app:tool_latency}, where adding a tool-latency penalty to the cost term progressively shifts workload away from agents that generate the most tool calls.

\section*{Broader Impacts}
This work contributes to the understanding of how task complexity mediates the effectiveness of language-model agents and proposes a coordination mechanism for efficiently allocating work across heterogeneous models. We discuss several dimensions of potential impact below.

\vspace{6pt}
\para{On the marketplace metaphor.} We employ auction-based coordination and freelancer-marketplace terminology as conceptual tools for organizing AI agents, not as prescriptions for labor markets. The analogy is strictly methodological: it motivates a mechanism design perspective on multi-agent systems in which bids, competition, and learning from feedback govern allocation among software components. Our framework neither models human workers nor recommends substituting them with AI. We emphasize that the ``agents'' in our system are language models executing computational tasks within sandboxed research environments, and the ``marketplace'' is a metaphor for principled resource allocation; distinct from, and not intended to inform, policies regarding human employment.

\vspace{6pt}
\para{Efficiency and environmental considerations.} By reducing reliance on the largest agent by approximately 52\% and lowering overall inference cost by 35\%, \textsc{sale} promotes more efficient use of computational resources. Given the substantial energy footprint of large-scale language-model inference, mechanisms that route simpler tasks to smaller models without degrading performance can contribute to more sustainable AI deployment.

\vspace{6pt}
\para{Democratizing access to capable AI.} Our findings suggest that carefully coordinated ensembles of smaller, less expensive models can match or exceed the performance of larger models on heterogeneous workloads. If such coordination mechanisms become practical, they may lower the cost barrier to deploying capable agentic systems, potentially broadening access beyond well-resourced institutions. We should note, however, that the benefits of efficiency gains are not automatically equitably distributed, and deployment contexts will shape who ultimately benefits.

\vspace{6pt}
\para{Dual-use considerations.} As with most advances in AI capability and efficiency, our contributions are dual-use. More efficient agentic systems could be applied to beneficial domains (e.g., scientific research, accessibility tools) or to applications with negative societal consequences. In particular, mechanisms that lower the cost of agentic capability, whether through cheaper base models or, as here, through more efficient coordination, can amplify the scale at which both beneficial and harmful applications become feasible. We do not foresee unique risks introduced by \textsc{sale} beyond those already present in the underlying language models and agentic frameworks; our contribution is to coordination, not to novel capabilities. Nonetheless, we encourage practitioners to consider deployment contexts carefully.

\vspace{24pt}
\section*{Acknowledgments}
We would like to thank Enrique Alfonseca, Misha Bilenko, Cheng Zhang, Yue Zhang, Igor Tufanov, Virginie Do, Emilien Garreau, Amine Benhalloum, Mathieu Rita, Romain Froger, Lovish Madaan, Anirudh Goyal, Iva Simon-Bubalo, Cindy Lee, Derek George Chan, Jordan Ward, and Joshua Lim for their valuable technical guidance and support in the development of this work. We are also grateful to Parag Jain, Amar Budhiraja, Graeme Nail, Thomas Scialom, Grégoire Mialon, Marin Vlastelica, Jenny Zhang, Md Rifat Arefin, Ulyana Piterbarg, Shashwat Goel, Philipp Mondorf, and Dulhan Jayalath for insightful discussions that helped shape and refine this research.

\clearpage
\newpage

\bibliographystyle{assets/plainnat}
\bibliography{paper}

\clearpage
\newpage
\beginappendix

\section{Dataset}\label{appendix:data}
We describe here how we construct \textsc{HST-Bench}, its composition across source datasets and complexity bins, and the details of the human solution–time annotation protocol.

\subsection{Data Composition}
\textsc{HST-Bench} is built from existing open-source benchmarks spanning deep search and coding. Concretely, we draw from SimpleQA~\citep{wei2024measuringshortformfactualitylarge}, PopQA~\citep{mallen-etal-2023-trust}, HotpotQA~\citep{yang-etal-2018-hotpotqa}, GAIA~\citep{mialon2024gaia}, Humanity's Last Exam (HLE)~\citep{phan2025humanitysexam}, MBPP~\citep{austin2021programsynthesislargelanguage}, and LeetCode~\citep{xia2025leetcodedatasettemporaldatasetrobust}. In addition, we construct a small corpus of multiple-choice coding questions, which we refer to as Coding-MCQ (see example questions in Appendix~\ref{appendix:cmcq}, to better populate the lowest-complexity bin for coding tasks. We randomly sample instances from the official test splits of each benchmark. In order to ensure label quality, we validate all samples, discarding and replacing those for which it is not possible to derive the provided ground-truth answer from the question. For HLE, we restrict to chemistry and biology questions that have been validated by domain experts~\citep{white2025hlewrong}. For GAIA, we sample from the validation split, which includes human solution times collected under comparable experimental conditions (timed, independent problem-solving by proficient users) and verified by the original authors; we directly reuse these annotations. After sampling, we annotate and aggregate human solution times for each instance and assign it to one of the five non-overlapping complexity bins defined in Section~\ref{sec:datasets}, based on its average human solution time. Table~\ref{tab:data_origin} reports, for each complexity bin, how many \textsc{HST-Bench} instances originate from each source dataset. This reveals a shift from short-form factual QA and \textsc{Coding-MCQ} in the lower-complexity bins toward tasks demanding extended agentic workflows: multi-source information retrieval, cross-referencing, and synthesis for reasoning benchmarks (e.g., HotpotQA, GAIA, HLE), and iterative implementation with intermediate testing and debugging for coding problems (e.g., LeetCode `Hard') in the higher-complexity bins.

\begin{table}[H]
\centering
\small
\setlength{\tabcolsep}{18pt} 
\renewcommand{\arraystretch}{1}%
\begin{tabular}{llclc}
\toprule
\textbf{Domain} & \textbf{Complexity bin} & \textbf{\# tasks} & \multicolumn{2}{c}{\textbf{Source}} \\
\cmidrule(l){4-5}
& & & \textbf{Dataset} & \textbf{Percentage} \\
\midrule
\multirow{11}{*}{\textit{Deep search}}
  & \multirow{3}{*}{$0 < \tau(t) \le 0.1$}
  & \multirow{3}{*}{80}
  & SimpleQA & 38\% \\
  & & & PopQA & 50\% \\
  & & & HotpotQA & 13\% \\
\cmidrule(l){2-5}
  & \multirow{3}{*}{$0.1 < \tau(t) \le 0.5$}
  & \multirow{3}{*}{80}
  & SimpleQA & 8\% \\
  & & & PopQA & 5\% \\
  & & & HotpotQA & 88\% \\
\cmidrule(l){2-5}
  & \multirow{2}{*}{$0.5 < \tau(t) \le 2.5$}
  & \multirow{2}{*}{80}
  & HotpotQA & 98\% \\
  & & & HLE & 3\% \\
\cmidrule(l){2-5}
  & \multirow{2}{*}{$2.5 < \tau(t) \le 12.5$}
  & \multirow{2}{*}{82}
  & HotpotQA & 2\% \\
  & & & GAIA & 98\% \\
\cmidrule(l){2-5}
  & $12.5 < \tau(t) \le 60$
  & 32
  & HLE & 100\% \\
\midrule
\multirow{9}{*}{\textit{Coding}}
  & $0 < \tau(t) \le 0.1$
  & 80
  & Coding-MCQ & 100\% \\
\cmidrule(l){2-5}
  & $0.1 < \tau(t) \le 0.5$
  & 79
  & MBPP & 100\% \\
\cmidrule(l){2-5}
  & \multirow{2}{*}{$0.5 < \tau(t) \le 2.5$}
  & \multirow{2}{*}{80}
  & MBPP & 99\% \\
  & & & LeetCode (Medium) & 1\% \\
\cmidrule(l){2-5}
  & \multirow{2}{*}{$2.5 < \tau(t) \le 12.5$}
  & \multirow{2}{*}{81}
  & MBPP & 2\% \\
  & & & LeetCode (Medium) & 98\% \\
\cmidrule(l){2-5}
  & $12.5 < \tau(t) \le 60$
  & 79
  & LeetCode (Hard) & 100\% \\
\bottomrule
\end{tabular}
\caption{Composition of \textsc{HST-Bench} by complexity bin (grouped by average human solution time $\tau(t)$, in minutes). We report the percentage of instances contributed from each source dataset to each bin, rounded to the nearest integer.}
\label{tab:data_origin}
\end{table}

The distribution of source datasets across complexity bins reflects the design intent of existing benchmarks, many of which target specific difficulty ranges. This naturally results in a greater proportion of certain datasets within specific bins.

In addition to the test split, we construct separate development sets for both domains. For deep search, the development set contains 68 instances sampled from SimpleQA; for coding, it comprises 88 instances drawn from 40 Coding-MCQ questions and 48 LeetCode `Easy' problems. These development sets are disjoint from the test data and reflect the need for instances on which models exhibit a balanced mix of successes and failures to enable effective validation and tuning.

\subsection{Data Annotation}
\label{appendix:data-annotation}

To obtain human solution times for \textsc{HST-Bench}, we recruited a pool of paid annotators who are graduates in computer science or closely related fields, with demonstrated expertise in programming and familiarity with the types of deep search and coding problems we study. This helps ensure that the reported solution times reflect the behavior of reasonably proficient users rather than novices. For each task, we collect solution-time annotations from at least three distinct annotators, who work independently and use only tools permitted by the task guidelines (e.g., a web browser and search engine for search tasks, or a local editor/IDE for coding), while refraining from language models or other assistants that could directly solve the task. Annotators are given written task-specific guidelines (Section~\ref{sec:guidelines}), read each task once in full, then start a stopwatch, solve the task as quickly as possible while maintaining accuracy, and finally submit both their measured solution time and final answer (or code). For LeetCode `Hard' tasks, due to annotation cost constraints, we do not collect new human timings and instead rely on published human time estimates reported by~\cite{siros2024githubcopilotperfectcode}. To verify consistency, we independently annotate a random subset of 8 tasks ($\sim$10\%) and confirm that all measured times fall within the published ranges.

All collected annotations undergo a subsequent quality-control pass. First, submitted solutions are checked for correctness. Once a minimum of three correct solution times has been collected for a given task, we lightly filter for outliers to reduce the influence of anomalous timings (e.g., due to interruptions or misunderstandings), and collect further annotations if necessary. Concretely, solution times associated with incorrect answers or that deviate by more than two standard deviations from the task-wise mean are removed from the dataset. Once quality control and any necessary data re-collection have concluded, the times for each task are averaged together. We find good inter-annotator agreement across \textsc{HST-Bench} (CCI $= 0.83$, 95\% CI $[0.81, 0.85]$; Krippendorff's $\alpha = 0.86$, 95\% CI $[0.84, 0.87]$, $p < 0.001$), indicating that human solution times are reliably reproducible.

\subsection{Annotators' Guidelines}\label{sec:guidelines}
Below we reproduce the instructions provided to annotators for both deep search and coding tasks. These guidelines specify the allowed tools, what constitutes a correct solution in each domain, and how annotators should measure and report their solution times, ensuring consistency across annotators and task types.

\para{Deep Search Guidelines.}\textcolor{darkgray}{\small{\textit{Read these instructions very carefully. Only after you have understood them well, navigate to the tasks in the next tab.\\
\\
\textbfrm{Goal}\\
\\
The goal of this annotation exercise is to label how much time it will take a human (not an LLM) to solve a given question.\\
\\
You will be provided with questions and you will need to solve each with web searches, using Google or Bing.\\
\\
You will need to use a stopwatch to measure your task completion time.\\
\\
Task completion time must be reported in the format <HH hours MM minutes SS seconds>, for example, 25 seconds would be written as <00 hours 00 minutes 25 seconds> . DO NOT REPORT MILLISECONDS, EVEN IF YOUR STOPWATCH SHOWS THEM.\\
\\
BE FAST: We are trying to measure a human’s *BEST* completion time, so please complete the task (correctly) as quickly as you can. While for most questions you will likely need web search, it is fine not to use it if you already know the answer. For some tasks, it is likely that you will need multiple, in-depth web searches.\\ 
\\
It is assumed that the search engine is already open in a tab. To avoid wasting time unnecessarily, please arrange the windows on your screen so that you can see both the question and the search engine side by side at the same time.\\
\\
Solve the task by following these steps:\\
\\
- Step 1: Read the question first, slowly and carefully.\\
- Step 2: Start the stopwatch.\\
- Step 3: Text can be copy-pasted to the search engine directly from the question. Indeed, for most questions this is advisable as it can save time. As soon as the answer to the question is found, stop the stopwatch (i.e., do not wait to type the answer) and record the completion time.\\
- Step 4: Provide the answer and the task completion time (as per the stopwatch).\\ 
\\
Note: You are allowed to read the question directly from the AI-generated summary at the top of the search engine page, if this is given. However, you are not allowed to copy-paste the question into an LLM chat interface. Use Google or Bing search.\\
\\
\textbfrm{Examples:}\\
\\
Question:\\
What is Miley Cyrus' occupation?\\
\\
Completion Time: 00 hours 00 minutes 04 seconds\\
Your Answer: Singer, songwriter, actress\\
\\
—\\
\\
Question:\\
Which came out first, Titanic or Clueless?\\
\\
Completion Time: 00 hours 00 minutes 17 seconds\\
Your Answer: Clueless}}}

\para{Coding Guidelines.}\textcolor{darkgray}{\small{\textit{Read these instructions very carefully. Only after you have understood them well, navigate to the tasks in the next tab.\\
\\
\textbfrm{Goal}\\
\\
The goal of this annotation exercise is to label how much time it will take a human (not an LLM) to solve a given coding question.\\
\\
You will be provided with coding questions and you will need to solve each by writing code.\\
\\
You will need to use a stopwatch to measure your task completion time.\\
\\
Task completion time must be reported in the format <HH hours MM minutes SS seconds>, for example, 25 seconds would be written as <00 hours 00 minutes 25 seconds> . DO NOT REPORT MILLISECONDS, EVEN IF YOUR STOPWATCH SHOWS THEM.\\
\\
BE FAST: We are trying to measure a human’s *BEST* completion time, so please complete the task (correctly) as quickly as you can. You are allowed to use web search to look up syntax, however please do not overuse web search unnecessarily, as it tends to increase the completion time.\\ 
\\
If the question requires writing code, you MUST use a Python shell which allows running code at the click of a button. For example, use Google Colab or https://pythonhow.com/python-shell . For code-writing questions, you will be provided with one single test to check your code. We will run your code on more tests later to validate its correctness.\\ 
\\
It is assumed that the Python shell and the search engine are already open in a window. To avoid wasting time unnecessarily, please arrange the windows on your screen so that you can see both the question text, the coding editor and the search engine side by side at the same time.\\
\\
Solve the task by following these steps:\\
\\
- Step 1: Read the question first, slowly and carefully.\\
- Step 1a: If the question requires writing a Python function, copy the function header and, at the bottom, the given test into your Python shell *BEFORE* you start the stop watch. The required function name and arguments will be clear from the test.\\
\hspace*{20pt}For example you may have:} \texttt{def my\_function(my\_arg):}\\
\\
\hspace*{145pt}\texttt{assert my\_function(3)==True}\\
\textit{\hspace*{20pt}So that when the stopwatch starts you will only need to write the function body.\\
- Step 2: Start the stopwatch.\\
- Step 3a: If the question is multiple choice, stop the stopwatch as soon as the correct answer has been identified (no need to type it anywhere) and record the completion time.\\ 
- Step 3b: If the answer requires writing code, stop the stopwatch as soon as you have completed and run the code, and record the completion time.\\
- Step 4: Provide the answer and the task completion time (as per the stopwatch).\\ 
\\
Note: You are allowed to use Google, but not allowed to use AI Assistants.\\
\\
\textbfrm{Examples:}\\
\\
Question:\\
Which of the following lines of code is the correct way to raise a to the power of b in python? Give only the number corresponding to the answer, and nothing else.}\\
\\
1: \texttt{a\textasciicircum b} \\
2: \texttt{a**b}\\
\\
\textit{Completion Time: 00 hours 00 minutes 02 seconds\\
Your Answer: 2\\
\\
—\\
\\
Question:\\
Write a python function to find the first even number in a given list of numbers.\\
\\
Your function should satisfy the following test:}\\
\texttt{assert first\_even ([1, 3, 5, 7, 4, 1, 6, 8]) == 4}\\
\\
\textit{Completion Time: 00 hours 00 minutes 39 seconds\\
Your Answer:}\\
\texttt{def first\_even(nums):\\
\hspace*{20pt}first\_even = next((el for el in nums if el\%2==0), -1)\\
\hspace*{20pt}return first\_even}}}

\subsection{Coding-MCQ Examples}\label{appendix:cmcq}

Below are representative multiple-choice questions from the Coding-MCQ dataset, designed to assess performance on short, low-complexity coding tasks that target core programming concepts.

\vspace{15pt}
\begin{tcolorbox}[
    colback=white,
    colframe=black,
    boxrule=0.5pt,
    arc=0pt,
    left=10pt,
    right=10pt,
    top=10pt,
    bottom=10pt
]
Which of the following lines of code prints the word `hello'? Give only the number corresponding to the answer, and nothing else. \\
\\
1: \texttt{print(`hello') if 1\%2==0 else print(`goodbye')}  \\
\\
2: \texttt{print(`goodbye') if 1\%2==0 else print(`helloworld'[:5])}
\end{tcolorbox}

\vspace{12.5pt}
\begin{tcolorbox}[
    colback=white,
    colframe=black,
    boxrule=0.5pt,
    arc=0pt,
    left=10pt,
    right=10pt,
    top=10pt,
    bottom=10pt
]
Which of the following lines of code will not throw an error in Python? Give only the number corresponding to the answer, and nothing else.  \\
\\
1: \texttt{100 \& 100}  \\
\\
2: \texttt{100.0 \& 100.0}
\end{tcolorbox}

\vspace{12.5pt}
\begin{tcolorbox}[
    colback=white,
    colframe=black,
    boxrule=0.5pt,
    arc=0pt,
    left=10pt,
    right=10pt,
    top=10pt,
    bottom=10pt
]
Which of the following files is a configuration file? Give only the number corresponding to the answer, and nothing else. \\
\\
1: \texttt{run\_agent.yaml}  \\
\\
2: \texttt{README.md} \\
\\
3: \texttt{run\_agent.py}
\end{tcolorbox}

\vspace{12.5pt}
\begin{tcolorbox}[
    colback=white,
    colframe=black,
    boxrule=0.5pt,
    arc=0pt,
    left=10pt,
    right=10pt,
    top=10pt,
    bottom=10pt
]
Which of the following lines of code returns an empty list in python? Give only the number corresponding to the answer, and nothing else. \\
\\
1: \texttt{[elem for elem in [2,3,4,5] if elem // 2 == 0]}  \\
\\
2: \texttt{[elem for elem in [2,3,4,5] if elem \% 2 == 0]}
\end{tcolorbox}

\vspace{12.5pt}
\begin{tcolorbox}[
    colback=white,
    colframe=black,
    boxrule=0.5pt,
    arc=0pt,
    left=10pt,
    right=10pt,
    top=10pt,
    bottom=10pt
]
Which of the following lines of code correctly replaces a character in a string in Python? Give only the number corresponding to the answer, and nothing else. \\
\\
1: \texttt{"a,b,d".replace("d", "c")}  \\
\\
2: \texttt{[char if char in "a,b" else "c" for char in "a,b,d"]}
\end{tcolorbox}

%%%%%%%%%%%%%%%%%%%%%%%%%%%%%%%%%%%%%%%%%%%%%%%%%%%%%%%%%%%%%%%%%%%%%%%%%

\newpage\section{Estimated Cost of Running Models}\label{appendix:cost}
Inference prices for Qwen3 models vary substantially across providers and deployment settings, reflecting differences in supported context length, geographical region, and commercial factors such as traffic volume and competition.\footnote{\url{https://huggingface.co/datasets/reach-vb/inference-provider-pricing}} To obtain a simple, reproducible cost model for our experiments, we adopt an empirically calibrated pricing schedule. Our approach is grounded in recent empirical analyses of inference markets demonstrating that, for dense models, per-token prices scale approximately linearly with the number of parameters~\citep{scher2025observations}.

Following this established relationship, we model cost as proportional to the number of parameters and anchor our schedule using publicly advertised prices from established inference providers. Specifically, at the time of writing Groq reports separate prices for input and output tokens for Qwen3 32B,\footnote{\url{https://groq.com/pricing}} listing
\[
\$0.29 / \text{Mt} \ \text{for input tokens} \quad+\quad \$0.59 / \text{Mt} \ \text{for output tokens},
\]
where \textit{Mt} denotes one million tokens. We use these figures as a reference anchor for a high-capacity Qwen3 model and scale costs for other sizes in proportion to their parameter counts.

In our agentic runs we consistently observe an average input-to-output token ratio of about 4:1 across task domains and horizons. Under this assumption, we take the expected cost per million \emph{total} tokens for an agent instantiated with Qwen3 32B to be
\[
\pi(a_{\text{32B}}) 
\;=\;
\frac{4 \cdot 0.29 + 1 \cdot 0.59}{5}
\;\approx\;
\$0.36 / \text{Mt}.
\]
Applying the same linear scaling in parameter count yields the effective prices per million total tokens used in our experiments:
\[
\pi(a_{\text{4B}}) = \$0.05,\quad
\pi(a_{\text{8B}}) = \$0.09,\quad
\pi(a_{\text{14B}}) = \$0.16,\quad
\]

where $\pi(a_{\text{4B}}) \approx 0.045$ is rounded to \$0.05. 

\para{Empirical Validation.} To verify that the linear scaling assumption holds in practice for the other Qwen3 sizes in our experiments, we compare our derived prices against independently advertised rates from major inference providers at the time of writing.\footnote{\url{https://nebius.com/token-factory/prices}}\footnote{\url{https://novita.ai/pricing}}\footnote{\url{https://www.alibabacloud.com/help/en/model-studio/models}} Since providers list separate prices for input and output tokens, we compute comparable per-million-total-token rates by applying the same 4:1 input-to-output weighting used in our estimates. Table~\ref{tab:price_validation} reports this comparison. The mean absolute deviation between our estimates and observed provider averages is within 6\%, confirming that the linear approximation is well-supported for this model family.

\vspace{5pt}
\begin{table}[h]
\centering
\setlength{\tabcolsep}{25pt}
\begin{tabular}{lccc}
\toprule
Model & Estimated (\$/Mt) & Provider Avg. (\$/Mt) & Deviation \\
\midrule
Qwen3-4B & 0.05 & 0.05 & 0\% \\
Qwen3-8B & 0.09 & 0.09 & 0\% \\
Qwen3-14B & 0.16 & 0.17 & 6\% \\
\bottomrule
\end{tabular}
\caption{Comparison of estimated prices (derived via linear scaling from Qwen3 32B) against average advertised prices across providers (Nebius, Novita, Alibaba Cloud). Deviations are within 6\%, validating the linear cost model.}
\label{tab:price_validation}
\end{table}

\para{Scope of the Cost Model.} Our cost model assumes access via third-party inference APIs, where infrastructure overhead, including hardware provisioning, energy consumption, and maintenance, is fully absorbed into the provider's per-token pricing. Under usage-based API billing, the user incurs costs only for tokens consumed, making \$/Mt the appropriate metric for our analysis. We note that latency and throughput vary substantially across providers, regions, and time of day, making them difficult to model consistently (though smaller models are typically also faster); we therefore leave them outside the scope of our study. Per-token pricing, by contrast, is publicly advertised and stable, providing a reproducible basis for cost comparison.

%%%%%%%%%%%%%%%%%%%%%%%%%%%%%%%%%%%%%%%%%%%%%%%%%%%%%%%%%%%%%%%%%%%%%%%%%

\newpage\section{Environment, Prompts and Hyperparameters}\label{appendix:are_setup}

All experiments are conducted within the open-source Agent Research Environment (ARE), which provides a standardized, tool-augmented interface for evaluating heterogeneous language-model agents on real-world tasks. Unless otherwise noted, agents, tools, and evaluation protocols follow the default ARE configuration. In the subsections below, we describe the exact model hyperparameters, environment prompts, and other implementation details needed to fully reproduce our setup.

\subsection{Model Hyperparameters}
All experiments were run on NVIDIA A100 and H100 GPU clusters with 40--80\,GB of HBM per accelerator. We use the same decoding configuration across all agents. The full set of model- and decoding-related hyperparameters used in our experiments is summarized in Table~\ref{tab:hyperparams}.

\begin{table}[H]
    \centering
    \setlength{\tabcolsep}{20pt}
    \begin{tabular}{lccccc}
        \toprule
        & \textbf{Max length} & \textbf{Temperature} & \textbf{Top-p} & \textbf{Top-k} & \textbf{Batch size} \\
        \midrule
        Values & 40,960 & 0.0 & 1.0 & 0 & 10 \\
        \bottomrule
    \end{tabular}
    \caption{Decoding and batching hyperparameters used for all Qwen3 agents.}
    \label{tab:hyperparams}
\end{table}

\subsection{Environment Hyperparameters}
We run all experiments under the default configuration of ARE. Each episode is terminated as soon as either the time or iteration budget is exhausted, and agents must return a single final solution (i.e., we report pass@1 under the environment's LLM-as-a-judge evaluation). Notably, the LLM-as-a-judge evaluation is straightforward in our setup: search outputs are directly matched against ground truth, while coding outputs, though potentially differing lexically, can be reliably compared for functional equivalence by models like GPT-4o. We refer readers to the ARE default configuration for LLM-as-a-judge prompts and other standard hyperparameters not explicitly mentioned here. We report environment hyperparameters in Table~\ref{tab:env_hyperparams}.

\begin{table}[H]
    \centering
    \setlength{\tabcolsep}{30pt}
    \begin{tabular}{lcc}
        \toprule
        \textbf{Hyperparameter} & \textbf{Value} & \textbf{Description} \\
        \midrule
        \texttt{task\_timeout\_seconds}          & 3600  & Maximum wall-clock time per task \\
        \texttt{max\_iterations}        & 100     & Maximum agent steps per task \\
        \texttt{llm\_judge} & GPT-4o & Base model for LLM-as-a-judge evaluation \\
        \bottomrule
    \end{tabular}
    \caption{Environment-level limits used for all tasks; an episode terminates when either limit is reached.}
    \label{tab:env_hyperparams}
\end{table}

\begin{table}[H]
    \centering
    \setlength{\tabcolsep}{9.5pt}
    \begin{tabular}{m{0.25\linewidth} m{0.65\linewidth}}
        \toprule
        \textbf{Tool} & \textbf{Description} \\
        \midrule
        \texttt{ask\_search\_agent} &
        Delegates a natural-language query to a web search agent and returns its response. \\
        \cmidrule(lr){1-2}
        \texttt{inspect\_file\_as\_text} &
        Reads a file from the workspace as markdown text and returns its contents for subsequent inspection and reasoning \\
        \cmidrule(lr){1-2}
        \texttt{final\_answer} &
        Submits the agent's final solution and terminates the episode. \\
        \cmidrule(lr){1-2}
        Python environment &
        Executes Python code for calculations, data manipulation, and lightweight scripting; it is preconfigured with the standard library and commonly used packages sufficient to solve the
        benchmark tasks. \\
        \bottomrule
    \end{tabular}
    \caption{Tools and execution environment available to agents. Deep search tasks use all tools; coding tasks use \texttt{ask\_search\_agent}, \texttt{final\_answer}, and the Python environment.}
    \label{tab:tools}
\end{table}

For deep search tasks, each episode begins with a \emph{fact extraction} pre-step, followed by a \emph{strategy planning} step. For coding tasks, the agent performs only a strategic planning step without explicit fact extraction. Table~\ref{tab:tools} summarizes the tools and Python execution environment available to the agent in each domain and provides brief descriptions of their functionality. For all remaining environment details (e.g., the exact tool interfaces, the format of observations returned to the agent, and error handling), we refer the reader to the original ARE paper~\citep{froger2025arescalingagentenvironments}.

\subsection{Retrieval Hyperparameters}\label{sec:retrieval_hyperparams}
We use embedding-based retrieval over the shared auction memory: at each episode, the agent retrieves strategies that both won and lost past auctions for the $k$ most similar tasks to the current one. Following established practice, we set $k = 8$, which prior studies commonly find to be a strong practical trade-off between coverage/diversity and context/latency overhead~\citep{dai2023promptagator, wang-etal-2024-learning, rashid-hakak-2025-fathom}. Table~\ref{tab:mem_hyperparams} summarizes the retrieval hyperparameters used in all experiments.

\begin{table}[H]
    \centering
    \setlength{\tabcolsep}{33pt}
    \begin{tabular}{lccc}
        \toprule
        & \textbf{Embedding model} & \textbf{Distance metric} & \textbf{Top-$k$} \\
        \midrule
        Values & \texttt{all-MiniLM-L6-v2} & cosine & 8 \\
        \bottomrule
    \end{tabular}
    \caption{Strategy retrieval hyperparameters used across all experiments.}
    \label{tab:mem_hyperparams}
\end{table}

\subsection{Prompts}\label{appendix:prompts}
We detail here the \emph{judge} prompt for scoring candidate strategies, the \emph{strategy} prompts used to generate $s_{t,i}$ for a task $t$, and the \emph{refinement} prompts to produce $s^{r}_{t,i}$. The judge prompt uses a 0--5 Likert scale rather than a wider range (e.g., 0--100), based on preliminary experiments where 0--5 yielded higher correlation with human scores, consistent with~\citet{li2026gradingscaleimpactllmasajudge}. For coding tasks, we omit fact extraction before planning as the task specification is self-contained. Strategy refinement uses a template similar to that introduced by~\citet{alazraki-etal-2025-need} to facilitate contrastive learning from prior outcomes. For all remaining prompts used to interact with the environment, tools, and task wrappers, refer to the standard ARE configuration.

\vspace{5pt}
\begin{figure}[H]
  \centering
  \begin{tcolorbox}[promptbox]
    \textbf{\large{Judge Prompt}}\\[0.4em]
    \ttfamily
    \\
    \\
    Provide an integer reward score between 0 and 5 (inclusive) for the quality of the provided plan steps, using strict evaluation standards. Ensure the reward reflects how effectively the plan contributes to progressing toward the correct solution.\\
\\
\\
Problem Statement:\\
***begin problem statement***\\
\{task\}\\
***end problem statement***\\
\\
Plan:\\
\{plan\}\\
\\
\\
Be harsh in your evaluation. Only plans that you are extremely confident will succeed should be assigned the maximum score.\\
\\
\\
Score: $[$Strictly provide an integer reward score between 0 and 5$]$\\
  \end{tcolorbox}
  \label{fig:judge-prompt}
\end{figure}

%%%%%%%%%%%%%%%%%%%%%%%%%%%%%%%%%%%%%%%%%%%%%%%%%%%%%%%%%%%%%%%%%%%%%%

\vspace*{64pt}
\begin{figure}[H]
  \centering
  \begin{tcolorbox}[promptbox]
    \textbf{\large{Strategy Prompt (Deep Search)}}\\[0.4em]
    \ttfamily
    \\
    \\
    You are a world expert at making efficient plans to solve any task using a set of carefully crafted tools.\\
\\
Now for the given task, develop a step-by-step high-level plan taking into account the following inputs and list of facts.\\
This plan should involve individual tasks based on the available tools, that if executed correctly will yield the correct answer.\\
Do not skip steps, do not add any superfluous steps. Only write the high-level plan, DO NOT DETAIL INDIVIDUAL TOOL CALLS.\\
After writing the final step of the plan, write the '<end\_plan>' tag and stop there.\\
Always search for the exact task at the beginning. If you are given an external file, always inspect it first to explore its content.\\
Do a very concise plan that only focus on the given task.\\
Do not attempt to answer the question without calling tools, even if you know the answer. You must always use at least one tool to find the answer.\\
\\
\\
\\
Here is your task:\\
\\
Task:\\

\textasciigrave\textasciigrave\textasciigrave\\
\{task\}\\
\textasciigrave\textasciigrave\textasciigrave\\
\\
Your plan can leverage any of these tools:\\
\{tool\_descriptions\}\\
\\
List of facts that you know:\\
\textasciigrave\textasciigrave\textasciigrave\\
\{answer\_facts\}\\
\textasciigrave\textasciigrave\textasciigrave\\
\\
Now begin! Write your plan below.\\
  \end{tcolorbox}
  \label{fig:strategy-prompt-search}
\end{figure}

%%%%%%%%%%%%%%%%%%%%%%%%%%%%%%%%%%%%%%%%%%%%%%%%%%%%%%%%%%%%%%%%%%%%%%

\vspace*{100pt}
\begin{figure}[H]
  \centering
  \begin{tcolorbox}[promptbox]
    \textbf{\large{Strategy Prompt (Coding)}}\\[0.4em]
    \ttfamily
    \\
    \\
    You are a world expert at making efficient plans to solve any task using a set of carefully crafted tools.\\
\\
Now for the given task, develop a step-by-step high-level plan taking into account the following inputs.\\
This plan should involve individual tasks based on the available tools, that if executed correctly will yield the correct answer.\\
Do not skip steps, do not add any superfluous steps. Only write the high-level plan, DO NOT DETAIL INDIVIDUAL TOOL CALLS.\\
After writing the final step of the plan, write the '<end\_plan>' tag and stop there.\\
Do a very concise plan that only focus on the given task.\\
Do not attempt to answer the question without calling tools, even if you know the answer. You must always use at least one tool to find the answer.\\
\\
\\
\\
Here is your task:\\
\\
Task:\\

\textasciigrave\textasciigrave\textasciigrave\\
\{task\}\\
\textasciigrave\textasciigrave\textasciigrave\\
\\
Your plan can leverage any of these tools:\\
\{tool\_descriptions\}\\
\\
Now begin! Write your plan below.\\
  \end{tcolorbox}
  \label{fig:strategy-prompt-code}
\end{figure}

%%%%%%%%%%%%%%%%%%%%%%%%%%%%%%%%%%%%%%%%%%%%%%%%%%%%%%%%%%%%%%%%%%%%%%

\vspace*{10pt}
\begin{figure}[H]
  \centering
  \begin{tcolorbox}[promptbox]
    \textbf{\large{Strategy Refinement Prompt (Deep Search)}}\\[0.4em]
    \ttfamily
    \\
    \\
    You are a world expert at making efficient plans to solve any task using a set of carefully crafted tools.\\
\\
Now for the given task, develop a step-by-step high-level plan taking into account the following inputs and list of facts.\\
This plan should involve individual tasks based on the available tools, that if executed correctly will yield the correct answer.\\
Do not skip steps, do not add any superfluous steps. Only write the high-level plan, DO NOT DETAIL INDIVIDUAL TOOL CALLS.\\
After writing the final step of the plan, write the '<end\_plan>' tag and stop there.\\
Always search for the exact task at the beginning. If you are given an external file, always inspect it first to explore its content.\\
Do a very concise plan that only focus on the given task.\\
Do not attempt to answer the question without calling tools, even if you know the answer. You must always use at least one tool to find the answer.\\
\\
\\
\\
Here is your task:\\
\\
Task:\\

\textasciigrave\textasciigrave\textasciigrave\\
\{task\}\\
\textasciigrave\textasciigrave\textasciigrave\\
\\
Your plan can leverage any of these tools:\\
\{tool\_descriptions\}\\
\\
List of facts that you know:\\
\textasciigrave\textasciigrave\textasciigrave\\
\{answer\_facts\}\\
\textasciigrave\textasciigrave\textasciigrave\\
\\
Below you will find some example tasks followed by two corresponding plans - one plan that lost in a previous plan competition and one that won. Use these examples to understand what makes a plan lose or win.\\
\\
\{retrieved\_tasks\_and\_plans\}\\
\\
Now apply what you have learned and given the task and a corresponding losing plan, write a winning plan.\\
\\
\{previous\_losing\_plan\}\\
Winning plan:\\
  \end{tcolorbox}
  \label{fig:refine-prompt-search}
\end{figure}

%%%%%%%%%%%%%%%%%%%%%%%%%%%%%%%%%%%%%%%%%%%%%%%%%%%%%%%%%%%%%%%%%%%%%%

\vspace*{50pt}
\begin{figure}[H]
  \centering
  \begin{tcolorbox}[promptbox]
    \textbf{\large{Strategy Refinement Prompt (Coding)}}\\[0.4em]
    \ttfamily
    \\
    \\
    You are a world expert at making efficient plans to solve any task using a set of carefully crafted tools.\\
\\
Now for the given task, develop a step-by-step high-level plan taking into account the following inputs.\\
This plan should involve individual tasks based on the available tools, that if executed correctly will yield the correct answer.\\
Do not skip steps, do not add any superfluous steps. Only write the high-level plan, DO NOT DETAIL INDIVIDUAL TOOL CALLS.\\
After writing the final step of the plan, write the '<end\_plan>' tag and stop there.\\
Do a very concise plan that only focus on the given task.\\
Do not attempt to answer the question without calling tools, even if you know the answer. You must always use at least one tool to find the answer.\\
\\
\\
\\
Here is your task:\\
\\
Task:\\

\textasciigrave\textasciigrave\textasciigrave\\
\{task\}\\
\textasciigrave\textasciigrave\textasciigrave\\
\\
Your plan can leverage any of these tools:\\
\{tool\_descriptions\}\\
\\
Below you will find some example tasks followed by two corresponding plans - one plan that lost in a previous plan competition and one that won. Use these examples to understand what makes a plan lose or win.\\
\\
\{retrieved\_tasks\_and\_plans\}\\
\\
Now apply what you have learned and given the task and a corresponding losing plan, write a winning plan.\\
\\
\{previous\_losing\_plan\}\\\\
Winning plan:\\
  \end{tcolorbox}
  \label{fig:refine-prompt-code}
\end{figure}

%%%%%%%%%%%%%%%%%%%%%%%%%%%%%%%%%%%%%%%%%%%%%%%%%%%%%%%%%%%%%%%%%%%%%%

%%%%%%%%%%%%%%%%%%%%%%%%%%%%%%%%%%%%%%%%%%%%%%%%%%%%%%%%%%%%%%%%%%%%%%%%%

\newpage\section{Cost--Value Optimization}
\label{appendix:minmax}

\vspace{10pt}
\subsection{Cost and Value Function Design}
\label{appendix:cost-value-design}

\para{Strategy Length in Cost Function.} In our setting, strategy length correlates with realized trajectory length (Spearman's $\rho=0.39$, Pearson's $r=0.36$, both $p<0.001$), consistent with \citet{goebel2025llmreasoningmodelsreplaceclassical}. The close alignment between rank-based and linear correlation measures indicates a stable monotonic relationship rather than an artifact driven by outliers, reinforcing that longer strategies generally yield longer trajectories. In agentic planning, such a stable and consistent proxy signal provides meaningful guidance for cost estimation. Although the correlation is moderate in absolute magnitude, this level of predictive strength is sufficient in practice: prior work on agentic orchestration demonstrates that proxies of relatively moderate strength can meaningfully inform decision-making, even in the presence of systematic over- or underestimation \citep{amin2026bayesianorchestrationmultillmagents}. Crucially, including $|s_{t,i}|$ in $C_{t,i}$ improves development set accuracy, and removing it degrades performance in ablations (Appendix~\ref{appendix:ablations}).

\para{Entropy and Jury in Value Function.} We use normalized entropy $H(s_{t,i})$, computed as the mean per-token entropy from generation log-probabilities, as a lightweight signal of informational content/non-redundancy. In addition to being motivated by prior work linking higher-entropy planning to better outcomes (see Section~\ref{sec:strategy-auctions}); this term is also validated by development set performance and confirmed by ablations. Finally, we score each strategy with a full jury (including self-judgment): this configuration performs best on validation, learned weights calibrate each judge’s influence, and ablations show that removing any judge reduces accuracy (Appendix~\ref{appendix:ablations}).

\vspace{7.5pt}
\subsection{Min-Max Formulation}
In Mathematical Program~\ref{prog:minmax}, we provide the full mixed-integer linear program (MILP) for learning the scoring weights $w = (w_c, w_h, \{w_j\}_{a_j \in \mathcal{A}})$. Let $\mathcal{T}$ denote the training set of tasks. 

\vspace{15pt}
\floatname{algorithm}{Mathematical Program}

\begin{algorithm}[H]
\caption{Min-Max Weight Optimization}
\label{prog:minmax}

\hrule
\vspace{0.9em}
\begin{align}
\min_{w, x, z, Q} \quad & Q \nonumber \\[6pt]
\textnormal{s.t.} \quad 
& \sum_{a_i \in \mathcal{A}} x_{t,i} = 1 && \forall t \in \mathcal{T} \nonumber \\[6pt]
& x_{t,i} \in \{0,1\} && \forall t, a_i \nonumber \\[6pt]
& z_t \geq C_{t,i} - V_{t,i} - M(1 - x_{t,i}) && \forall t, a_i \nonumber \\[6pt]
& z_t \leq C_{t,i} - V_{t,i} + M(1 - x_{t,i}) && \forall t, a_i \nonumber \\[6pt]
& z_t \leq Q && \forall t \in \mathcal{T} \nonumber \\[6pt]
& w \in \mathbb{R}^{2 + |\mathcal{A}|} && \nonumber
\end{align}
\vspace{-0.1em}
\hrule
\end{algorithm}

\vspace{5pt}
The big-$M$ constraints ensure that $z_t = C_{t,i} - V_{t,i}$ only for the selected agent (i.e., when $x_{t,i} = 1$), while remaining inactive otherwise. We set $M = 10^4$, chosen to exceed the observed range of $C_{t,i} - V_{t,i}$ scores across all task-agent pairs in the training set. Specifically, $M$ exceeds $\max_{t,i} |C_{t,i} - V_{t,i}|$ by at least two orders of magnitude, ensuring numerical stability without introducing solver issues.

\subsection{Implementation Details}
We solve the MILP using PuLP~\citep{mitchell2011pulp} with the CBC solver. Weights are unconstrained. The optimization is performed separately on the development sets for deep search and coding tasks (see Appendix~\ref{appendix:data} for details), with $|\mathcal{A}| = 4$ agents.

The learned weights exhibit consistent patterns across both domains. Entropy receives the highest weight in both cases, suggesting that information density in the generated strategy is a strong indicator of expected performance. This aligns with prior findings on the informativeness of high-entropy reasoning~\citep{li2025compressingchainofthoughtllmsstep} and is further supported by our ablation study in Appendix~\ref{appendix:ablations}. The relative importance of individual judge models varies by domain: for deep search tasks, smaller model judgments receive higher weight, whereas for coding tasks, the largest agent (32B) contributes most to the value estimate.

\subsection{Alternative Formulations}
\label{appendix:alternative-formulations}

\para{Rationale for the Optimization Objective.} The difference $C_{t,i} - V_{t,i}$ can be interpreted as a \emph{negative net utility}: lower cost and higher value both reduce this quantity, so minimizing it naturally favors agents that are both efficient and effective. This formulation mirrors classical economic frameworks where profit is defined as revenue minus cost. Crucially, because the weights $(w_c, w_h, \{w_j\})$ are learned jointly and left unconstrained, the optimization implicitly calibrates the relative scales of cost and value components without requiring manual normalization or threshold selection.

\para{Normalized Features.} An alternative approach is to normalize $C$ and $V$ before combining them, for instance by Z-score, standardizing all input features prior to learning the weights. While this ensures that cost and value terms are on comparable scales a priori, it discards meaningful structural information. In our setting, cost and value features operate on different scales, entropy in particular is orders of magnitude smaller than cost, while judge scores fall in between. These differences are not arbitrary; instead, they reflect the heterogeneous nature of the signals: token-based costs, normalized entropy, and Likert-scale judgments each carry information at their natural scale. Our learned weights implicitly calibrate these differences, ensuring that smaller-scale features (particularly entropy, which receives the largest weight magnitude) contribute meaningfully to routing decisions.

To validate this intuition empirically, we evaluated a variant in which all input features are Z-score normalized using training-set statistics before weight optimization. This variant achieves lower accuracy while disproportionately selecting cheaper models. Inspection of the learned weights reveals the cause: under normalization, the optimizer discovers a degenerate solution in which several value-related weights become negative (e.g., $w_h < 0$, $w_{32B} < 0$), effectively \emph{penalizing} quality signals rather than rewarding them. This occurs because, once all features are rescaled to comparable magnitudes, the optimizer can minimize $C - V$ simply by selecting cheap models; there is no structural pressure to preserve value contributions. In contrast, our unnormalized formulation forces the learned weights to jointly encode both relative importance and scale calibration, preventing such degenerate solutions.

\para{Constrained Optimization.} Another alternative is constrained optimization, for example minimizing cost subject to $V > \eta$ for some threshold $\eta$. However, this formulation requires choosing $\eta$ a priori, which is difficult in practice: the appropriate threshold may vary across task distributions, agent pools, and deployment settings. Setting $\eta$ too high excludes cost-efficient agents that would have succeeded on easier tasks, while setting it too low allows poor-quality assignments to slip through on harder ones. Our unconstrained formulation sidesteps this problem altogether. The learned weights end up implicitly encoding the right cost–value trade-off for the training distribution, and because the min–max objective optimizes against worst-case tasks, we do not need to hand-pick thresholds for robustness.

\vspace{5pt}
In conclusion, normalization and constrained optimization both introduce additional design choices (normalization statistics, threshold values) that need to be tuned and often do not transfer well across settings. By minimizing $C - V$ with unconstrained, jointly learned weights, we avoid this overhead, while still obtaining strong empirical results. The natural scale differences between cost and value features are informative on their own, and the learned weights can exploit this, as confirmed by ablation experiments showing that adding explicit normalization hurts routing quality.

\subsection{Tool-Latency Cost Extension}
\label{app:tool_latency}

\vspace{15pt}
To demonstrate that the modular cost function naturally accommodates non-token costs, we simulate a deployment where tool calls contribute to overall latency. We augment the cost term with a tool-call penalty:
\vspace{10pt}
\begin{equation*}
C_{t,i} = w_c \cdot \pi(a_i) \cdot |s_{t,i}| + \beta \cdot n_{\text{tools}}(s_{t,i}) \cdot \ell,\vspace{10pt}
\end{equation*}
where $n_{\text{tools}}(s_{t,i})$ counts the tool calls extracted from the strategy, $\ell$ is a synthetic average per-call latency, and $\beta$ controls the penalty strength. Note that the value function is unchanged. We sweep $\beta$ from $0$ (original cost function) to $5$ (strong tool penalty) on a 100-task subset evenly stratified by type and complexity. Results are reported in Table~\ref{tab:tool_latency}.

\vspace{5pt}
As $\beta$ grows, the 32B agent's overall share drops from 23\% to 11\% (a 52\% relative reduction), reflecting its higher per-token price combined with the highest tool density in its generated strategies. The 14B agent rises from 18\% to 28\% overall, offering the best cost--value trade-off under tool latency. The effect is also domain-sensitive: when we consider only deep search tasks, where tool use is inherent, the 32B share drops from 20\% to 8\%. The MILP converges to optimal at every $\beta$, confirming that the cost function accepts this extension without modifications to the auction mechanism.

\begin{table}[h]
\centering
\vspace{25pt}
\setlength{\tabcolsep}{26pt}
\caption{Agent allocation under increasing tool-latency penalty $\beta$, overall and broken down by domain. The 32B agent, which generates the highest mean tool-call count (2.48 vs.\ 0.64 for the 4B), is progressively deselected as $\beta$ grows, with the sharpest decline observed in the deep-search subset where tool use is inherent.}
\label{tab:tool_latency}
  \begin{NiceTabular}{cl rrrr}[cell-space-limits=2pt]
  \CodeBefore
    \rectanglecolor{metabg}{2-2}{2-6}
    \rectanglecolor{metabg}{5-2}{5-6}
    \rectanglecolor{metabg}{8-2}{8-6}
  \Body
  \toprule
  $\beta$ & Subset & 4B & 8B & 14B & 32B \\
  \midrule
  \multirow{3}{*}{0.0} & All & 26\% & 33\% & 18\% & 23\% \\
                       & Deep search & 28\% & 34\% & 18\% & 20\% \\
                       & Coding      & 24\% & 32\% & 18\% & 26\% \\
  \midrule
  \multirow{3}{*}{1.0} & All & 29\% & 34\% & 21\% & 16\% \\
                       & Deep search & 30\% & 38\% & 18\% & 14\% \\
                       & Coding      & 28\% & 30\% & 24\% & 18\% \\
  \midrule
  \multirow{3}{*}{5.0} & All & 32\% & 29\% & 28\% & 11\% \\
                       & Deep search & 34\% & 32\% & 26\% & \phantom{0}8\% \\
                       & Coding      & 30\% & 26\% & 30\% & 14\% \\
  \bottomrule
  \end{NiceTabular}
  \label{tab:tool_latency}
\end{table}

%%%%%%%%%%%%%%%%%%%%%%%%%%%%%%%%%%%%%%%%%%%%%%%%%%%%%%%%%%%%%%%%%%%%%%%%%

\newpage\section{Strategy Refinement}\label{appendix:strategy_refinement}

\subsection{Algorithm}
In Algorithm~\ref{alg:refinement}, we provide a complete description of the strategy refinement procedure outlined in Section~\ref{sec:strategy-auctions}. The algorithm details how initial bids are collected, how the provisional winner is selected, and how cheaper agents refine their strategies using retrieved contrastive examples from the auction memory $\mathcal{M}$.

\floatname{algorithm}{Algorithm}
\setcounter{algorithm}{0}
\begin{algorithm}[H]
\caption{Strategy Refinement from Auction Memory}
\label{alg:refinement}
\begin{algorithmic}
\hrule
\vspace{0.5em}
\Require Task $t$, agents $\mathcal{A}$, memory $\mathcal{M}$, learned weights $w$, retrieval size $k$
\Ensure Final agent assignment $i^*(t)$
\\
\For{each agent $a_i \in \mathcal{A}$}
    \State Generate initial strategy $s_{t,i}$
    \State Compute cost $C_{t,i} \gets w_c \cdot \pi(a_i) \cdot |s_{t,i}|$
    \State Compute value $V_{t,i} \gets w_h \cdot H(s_{t,i}) + \sum_{a_j \in \mathcal{A}} w_j \cdot \gamma_j(s_{t,i})$
\EndFor
\\
\State $\hat{i}(t) \gets \arg\min_{i} (C_{t,i} - V_{t,i})$
\State $i^*(t) \gets \hat{i}(t)$
\\
\For{each agent $a_i \in \mathcal{A}$ where $\pi(a_i) < \pi(a_{\hat{i}(t)})$}
    \State $\tilde{k} \gets \min(k, |\mathcal{T}_{\mathcal{M}}|)$
    \State Retrieve $\mathcal{T}' \gets \operatorname{top\text{-}\tilde{k}}_{t' \in \mathcal{T}_{\mathcal{M}}} \mathrm{sim}(t, t')$
    \State $\mathcal{M}_{t,i} \gets \{(s^{\text{lose}}_{t'}, s^{\text{win}}_{t'})_i \mid t' \in \mathcal{T}'\}$
    \State Format $\mathcal{M}_{t,i}$ as contrastive examples
    \State Generate refined strategy $s^r_{t,i}$ conditioned on $s_{t,i}$ and $\mathcal{M}_{t,i}$
    \State Compute $C^r_{t,i} \gets w_c \cdot \pi(a_i) \cdot |s^r_{t,i}|$
    \State Compute $V^r_{t,i} \gets w_h \cdot H(s^r_{t,i}) + \sum_{a_j \in \mathcal{A}} w_j \cdot \gamma_j(s^r_{t,i})$
    \If{$C^r_{t,i} - V^r_{t,i} < C_{t,i^*(t)} - V_{t,i^*(t)}$}
        \State $i^*(t) \gets i$
    \EndIf
\EndFor
\\
\State \Return $i^*(t)$
\end{algorithmic}
\vspace{0.5em}
\hrule
\end{algorithm}
\subsection{Contrastive Example Selection}
The refinement step exposes agents to contrastive pairs of winning and losing strategies from past auctions. Since the auction objective minimizes cost-minus-value, winning strategies tend to originate from cost-efficient agents---typically smaller models achieving competitive value. This aligns naturally with our refinement setting, where only cheaper agents refine: the positive examples shown to a refining agent come from models of comparable capability. Prior work has shown that smaller models can benefit from exposure to high-quality plans~\citep{sun-etal-2024-enhancing-code, kang2025distillingllmagentsmall, xiong-etal-2025-mpo}, yet often struggle to execute strategies designed by much larger models when the required reasoning exceeds their capacity~\citep{chen-etal-2025-unveiling-key, lee2025efficientllmcollaborationplanning}. By using auction winners, which inherently balance quality with executability, as positive examples, our approach provides smaller models with effective yet feasible strategies to learn from.

An alternative design would label contrastive examples by downstream task success rather than auction selection. However, this assumes access to a reliable automated evaluator, an assumption that often fails in practice, where ground-truth feedback may be sparse, noisy, or available only through human assessment or ground truth labels. The auction selection signal, by contrast, emerges naturally from the mechanism itself and requires no external oracle, making our refinement procedure broadly applicable.

\subsection{Refinement Patterns}
Table~\ref{tab:refinement_patterns} reports, for both search and coding tasks, how often each refinement pattern appears in selected winning strategies. These statistics show that the observed structural elements recur consistently across tasks.
\begin{table}[H]
\centering
\small
\setlength{\tabcolsep}{30pt}
\begin{tabular}{llc}
\toprule
\textbf{Domain} & \textbf{Refinement pattern} & \textbf{Prevalence} \\
\midrule
\multirow{4}{*}{\textit{Deep search}} & Improved tool and argument specification          & 68\% \\
       & Provides (additional) trusted sources for search & 15\% \\
       & Intermediate cross-reference/consistency checks & 32\% \\
       & Clearer/more structured layout                     & 49\% \\
\midrule
\multirow{4}{*}{\textit{Coding}} & Precise function and helper signatures            & 33\% \\
       & Explicit focus on returning the correct artifact  & 30\% \\
       & Systematic testing and edge-case coverage         & 41\% \\
       & Clearer/more structured layout                    & 26\% \\
\bottomrule
\end{tabular}
\caption{Refinement patterns and their prevalence in chosen refined plans for search and coding tasks. Note that patterns are not mutually exclusive; a single refined plan may exhibit multiple patterns simultaneously.}
\label{tab:refinement_patterns}
\end{table}

\subsection{Representative Examples}\label{appendix:examples}
Below we show side-by-side representative examples of initial and refined strategies for individual tasks, together with the refinement patterns (see Table~\ref{tab:refinement_patterns}) identified in each. These examples illustrate how refinement turns loosely specified or linear strategies into more constrained, check-pointed procedures that score higher in the auction. In all examples, the unrefined strategy is rejected and the refined one is selected, yielding a successful downstream answer.

\subsubsection{Deep Search Examples}
\vspace{5pt}
\PlanTriple{%
  Do The Importance of Being Icelandic and The Five Obstructions belong to different film genres?
}{%
  1. Use the \texttt{ask\_search\_agent} to determine the genre of ``The Importance of Being Icelandic'' by querying a reliable film database or source. Provide the query: ``What is the primary film genre of `The Importance of Being Icelandic'? Provide the genre name.''\\  
  2. Use the \texttt{ask\_search\_agent} again to determine the genre of ``The Five Obstructions'' with the query: ``What is the primary film genre of `The Five Obstructions'? Provide the genre name.''\\  
  3. Compare the two genre names obtained from the searches. If they are distinct, conclude that the films belong to different genres; otherwise, state they belong to the same genre.
}{%
  1. Use \texttt{ask\_search\_agent} to determine the primary film genre of ``The Importance of Being Icelandic'' by querying a reliable source like IMDb or Wikipedia. Provide \texttt{output\_content} as the genre name.\\  
  2. Use \texttt{ask\_search\_agent} to determine the primary film genre of ``The Five Obstructions'' using the same method. Provide \texttt{output\_content} as the genre name.\\ 
  3. Compare the two genre names obtained from the searches.\\  
  4. Use \texttt{final\_answer} to state whether the films belong to different genres based on the comparison.  
}{%
  Improved tool and argument specification, provides (additional) trusted sources for search
}\label{ex:genres}

\newpage
\PlanTriple{%
  According to Google Finance, when was the first year the Apple stock went above \$50 (without adjusting for stock split)?%
}{%
1. Use the \texttt{ask\_search\_agent} to query Google Finance for the first date when Apple's stock (AAPL) exceeded \$50 without adjusting for stock splits. Specify the query as: ``According to Google Finance, what was the first date Apple's stock (AAPL) went above \$50 without adjusting for stock splits? Provide the date.''\\
2. Extract the year from the identified date.\\
3. Verify the result by cross-checking Apple's stock split history to ensure the price threshold is unadjusted.\\  
4. Use the \texttt{final\_answer} tool to confirm the first year. 
}{%
  1. Use \texttt{ask\_search\_agent} to query Google Finance for the earliest date Apple's stock (AAPL) exceeded \$50 **without adjusting for stock splits**. Query: ``According to Google Finance, what was the first date Apple's stock (AAPL) closed above \$50 without adjusting for stock splits? Provide the date.''\\  
  2. Extract the year from the retrieved date.\\  
  3. Use \texttt{ask\_search\_agent} to confirm Apple's stock split history to ensure the identified date is **pre-split** (e.g., check split dates and ratios). Query: ``When did Apple (AAPL) undergo stock splits? Provide split dates and ratios.''\\ 
  4. Cross-reference the split dates with the retrieved date to verify it is unadjusted.\\  
  5. Use \texttt{final\_answer} to submit the confirmed year.
}{%
  Improved tool and argument specification, intermediate cross-reference/consistency checks
}\label{ex:apple}

\vspace{15pt}
\PlanTriple{%
  How many pages of the 2023 IPCC report (85 pages version) mention nuclear energy?
}{%
  1. Use \texttt{ask\_search\_agent} to search for the 2023 IPCC report's 85-page version text, specifying the need for the full content to analyze mentions of nuclear energy. Provide the query: ``What is the full text of the 2023 IPCC report (85 pages version)? Provide the complete document content.''\\  
  2. Use \texttt{inspect\_file\_as\_text} to analyze the retrieved report text (assuming the search provides a file path or URL) and identify all instances of terms related to nuclear energy (e.g., ``nuclear energy,'' ``nuclear power,'' ``nuclear fission'').\\  
  3. Count the unique pages containing these terms from the inspected text.\\  
  4. Use \texttt{final\_answer} to report the total number of pages mentioning nuclear energy.  
}{%
  1. Use \texttt{ask\_search\_agent} to locate the official 2023 IPCC report (85 pages version) and retrieve its full text or a downloadable PDF link from the IPCC website (https://www.ipcc.ch/). Provide the query: ``Where can I access the full text of the 2023 IPCC report (85 pages version)? Provide a direct download link.''\\
  2. Use \texttt{inspect\_file\_as\_text} to analyze the retrieved report text (via the provided link) and identify all instances of terms related to nuclear energy (e.g., ``nuclear energy,'' ``nuclear power,'' ``nuclear fission'').\\  
  3. Count the number of unique pages containing these terms, ensuring no duplicates.\\  
  4. Use \texttt{final\_answer} to submit the total count of pages mentioning nuclear energy. 
}{%
  Provides (additional) trusted sources for search
}\label{ex:nuclear}

\newpage
\vspace*{1pt}
\PlanTriple{%
  Who was the director of Chinese Coffee?
}{%
  Use \texttt{ask\_search\_agent} to search for ``Who is the director of the movie Chinese Coffee? Provide the name.'' on IMDb (https://www.imdb.com).\\  
  Extract the director's name from the search result.\\  
  Use \texttt{final\_answer} to provide the identified director's name.
}{%
  1. Use \texttt{ask\_search\_agent} to search for ``Who directed the movie Chinese Coffee?'' and request the output to provide the director's name.\\
  2. Analyze the detailed search results to confirm the director's identity from reliable sources (e.g., IMDb, Wikipedia).\\  
  3. Cross-check for title variations or potential ambiguities (e.g., ``Chinese Coffee'' vs. similar titles) in the search results.\\  
  4. Use \texttt{final\_answer} to provide the confirmed director's name.
}{%
  Provides (additional) trusted sources for search, intermediate cross-reference/consistency checks, clearer/more structured layout
}\label{ex:coffee}
\vspace{7.5pt}

\subsubsection{Coding Examples}
\vspace{5pt}
\PlanTriple{%
  Write a function to verify validity of a string of parentheses. Your code should pass the following test: \texttt{assert is\_valid\_parenthese("(){}[]")==True}
}{%
  1. Implement a function using a stack to track opening brackets.\\
  2. Create a mapping of closing brackets to their corresponding opening brackets.\\
  3. Iterate through each character in the input string.\\
  4. Push opening brackets onto the stack.\\
  5. For closing brackets, check if the stack is empty or the top element does not match the expected opening bracket.\\
  6. Return True if the stack is empty after processing all characters, otherwise return False.
}{%
  1. Analyze the task: Create a function to validate balanced parentheses using a stack-based approach. \\ 
  2. Define a mapping of closing brackets to their corresponding opening brackets.\\  
  3. Iterate through each character in the input string, pushing opening brackets onto the stack and checking for matches when encountering closing brackets.\\  
  4. Return \texttt{True} if the stack is empty after processing all characters (indicating balanced parentheses), otherwise \texttt{False}.\\  
  5. Validate the function with the test case \texttt{assert is\_valid\_parenthese("(){}[]") == True}.\\  
  6. Use the \texttt{final\_answer} tool to submit the implemented code.  
}{%
  Explicit focus on returning the correct artifact, systematic testing and edge-case coverage 
}\label{ex:parentheses}

\newpage
\PlanTriple{%
  Write a function to calculate magic square. Your code should pass the following test: \texttt{assert magic\_square\_test([[7, 12, 1, 14], [2, 13, 8, 11], [16, 3, 10, 5], [9, 6, 15, 4]])==True}
}{%
  1. Verify the input matrix is a square (all rows have the same length as the number of rows).\\
  2. Check all elements are unique and within the range 1 to $n^2$.\\
  3. Calculate the magic constant as the sum of the first row.\\
  4. Validate all rows, columns, and both diagonals sum to the magic constant.\\
  5. Return True if all checks pass, otherwise False.
}{%
  1. Define the function \texttt{magic\_square\_test} that takes a matrix as input.
  2. Verify the matrix is a square (all rows have the same length as the number of rows).\\
  3. Check all elements are unique and within the range 1 to $n^2$ (where $n$ is the size of the matrix).\\
  4. Calculate the magic constant using the formula \texttt{n * (n**2 + 1) // 2}.\\
  5. Validate all rows, columns, and both diagonals sum to the magic constant.\\
  6. Return \texttt{True} if all checks pass, otherwise \texttt{False}.\\
  7. Use the \texttt{final\_answer} tool to submit the implemented function as the solution. 
}{%
  Precise function and helper signatures, explicit focus on returning the correct artifact
}\label{ex:magic_square}

\vspace{20pt}
\PlanTriple{%
  Write a function to sort a list of elements using comb sort. Your code should pass the following tests: assert \texttt{comb\_sort([5, 15, 37, 25, 79]) == [5, 15, 25, 37, 79]}
}{%
  1. Create a copy of the input list to avoid modifying the original.\\
  2. Initialize the gap as the length of the list divided by 1.3, then cast to an integer.\\
  3. While the gap is greater than 1, iterate through the list with the current gap, comparing elements i and i + gap. If they are out of order, swap them.\\
  4. After each pass, update the gap by dividing it by 1.3 and casting to an integer.\\
  5. Once the gap is 1, perform a final bubble sort pass to ensure the list is fully sorted.\\
  6. Return the sorted list.
}{%
  1. Implement the \texttt{comb\_sort} function to sort a list using the comb sort algorithm.\\  
  2. Initialize the gap as the length of the list divided by 1.3, then cast to an integer.\\  
  3. While the gap is greater than 1, iterate through the list and compare elements \texttt{i} and  \texttt{i + gap}. Swap them if they are out of order.\\  
  4. After each pass, reduce the gap by dividing it by 1.3 and cast to an integer, ensuring it does not fall below 1.\\  
  5. Once the gap is 1, perform a final pass to ensure the list is fully sorted (similar to bubble sort).\\  
  6. Return the sorted list.\\  
  7. Use the \texttt{final\_answer} tool to submit the implemented function as the solution.
}{%
  Precise function and helper signatures, explicit focus on returning the correct artifact
}\label{ex:comb_sort}

\newpage
\PlanTriple{%
  Write a python function to find the sum of common divisors of two given numbers. Your code should pass the following test: \texttt{assert sum(10,15) == 6}
}{%
  1. Compute the greatest common divisor (GCD) of the two input numbers.\\
  2. Find all divisors of the GCD by iterating from 1 to the GCD and checking for divisibility.\\
  3. Sum all the divisors found in step 2.\\
  4. Return the computed sum as the result of the function.
}{%
  1. **Understand the problem**: Create a Python function \texttt{sum(a, b)} that returns the sum of all common divisors of \texttt{a} and \texttt{b}.\\
  \\
  2. **Compute GCD**: Use Euclidean algorithm to find the greatest common divisor (GCD) of the two numbers, as all common divisors of \texttt{a} and \texttt{b} are divisors of their GCD.\\
  \\
  3. **Find divisors of GCD**: Efficiently find all divisors of the GCD by iterating up to its square root. For each divisor \texttt{i}, add both \texttt{i} and \texttt{gcd // i} to the sum (if they are distinct).\\
  \\
  4. **Handle edge cases**: Ensure the function works for inputs where GCD is 0 (though inputs are positive integers as implied by the test case).\\
  \\
  5. **Implement the function** in Python using the described logic.\\
  \\
  6. **Verify correctness** with the provided test case \texttt{sum(10, 15) == 6}.\\
  \\
  7. **Provide the final answer** using the \texttt{final\_answer} tool.
}{%
  Precise function and helper signatures, systematic testing and edge-case coverage, clearer/more structured layout 
}\label{ex:common_divisors}

\subsection{Refinement and Auction Overhead}
\label{appendix:overhead}

We quantify the overhead introduced by the auction phase, including refinement, along two axes: tokens consumed and wall-clock time.

\para{Token overhead.} On average, \textsc{sale} generates 669 additional tokens per task for deep search and 1042 for coding, including strategy generation, refinement, and jury-vote tokens across all agents. By comparison, full agentic trajectories span tens of thousands to millions of tokens (Figure~\ref{fig:plots:b}), so the auction phase represents under 1\% of total token consumption.

\para{Wall-clock overhead.} We measure average wall time of the full auction phase, including refinement, on a stratified subset of 100 \textsc{HST-Bench} tasks. Agents are deployed as vLLM servers and queried concurrently within each phase, so wall time is bounded by the slowest agent. Table~\ref{tab:wall_time} reports the per-task wall time under our setup. By comparison, executing the final agentic trace of the best-performing agent takes 5--10 minutes on average, so the auction phase represents approximately 1\% of total wall time.

\begin{table}[h]
\centering
\vspace{15pt}
\setlength{\tabcolsep}{40pt}%
\caption{Average wall-clock time of the auction phase per task, including strategy generation, jury scoring, memory-based refinement, and post-refinement jury scoring.}
\label{tab:wall_time}
\begin{tabular}{ccc}
\toprule
GPU type & GPUs per node & Avg. wall time per task \\
\midrule
H100 & 8 & 3.59s \\
\bottomrule
\end{tabular}
\end{table}

%%%%%%%%%%%%%%%%%%%%%%%%%%%%%%%%%%%%%%%%%%%%%%%%%%%%%%%%%%%%%%%%%%%%%%%%%

\newpage\section{Significance Tests}\label{appendix:tests}

Methods with memory are sensitive to test-set ordering, so we evaluate memory-based methods across five runs with different random permutations; baselines without memory use a single run. To assess significance, we employ two-tailed one-sample $t$-tests and bootstrap confidence intervals (CIs), which provide robustness against distributional assumptions.

We compare against five baselines: the best single agent, \textsc{wtp}, \textsc{carrot}, TO-Router, and FrugalGPT (Tables~\ref{tab:t-tests} and~\ref{tab:bootstrap-ci}). On the full test set, \textsc{sale} achieves significantly higher pass@1 than all baselines across both task types, with large effect sizes (all $|t| > 5$, $p < .005$); bootstrap CIs confirm these results. Within individual bins, a few comparisons yield mixed results across the two tests, potentially reflecting reduced power in relatively smaller samples; where they diverge, at least one test typically remains significant: FrugalGPT on coding $\tau(t) \le 0.1$ ($p = .208$ for pass@1, though \$/Mt remains highly significant); best single agent and TO-Router on coding $\tau(t) \le 60$ (borderline $p \approx .057$, with \$/Mt strongly significant); and \textsc{carrot} on coding $\tau(t) \le 0.5$ ($p = .339$), where \textsc{carrot}'s marginal advantage over \textsc{sale} is not statistically significant. For cost, \textsc{sale} significantly reduces \$/Mt versus the best single agent, \textsc{carrot}, TO-Router, and FrugalGPT across all conditions. Compared to \textsc{wtp}, \textsc{sale} dominates on deep search (both higher pass@1 and lower \$/Mt); on coding, \textsc{wtp} achieves lower cost but at substantially degraded performance.

\vspace{5pt}
\begin{table}[H]
    \centering
    \setlength{\tabcolsep}{1pt}%
    \renewcommand{\arraystretch}{1.0}%
    \scalebox{0.775}{%
    \begin{NiceTabular}{lc*{20}{c}}[cell-space-limits=2pt]
    \CodeBefore
      \rectanglecolor{metabg}{9-2}{9-22}
      \rectanglecolor{metabg}{15-2}{15-22}
    \Body
    \toprule
      \Block{3-1}{\textbfrm{Task} \\\textbfrm{type}} &
      \Block{3-1}{{$\tau(t)$}} &
      \Block{1-4}{\textbfrm{\small{Best single agent}}} & & & &
      \Block{1-4}{\textbfrm{\textsc{wtp}}} & & & &
      \Block{1-4}{\textbfrm{\textsc{carrot}}} & & & &
      \Block{1-4}{\textbfrm{\footnotesize{TO-Router}}} & & & &
      \Block{1-4}{\textbfrm{\footnotesize{FrugalGPT}}} & & & \\
      \cmidrule(lr){3-6} \cmidrule(lr){7-10} \cmidrule(lr){11-14} \cmidrule(lr){15-18} \cmidrule(lr){19-22}
      & & \Block{1-2}{Pass@1} & & \Block{1-2}{\$/Mt} & &
          \Block{1-2}{Pass@1} & & \Block{1-2}{\$/Mt} & &
          \Block{1-2}{Pass@1} & & \Block{1-2}{\$/Mt} & &
          \Block{1-2}{Pass@1} & & \Block{1-2}{\$/Mt} & &
          \Block{1-2}{Pass@1} & & \Block{1-2}{\$/Mt} & \\
      \cmidrule(lr){3-4} \cmidrule(lr){5-6} \cmidrule(lr){7-8} \cmidrule(lr){9-10} \cmidrule(lr){11-12} \cmidrule(lr){13-14} \cmidrule(lr){15-16} \cmidrule(lr){17-18} \cmidrule(lr){19-20} \cmidrule(lr){21-22}
      & & $t$ & $p$ & $t$ & $p$
        & $t$ & $p$ & $t$ & $p$
        & $t$ & $p$ & $t$ & $p$
        & $t$ & $p$ & $t$ & $p$
        & $t$ & $p$ & $t$ & $p$ \\
    \midrule
    \Block{6-1}{\textit{Deep} \\\textit{search}} & $\le0.1\phantom{0}$
      & — & — & -37.95 & {<}.001
      & — & — & -27.26 & {<}.001
      & — & — & -13.90 & {<}.001
      & — & — & -16.57 & {<}.001 
      & — & — & -67.35 & {<}.001 \\
    & $\le0.5\phantom{0}$
      & 4.00 & .016 & -36.88 & {<}.001
      & 8.80 & {<}.001 & -28.86 & {<}.001
      & 8.80 & {<}.001 & -15.50 & {<}.001
      & 8.80 & {<}.001 & -26.19 & {<}.001 
      & 28.80 & {<}.001 & -68.95 & {<}.001 \\
    & $\le2.5\phantom{0}$
      & 7.76 & .002 & -35.28 & {<}.001
      & 9.80 & {<}.001 & -21.92 & {<}.001
      & 11.76 & {<}.001 & -16.57 & {<}.001
      & 9.80 & {<}.001 & -29.93 & {<}.001 
      & 11.76 & {<}.001 & -80.71 & {<}.001 \\
    & $\le12.5$
      & 4.54 & .011 & -51.31 & {<}.001
      & 3.15 & .035 & -40.62 & {<}.001
      & 8.52 & .001 & -40.62 & {<}.001
      & 4.57 & .010 & -51.31 & {<}.001 
      & 7.20 & .002 & -88.73 & {<}.001 \\
    & $\le60\phantom{.}\phantom{0}$
      & 6.00 & .004 & -11.09 & {<}.001
      & 10.96 & {<}.001 & -6.95 & .002
      & 10.96 & {<}.001 & -4.47 & .011
      & 6.00 & .004 & -11.09 & {<}.001 
      & 6.00 & .004 & -30.95 & {<}.001 \\
    & All
      & 14.36 & {<}.001 & -47.43 & {<}.001
      & 20.12 & {<}.001 & -34.79 & {<}.001
      & 24.64 & {<}.001 & -22.14 & {<}.001
      & 17.65 & {<}.001 & -37.95 & {<}.001
      & 25.88 & {<}.001 & -94.87 & {<}.001 \\
    \midrule
    \Block{6-1}{\textit{Coding}} & $\le0.1\phantom{0}$
      & 6.50 & .003 & -91.00 & {<}.001
      & 8.90 & {<}.001 & 9.00 & {<}.001
      & 6.50 & .003 & -91.00 & {<}.001
      & 6.50 & .003 & -91.00 & {<}.001 
      & 1.50 & .208 & -106.00 & {<}.001 \\
    & $\le0.5\phantom{0}$
      & 9.00 & {<}.001 & -23.52 & {<}.001
      & 23.80 & {<}.001 & 32.61 & {<}.001
      & -1.09 & .339 & 5.88 & .004
      & 9.19 & {<}.001 & -23.52 & {<}.001 
      & 49.08 & {<}.001 & -90.33 & {<}.001 \\
    & $\le2.5\phantom{0}$
      & 6.00 & .004 & -26.94 & {<}.001
      & 36.00 & {<}.001 & 58.79 & {<}.001
      & 36.00 & {<}.001 & 13.88 & {<}.001
      & 6.00 & .004 & -26.94 & {<}.001 
      & 50.80 & {<}.001 & -129.01 & {<}.001 \\
    & $\le12.5$
      & 2.98 & .041 & -6.42 & .003
      & 14.47 & {<}.001 & 29.14 & {<}.001
      & 2.95 & .042 & -6.42 & .003
      & 2.95 & .042 & -6.42 & .003 
      & 11.03 & {<}.001 & -35.10 & {<}.001 \\
    & $\le60\phantom{.}\phantom{0}$
      & 2.65 & .057 & -14.70 & {<}.001
      & 15.95 & {<}.001 & 48.58 & {<}.001
      & 3.69 & .021 & -12.66 & {<}.001
      & 2.64 & .058 & -14.70 & {<}.001 
      & 12.88 & {<}.001 & -65.73 & {<}.001 \\
    & All
      & 5.90 & .004 & -38.38 & {<}.001
      & 24.02 & {<}.001 & 63.69 & {<}.001
      & 8.74 & {<}.001 & -17.96 & {<}.001
      & 5.90 & .004 & -38.38 & {<}.001
      & 23.37 & {<}.001 & -124.11 & {<}.001 \\
    \bottomrule
    \end{NiceTabular}
    }
\caption{
One-sample $t$-test results comparing \textsc{sale} against each baseline. We report $t$-statistics and $p$-values for pass@1 and \$/Mt across task types and complexity bins. Dashes (—) indicate undefined tests due to zero variance.}
\label{tab:t-tests}
\end{table}

\begin{table}[H]
    \centering
    \setlength{\tabcolsep}{1.5pt}%
    \renewcommand{\arraystretch}{1.0}%
    \scalebox{0.745}{%
    \begin{NiceTabular}{lc*{10}{c}}[cell-space-limits=2pt]
    \CodeBefore
      \rectanglecolor{metabg}{8-2}{8-12}
      \rectanglecolor{metabg}{14-2}{14-12}
    \Body    
    \toprule
      \Block{2-1}{\textbfrm{Task}\\\textbfrm{type}} &
      \Block{2-1}{{$\tau(t)$}} &
      \Block{1-2}{\textbfrm{\small{Best single agent}}} & &
      \Block{1-2}{\textbfrm{\textsc{wtp}}} & &
      \Block{1-2}{\textbfrm{\textsc{carrot}}} & &
      \Block{1-2}{\textbfrm{\footnotesize{TO-Router}}} & &
      \Block{1-2}{\textbfrm{\footnotesize{FrugalGPT}}} & \\
      \cmidrule(lr){3-4} \cmidrule(lr){5-6} \cmidrule(lr){7-8} \cmidrule(lr){9-10} \cmidrule(lr){11-12}
      & & Pass@1 & \$/Mt & Pass@1 & \$/Mt & Pass@1 & \$/Mt & Pass@1 & \$/Mt & Pass@1 & \$/Mt \\
    \midrule
    \Block{5-1}{\textit{Deep}\\\textit{search}} & $\le0.1\phantom{0}$
      & [3.75, 3.75] & [-0.15, -0.14] & [7.45, 7.45] & [-0.11, -0.10] & [6.25, 6.25] & [-0.06, -0.05] & [4.95, 4.95] & [-0.07, -0.06] & [4.95, 4.95] & [-0.26, -0.25] \\
    & $\le0.5\phantom{0}$
      & [0.50, 1.25] & [-0.14, -0.13] & [1.70, 2.45] & [-0.11, -0.10] & [1.70, 2.45] & [-0.06, -0.05] & [1.70, 2.45] & [-0.10, -0.09] & [6.70, 7.45] & [-0.26, -0.25] \\
    & $\le2.5\phantom{0}$
      & [3.75, 5.75] & [-0.14, -0.13] & [5.00, 7.00] & [-0.09, -0.08] & [6.20, 8.20] & [-0.07, -0.06] & [5.00, 7.00] & [-0.12, -0.11] & [6.20, 8.20] & [-0.31, -0.30] \\
    & $\le12.5$
      & [2.68, 5.61] & [-0.20, -0.19] & [1.41, 4.34] & [-0.16, -0.15] & [6.31, 9.24] & [-0.16, -0.15] & [2.71, 5.64] & [-0.20, -0.19] & [5.11, 8.04] & [-0.34, -0.33] \\
    & $\le60\phantom{.}\phantom{0}$
      & [3.13, 5.00] & [-0.16, -0.12] & [6.23, 8.10] & [-0.11, -0.07] & [6.23, 8.10] & [-0.08, -0.04] & [3.13, 5.00] & [-0.16, -0.12] & [3.13, 5.00] & [-0.40, -0.36] \\
    & All
      & [3.04, 3.83] & [-0.16, -0.14]
      & [4.44, 5.23] & [-0.12, -0.10] 
      & [5.54, 6.33] & [-0.08, -0.06]
      & [3.84, 4.63] & [-0.13, -0.11]
      & [5.84, 6.63] & [-0.31, -0.29] \\
    \midrule
    \Block{5-1}{\textit{Coding}} & $\le0.1\phantom{0}$
      & [2.25, 3.75] & [-0.19, -0.18] & [3.45, 4.95] & [0.01, 0.02] & [2.25, 3.75] & [-0.19, -0.18] & [2.25, 3.75] & [-0.19, -0.18] & [-0.25, 1.25]$^\dagger$ & [-0.22, -0.21] \\
    & $\le0.5\phantom{0}$
      & [1.77, 2.53] & [-0.09, -0.08] & [5.52, 6.28] & [0.12, 0.13] & [-0.78, -0.02] & [0.02, 0.03] & [1.82, 2.58] & [-0.09, -0.08] & [11.92, 12.68] & [-0.34, -0.33] \\
    & $\le2.5\phantom{0}$
      & [1.25, 2.00] & [-0.07, -0.06] & [8.75, 9.50] & [0.14, 0.15] & [8.75, 9.50] & [0.03, 0.04] & [1.25, 2.00] & [-0.07, -0.06] & [12.45, 13.20] & [-0.32, -0.31] \\
    & $\le12.5$
      & [1.48, 4.94] & [-0.07, -0.04] & [13.84, 17.30] & [0.24, 0.27] & [1.44, 4.90] & [-0.07, -0.04] & [1.44, 4.90] & [-0.07, -0.04] & [10.14, 13.60] & [-0.32, -0.29] \\
    & $\le60\phantom{.}\phantom{0}$
      & [1.27, 5.32] & [-0.08, -0.06] & [17.75, 21.80] & [0.23, 0.25] & [2.55, 6.60] & [-0.07, -0.05] & [1.25, 5.30] & [-0.08, -0.06] & [13.95, 18.00] & [-0.33, -0.31] \\
    & All
      & [1.95, 3.55] & [-0.10, -0.09]
      & [10.25, 11.85] & [0.15, 0.16]
      & [3.25, 4.85] & [-0.05, -0.04]
      & [1.95, 3.55] & [-0.10, -0.09]
      & [9.95, 11.55] & [-0.31, -0.30] \\
    \bottomrule
    \end{NiceTabular}
    }
\caption{
Bootstrap 95\% confidence intervals (10,000 resamples) for the difference between \textsc{sale} and each baseline. Positive pass@1 and negative \$/Mt indicate \textsc{sale} outperforms the baseline. CIs are constructed by resampling \textsc{sale}'s five runs against each baseline's single-run reference value. $^\dagger$Not significant (CI includes 0).}
\label{tab:bootstrap-ci}
\end{table}

%%%%%%%%%%%%%%%%%%%%%%%%%%%%%%%%%%%%%%%%%%%%%%%%%%%%%%%%%%%%%%%%%%%%%%%%%

\newpage\section{Baselines}\label{appendix:baselines}

\subsection{Baseline Implementation Details}
We provide implementation details and hyperparameter configurations for the baseline routing methods evaluated in Section~\ref{sec:results}.

\vspace{15pt}
\para{WTP Router.} The WTP router predicts model performance using a $K$-Nearest Neighbors classifier, then selects models based on a performance–cost trade-off controlled by the willingness-to-pay parameter. Following~\citet{hu2024routerbench}, we use $k=50$ neighbors with cosine distance over sentence embeddings. We tune the willingness-to-pay parameter on the deep search and coding training sets. Table~\ref{tab:wtp-hyperparams} presents the hyperparameter configuration.
\vspace{5pt}
\begin{table}[H]
\centering
\setlength{\tabcolsep}{60pt} 
\renewcommand{\arraystretch}{1.2}
\begin{tabular}{ll}
\toprule
\textbf{Hyperparameter} & \textbf{Value} \\
\midrule
$k$ (neighbors) & 50 \\
Embedding model & \texttt{all-MiniLM-L6-v2} \\
Distance metric & Cosine \\
\midrule
Willingness-to-pay (deep search) & 5.0 \\
Willingness-to-pay (coding) & 5.0 \\
\bottomrule
\end{tabular}
\caption{WTP router hyperparameters.}
\label{tab:wtp-hyperparams}
\end{table}

\vspace{15pt}
\para{CARROT Router.} \textsc{CARROT}~\citep{somerstep2025carrotcostawarerate} fine-tunes a RoBERTa encoder as a multi-label classifier to predict the probability of success for each candidate model given an input query. At inference time, routing decisions are made by selecting the model that maximizes a performance–cost trade-off score: $(1 - \mu) \cdot p_i - \mu \cdot c_i$, where $p_i$ is the predicted success probability and $c_i$ is the normalized cost of model $i$. The hyperparameter $\mu \in [0, 1]$ controls the trade-off between performance ($\mu = 0$) and cost ($\mu = 1$). Due to the limited size of our training set, we freeze the RoBERTa backbone and only train the classification head, which we found to yield more stable predictions. We set $\mu = 0.0$ based on development set performance; higher values of $\mu$ caused the router to route almost exclusively to the smallest agent, degrading task performance without meaningful cost savings. Table~\ref{tab:carrot-hyperparams} presents the hyperparameter configuration.

\vspace{5pt}
\begin{table}[H]
\centering
\setlength{\tabcolsep}{70pt} 
\renewcommand{\arraystretch}{1.2}
\begin{tabular}{ll}
\toprule
\textbf{Hyperparameter} & \textbf{Value} \\
\midrule
\multicolumn{2}{l}{\textit{Performance Predictor}} \\
Base model & \texttt{roberta-base} \\
Backbone & Frozen \\
Optimizer & AdamW \\
Learning rate & $1 \times 10^{-3}$ \\
Weight decay & $1 \times 10^{-2}$ \\
Batch size & 16 \\
Epochs & 10 \\
\midrule
\multicolumn{2}{l}{\textit{Routing}} \\
Cost-performance tradeoff ($\mu$) & 0.0 \\
\bottomrule
\end{tabular}
\caption{CARROT hyperparameters.}
\label{tab:carrot-hyperparams}
\end{table}

\para{TO-Router.} The TO-Router baseline fine-tunes a BERT encoder to predict model performance scores, using soft labels and inverse class frequency weighting to handle imbalanced performance distributions. We follow the implementation of \citet{stripelis-etal-2024-tensoropera}. Table~\ref{tab:torouter-hyperparams} presents the hyperparameter configuration.
\vspace{5pt}
\begin{table}[H]
\centering
\setlength{\tabcolsep}{67pt} 
\renewcommand{\arraystretch}{1.2}
\begin{tabular}{ll}
\toprule
\textbf{Hyperparameter} & \textbf{Value} \\
\midrule
Base model & \texttt{bert-base-uncased} \\
\midrule
Optimizer & AdamW \\
Learning rate & $5 \times 10^{-5}$ \\
Weight decay & $1 \times 10^{-4}$ \\
Batch size & 8 \\
Epochs & 5 \\
\midrule
Soft label temperature & 10 \\
Sample weighting & Inverse class frequency \\
Loss function & Cross-entropy \\
\bottomrule
\end{tabular}
\caption{TO-Router hyperparameters.}
\label{tab:torouter-hyperparams}
\end{table}

\para{FrugalGPT.} For the FrugalGPT baseline, we follow the implementation of \citet{chen2024frugalgpt}, which consists of two components: (i) a scoring function that predicts answer correctness, and (ii) a cascade optimizer that determines the optimal model ordering and acceptance thresholds. Table~\ref{tab:frugalgpt-hyperparams} presents the hyperparameter configuration.
\vspace{5pt}
\begin{table}[H]
\centering
\setlength{\tabcolsep}{65pt} 
\renewcommand{\arraystretch}{1.2}
\begin{tabular}{ll}
\toprule
\textbf{Hyperparameter} & \textbf{Value} \\
\midrule
\multicolumn{2}{l}{\textit{Scoring Function}} \\
Base model & \texttt{distilbert-base-uncased} \\
Optimizer & AdamW \\
Learning rate & $2 \times 10^{-5}$ \\
Weight decay & $1 \times 10^{-2}$ \\
Warmup steps & 500 \\
Batch size & 8 \\
Epochs & 8 \\
Gradient clipping & 1.0 \\
\midrule
\multicolumn{2}{l}{\textit{Cascade Optimization}} \\
Max cascade length & 3 \\
Quantile grid steps & 20 \\
Optimizer & \texttt{scipy.brute} + Nelder-Mead \\
\bottomrule
\end{tabular}
\caption{FrugalGPT hyperparameters.}
\label{tab:frugalgpt-hyperparams}
\end{table}

\subsection{Pareto Frontier vs. Baseline Routers}
\begin{figure}
    \centering
    \includegraphics[width=0.99\linewidth]{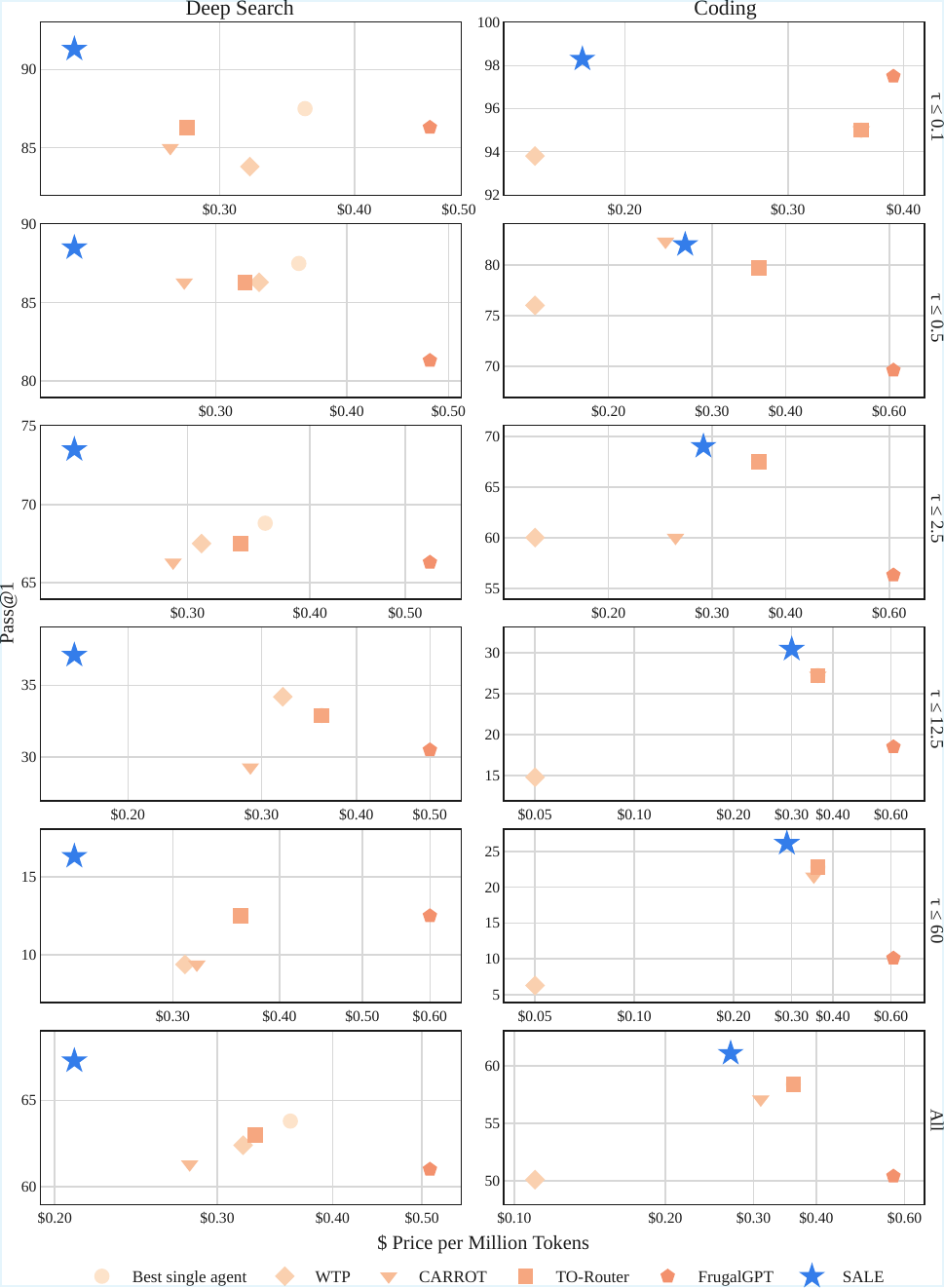}
    \caption{\textsc{sale} vs. baselines on the accuracy–cost trade-off. The upper-left envelope is Pareto-optimal.}
    \label{fig:pareto-scatter}
\end{figure}

Figure~\ref{fig:pareto-scatter} visualizes the accuracy--cost trade-offs reported in Section~\ref{sec:results} (Table~\ref{tab:results}). On deep search tasks, \textsc{sale} shifts the Pareto frontier outward for every value of $\tau$. On coding, \textsc{sale} improves the frontier in all bins except $\tau \le 0.5$, where \textsc{CARROT} attains a comparable trade-off. We also note that \textsc{WTP} can achieve lower cost on coding tasks, but only with substantial accuracy degradation.

%%%%%%%%%%%%%%%%%%%%%%%%%%%%%%%%%%%%%%%%%%%%%%%%%%%%%%%%%%%%%%%%%%%%%%%

\newpage\section{Oracle Results and Agent Selection Analysis}\label{appendix:oracle}

\subsection{Oracle Router}\label{appendix:oracle-router}
Table~\ref{tab:oracle} reports the oracle router results on \textsc{HST-Bench}, as an upper bound on what is achievable from per-task model selection alone. For each task $t$ we run all candidate agents and select the smallest agent that produces a correct final answer. If all answers are incorrect, we select the smallest agent in the pool. This yields the minimum possible spend subject to achieving correctness whenever any agent in the pool can solve the task. We note, however, that in complexity bins where no agent solves a large fraction of instances, the oracle's effective cost can be artificially low: on those unsolved tasks the choice of agent does not affect accuracy, and our definition defaults to the cheapest model, thereby minimizing expenditure in a regime where additional compute would not change outcomes. As a result, oracle cost in low-accuracy bins should be interpreted cautiously, primarily as a lower bound on spend rather than a representative operating point.

Relative to this hindsight upper bound, \textsc{sale} captures a substantial fraction of the available routing gains on easier tasks, but the gap to oracle performance grows with task complexity. For deep search, at $\tau(t) \le 0.1$ the oracle attains 97.5 pass@1 versus 91.3 for \textsc{sale} (Table~\ref{tab:results}), while for coding the gap is only 0.5 points (98.8 vs. 98.3). At higher complexity, the oracle–\textsc{sale} gap widens (deep search $\tau(t) \le 60$: 25.0 vs. 16.3; coding $\tau(t) \le 60$: 34.2 vs. 26.1). We also observe a growing gap between the oracle and the best single agent as task complexity increases, peaking in the medium–high complexity bins. This implies increased opportunity for routing in these instances.

While a gap between \textsc{sale} and the oracle remains, \textsc{sale} consistently exhibits the smallest accuracy gap to the oracle among all methods. On deep search, \textsc{sale}'s gap to the oracle is 10.6 points (67.3 vs. 77.9), compared to 14.1–16.9 points for the baselines (WTP: 15.5; \textsc{CARROT}: 16.6; TO-Router: 14.9; FrugalGPT: 16.9; best single agent: 14.1). On coding, \textsc{sale}'s accuracy gap is 7.3 points (61.1 vs. 68.4), whereas baselines range from 10.0 to 18.3 points behind the oracle (WTP: 18.3; FrugalGPT: 18.0; \textsc{CARROT}: 11.3; TO-Router and best single agent: 10.0). In cost, \textsc{sale} also closes more of the gap on deep search ($0.21$ vs. oracle $0.07$) than any baseline ($0.28$–$0.51$); on coding, WTP achieves lower cost (\$0.11) but at the expense of a substantially larger accuracy gap (18.3 points vs. \textsc{sale}'s 7.3 points). In short, the oracle–\textsc{sale} gap should be interpreted relative to the considerably wider oracle–baseline gaps: \textsc{sale} captures more of the available routing gains than any existing method, and the residual headroom reflects the inherent difficulty of task-complexity prediction rather than a limitation unique to our approach.

\vspace{15pt}
\begin{table}[H]
    \centering
    \setlength{\tabcolsep}{39pt}%
    \renewcommand{\arraystretch}{1.0}%
    \scalebox{0.98}{%
    \begin{NiceTabular}{lccc}[cell-space-limits=2pt]
    \CodeBefore
      \rectanglecolor{metabg}{8-2}{8-4}
      \rectanglecolor{metabg}{14-2}{14-4}
    \Body
    \toprule
      \Block{2-1}{\textbfrm{Task} \textbfrm{type}} &
      \Block{2-1}{{$\tau(t)$}} &
      \Block{1-2}{\textbfrm{Oracle router}} \\
      \cmidrule(lr){3-4}
      & & Pass@1($\uparrow$) & $\$$/Mt($\downarrow$) \\
    \midrule
    \Block{6-1}{\textit{Deep} \textit{search}} & $\le0.1\phantom{0}$ & 97.5 & 0.06 \\
    & $\le0.5\phantom{0}$ & 93.8 & 0.06 \\
    & $\le2.5\phantom{0}$ & 86.3 & 0.08 \\
    & $\le12.5$           & 56.1 & 0.08 \\
    & $\le60\phantom{.}\phantom{0}$  & 25.0 & 0.09 \\
    & All                 & 77.9 & 0.07 \\
    \midrule
    \Block{6-1}{\textit{Coding}} & $\le0.1\phantom{0}$ & 98.8 & 0.05 \\
    & $\le0.5\phantom{0}$ & 89.9 & 0.08 \\
    & $\le2.5\phantom{0}$ & 76.3 & 0.08 \\
    & $\le12.5$           & 43.2 & 0.08 \\
    & $\le60\phantom{.}\phantom{0}$  & 34.2 & 0.09 \\
    & All                 & 68.4 & 0.08 \\
    \bottomrule
    \end{NiceTabular}
    }
    \caption{
    Oracle-router performance (pass@1) and price per million tokens ($\$$/Mt) across task complexity bins for deep search and coding. The oracle selects, for each task, the smallest agent that returns a correct final answer; if no agent is correct, it defaults to the smallest agent.
    }
    \label{tab:oracle}
\end{table}\vspace{-12pt}

\subsection{Routing Diagnostics}
To better understand \textsc{sale}'s routing behavior and identify opportunities for improvement, we conduct a systematic error analysis comparing \textsc{sale}'s agent selections over a single run against the oracle router described in Appendix~\ref{appendix:oracle-router}. We categorize each routing decision into one of four outcomes:

\begin{enumerate}
    \item \textit{Correct:} \textsc{sale} selects the same agent as the oracle.
    \item \textit{Over-escalation:} \textsc{sale} selects a larger (more expensive) agent than necessary.
    \item \textit{Under-escalation}: \textsc{sale} selects a smaller agent that fails when a larger one would succeed.
    \item \textit{Unavoidable:} No agent in the pool produces a correct answer.
\end{enumerate}

\begin{figure}
    \centering
    \includegraphics[width=\linewidth]{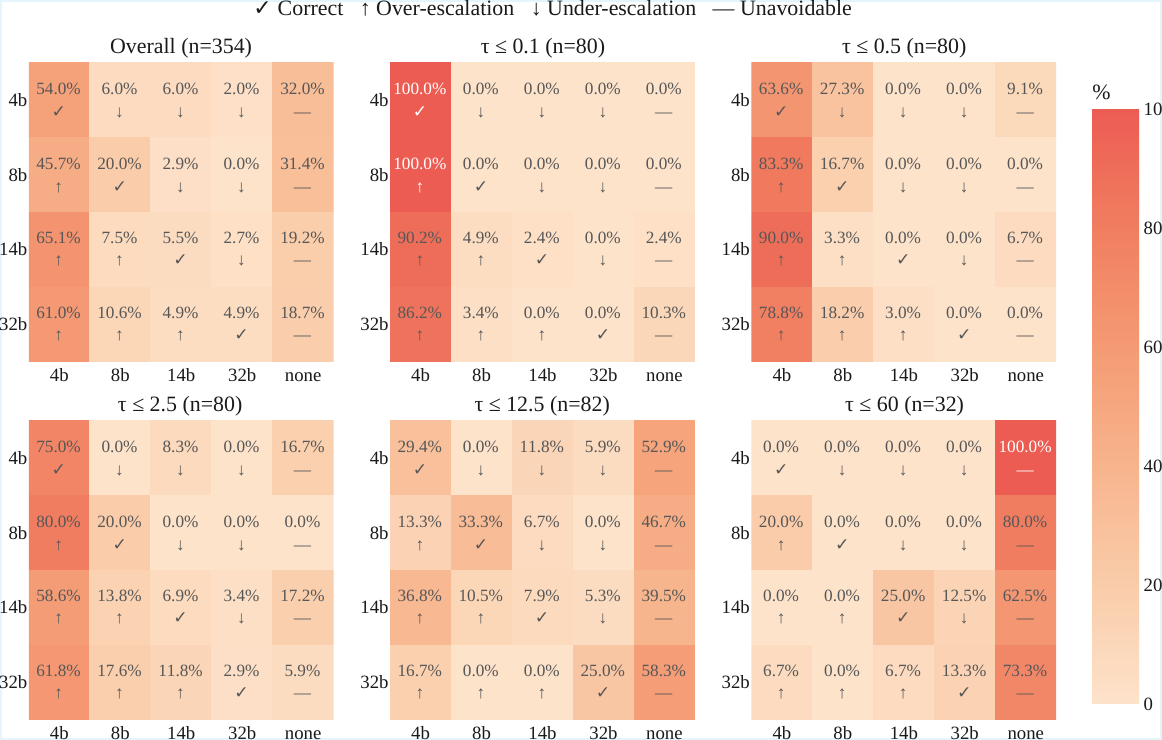}
    \caption{Deep search routing confusion matrices (\textsc{sale} vs.\ oracle). Cells show row-normalized percentages.}
    \label{fig:error_ds}
\end{figure}

\vspace{10pt}
\para{Deep Search Diagnostics.} Figure~\ref{fig:error_ds} presents row-normalized confusion matrices comparing \textsc{sale}'s agent selections against the oracle router described in Appendix~\ref{appendix:oracle-router}, both overall and stratified by complexity bin. Each cell shows the percentage of tasks where \textsc{sale} selected a given agent (row) and the oracle selected another (column), annotated as: correct ($\checkmark$, diagonal), over-escalation ($\uparrow$, \textsc{sale} selects larger than necessary), under-escalation ($\downarrow$, \textsc{sale} selects an agent that fails when a larger one succeeds), or unavoidable (---, no agent succeeds).

In the aggregate matrix ($n=354$), the dominant off-diagonal mass lies in the leftmost column, indicating frequent over-escalation: when \textsc{sale} selects the 14B or 32B agents, 65.1\% and 61.0\% of those cases could have been handled by the 4B agent. Under-escalation (↓) appears infrequently across all rows, confirming that \textsc{sale} rarely sacrifices accuracy by selecting an insufficiently capable agent. Notably, when \textsc{sale} does route to the 4B agent, it achieves the highest diagonal accuracy---54.0\% overall, and even higher within individual bins (100\% at $\tau \le 0.1$, 75.0\% at $\tau \le 2.5$)---suggesting the system correctly identifies 4B-suitable tasks but triggers such routing too conservatively. 

Across complexity bins, over-escalation decreases systematically: at $\tau \le 0.1$, over-escalation from 32B selections to the 4B oracle accounts for 86.2\%, dropping to 16.7\% at $\tau \le 12.5$ and near-zero at $\tau \le 60$. This shift reflects a structural change in the error landscape: as complexity grows, the ``none'' column increasingly dominates (reaching 73–100\% at $\tau \le 60$), indicating that most failures become unavoidable regardless of routing.

\begin{figure}
    \centering
    \includegraphics[width=\linewidth]{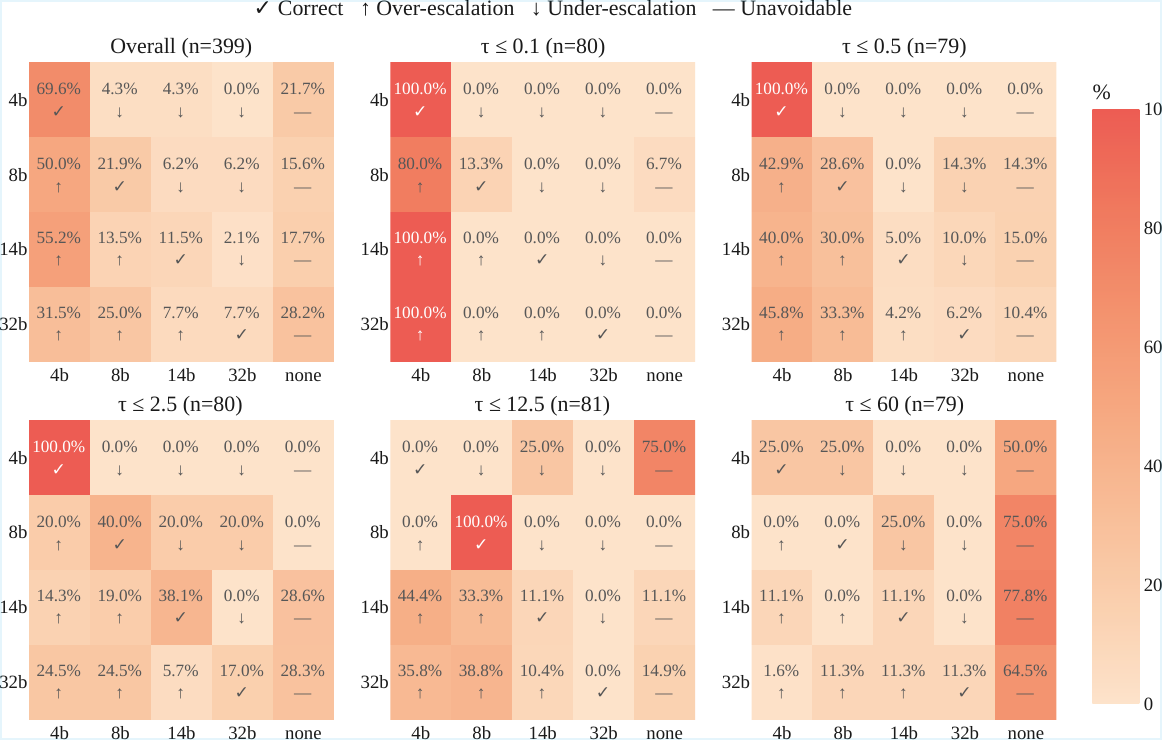}
    \caption{Coding routing confusion matrices (\textsc{sale} vs.\ oracle). Cells show row-normalized percentages.}
    \label{fig:error_co}
\end{figure}

\vspace{10pt}
\para{Coding Diagnostics.} Figure~\ref{fig:error_co} shows row-normalized confusion matrices for coding-agent selections. Mirroring the deep-search pattern, over-escalation dominates the aggregate matrix ($n=399$): when \textsc{sale} selects 14B or 32B, the oracle often prefers smaller agents: 55.2\% of 14B selections could have been handled by 4B, and 31.5\% of 32B selections map to 4B (with an additional 25.0\% mapping to 8B). Under-escalation remains similarly limited: \textsc{sale}'s 4B selections are predominantly correct (69.6\% diagonal), with most remaining errors unavoidable (21.7\%). This asymmetry where \textsc{sale} prioritizes accuracy via conservative routing at the cost of compute closely matches the deep-search behavior.

Stratifying by complexity reveals a parallel transition. At $\tau \le 0.1$, small-agent routing is perfect (4B achieves 100\%), but over-escalation is extreme: all 14B and 32B selections could have been routed to 4B (100\% in both rows). As complexity increases ($\tau \le 0.5$, $\tau \le 2.5$), over-escalation persists while the ``none'' column grows. However, the hardest bin ($\tau \le 60$) diverges from deep search: rather than collapsing to near-total unavoidable outcomes, coding shows more heterogeneous behavior, i.e., the ``none'' column reaches 50--78\% across rows, but under-escalation also emerges (e.g., 4B shows 25\% under-escalation to 8B). This indicates that high-complexity coding tasks retain meaningful variation in required agent capability, whereas deep-search tasks become predominantly unsolvable.

\vspace{10pt}
Overall, while \textsc{sale} already achieves the smallest oracle gap among all evaluated routers (Section~\ref{sec:results}), the error analysis reveals that further improvements require further reducing over-escalation on low-complexity tasks where the oracle favors 4B yet \textsc{sale} triggers larger agents. Importantly, this failure mode is conservative: \textsc{sale} errs toward preserving accuracy rather than sacrificing correctness, and under-escalation remains rare across both domains. Moreover, when \textsc{sale} does route to 4B, diagonal accuracy is high (54.0\% for deep search, 69.6\% for coding), suggesting the system can correctly identify easy tasks but currently triggers such routing too infrequently.

%%%%%%%%%%%%%%%%%%%%%%%%%%%%%%%%%%%%%%%%%%%%%%%%%%%%%%%%%%%%

%%%%%%%%%%%%%%%%%%%%%%%%%%%%%%%%%%%%%%%%%%%%%%%%%%%%%%%%%%%%

%%%%%%%%%%%%%%%%%%%%%%%%%%%%%%%%%%%%%%%%%%%%%%%%%%%%%%%%%%%%%%%%%%%%%%%%%

\newpage\section{Further Ablations}\label{appendix:ablations}
\vspace{10pt}

\subsection{Cost--Value Function Ablations}
Table~\ref{tab:cv_ablations} reports the results of ablating individual terms in the cost--value function. We separately remove the price per million tokens $\pi(a_i)$, the strategy length $|s_{t,i}|$, the normalized entropy $H(s_{t,i})$, and the entire jury term $\sum w_j \cdot \gamma(s_{t,i})$. For each ablated configuration, we re-optimize the remaining weights to best fit that specific setup.

Across both deep search and coding tasks, ablating any of the four terms leads to lower pass@1 scores, indicating that each component contributes meaningfully to \textsc{sale}'s overall performance. For deep search, all ablations also result in higher average cost per million tokens, suggesting that the full cost–value function more effectively balances accuracy and efficiency. In contrast, for coding tasks, ablating $\pi(a_i)$ and $H(s_{t,i})$ yields slightly lower costs, reflecting an increased selection of smaller models, though this comes at the expense of accuracy. This pattern is consistent with the findings in Section~\ref{sec:alloc}, where we show that \textsc{sale} already favors larger models for coding tasks even in its full form, whereas model selection is more varied for deep search.

The impact of each ablation also differs by task type and complexity. Removing the jury assessment causes the largest performance drop for deep search ($-3.6$ pass@1 on average), with the gap widening substantially on more complex tasks ($-7.6$ at $\tau \le 12.5$; $-6.3$ at $\tau \le 60$), underscoring the importance of jury-based scoring in challenging settings. For coding, \textsc{sale} is more robust to jury removal; instead, ablating any of the cost terms (price and strategy length) results in the steepest decline, particularly at mid-complexity levels ($-4.2$ and $-5.5$, respectively, at $\tau \le 2.5$), reflecting their role in routing beyond cost alone. Entropy ablation presents a nuanced trade-off: at $\tau \le 0.1$ for deep search, it achieves the lowest cost but sacrifices accuracy, suggesting entropy helps prevent under-spending on deceptively simple tasks. We note that a few ablated configurations marginally outperform \textsc{sale} in isolated bins, though these gains are offset by larger losses in other bins.

\vspace{30pt}
\begin{table}[H]
    \centering
    \setlength{\tabcolsep}{4pt}%
    \renewcommand{\arraystretch}{1.1}%
    \scalebox{0.95}{%
    \begin{NiceTabular}{lc*{8}{c}}[cell-space-limits=2pt]
    \CodeBefore
      \rectanglecolor{metabg}{8-2}{8-10}
      \rectanglecolor{metabg}{14-2}{14-10}
    \Body
    \toprule
      \multirow{2}{*}{\textbfrm{Task type}} &
      \multirow{2}{*}{{$\tau(t)$}} &
      \multicolumn{2}{c}{\textbfrm{w/o price}} &
      \multicolumn{2}{c}{\textbfrm{w/o length}} &
      \multicolumn{2}{c}{\textbfrm{w/o entropy}} &
      \multicolumn{2}{c}{\textbfrm{w/o jury}}
      \\
      \cmidrule(lr){3-4} \cmidrule(lr){5-6} \cmidrule(lr){7-8} \cmidrule(lr){9-10}
      & & Pass@1($\uparrow$) & $\$$/Mt($\downarrow$)
        & Pass@1($\uparrow$) & $\$$/Mt($\downarrow$)
        & Pass@1($\uparrow$) & $\$$/Mt($\downarrow$)
        & Pass@1($\uparrow$) & $\$$/Mt($\downarrow$) \\
    \midrule
    \multirow{6}{*}{\textit{Deep search}}
      & $\le0.1\phantom{0}$
      & 89.8$_{0.5}$ & 0.22$_{0.01}$
      & 91.0$_{0.5}$ & 0.28$_{0.00}$
      & 89.8$_{1.5}$ & 0.19$_{0.00}$
      & 86.3$_{0.0}$ & 0.28$_{0.00}$ \\
    & $\le0.5\phantom{0}$
      & 88.5$_{0.9}$ & 0.26$_{0.01}$
      & 88.8$_{1.0}$ & 0.31$_{0.01}$
      & 85.5$_{1.7}$ & 0.25$_{0.00}$
      & 89.0$_{1.7}$ & 0.23$_{0.01}$ \\
    & $\le2.5\phantom{0}$
      & 69.7$_{1.5}$ & 0.25$_{0.01}$
      & 71.8$_{1.7}$ & 0.30$_{0.01}$
      & 73.0$_{1.9}$ & 0.26$_{0.01}$
      & 72.2$_{0.9}$ & 0.21$_{0.00}$ \\
    & $\le12.5$
      & 33.7$_{2.3}$ & 0.19$_{0.01}$ 
      & 33.4$_{1.8}$ & 0.26$_{0.01}$
      & 35.1$_{0.9}$ & 0.22$_{0.01}$
      & 29.5$_{0.5}$ & 0.19$_{0.00}$ \\
    & $\le60\phantom{.}\phantom{0}$
      & 16.9$_{1.5}$ & 0.25$_{0.01}$
      & 15.6$_{2.0}$ & 0.29$_{0.01}$
      & 14.4$_{1.5}$ & 0.25$_{0.01}$
      & 10.0$_{2.3}$ & 0.20$_{0.01}$ \\
    & All
      & 65.4$_{0.7}$ & 0.23$_{0.00}$
      & 66.0$_{0.6}$ & 0.29$_{0.00}$
      & 65.5$_{0.7}$ & 0.23$_{0.00}$ 
      & 63.7$_{0.5}$ & 0.23$_{0.00}$ \\
    \midrule
    \multirow{6}{*}{\textit{Coding}}
      & $\le0.1\phantom{0}$
      & 98.3$_{1.0}$ & 0.18$_{0.00}$
      & 98.8$_{0.0}$ & 0.28$_{0.00}$
      & 97.2$_{0.5}$ & 0.17$_{0.00}$
      & 96.8$_{0.6}$ & 0.17$_{0.00}$ \\
    & $\le0.5\phantom{0}$
      & 79.2$_{1.5}$ & 0.28$_{0.00}$
      & 78.7$_{2.0}$ & 0.29$_{0.00}$
      & 78.0$_{2.2}$ & 0.28$_{0.01}$ 
      & 79.2$_{0.6}$ & 0.30$_{0.00}$ \\
    & $\le2.5\phantom{0}$
      & 64.8$_{2.2}$ & 0.28$_{0.01}$ 
      & 63.5$_{1.2}$ & 0.28$_{0.00}$
      & 64.8$_{2.5}$ & 0.25$_{0.00}$  
      & 67.5$_{1.9}$ & 0.28$_{0.00}$ \\
    & $\le12.5$
      & 30.1$_{2.4}$ & 0.27$_{0.00}$
      & 28.1$_{2.4}$ & 0.29$_{0.01}$
      & 29.9$_{2.0}$ & 0.29$_{0.00}$
      & 29.1$_{1.0}$ & 0.29$_{0.00}$ \\
    & $\le60\phantom{.}\phantom{0}$
      & 23.8$_{0.5}$ & 0.26$_{0.01}$
      & 24.8$_{1.6}$ & 0.28$_{0.01}$
      & 25.6$_{3.0}$ & 0.28$_{0.00}$ 
      & 27.3$_{0.6}$ & 0.29$_{0.00}$ \\
    & All
      & 59.4$_{0.8}$ & 0.26$_{0.00}$
      & 58.7$_{0.7}$ & 0.28$_{0.00}$
      & 59.1$_{1.0}$ & 0.26$_{0.00}$ 
      & 59.9$_{0.5}$ & 0.27$_{0.00}$ \\
    \bottomrule
    \end{NiceTabular}}
    \caption{
    Deep search and coding pass@1 and price per million tokens ($\$$/Mt) across task-complexity bins for \textsc{sale} in its full form and under four ablation conditions. We remove the price term $\pi(a_i)$ (w/o price), the strategy length $|s_{t,i}|$ (w/o length), the entropy term $H(s_{t,i})$ (w/o entropy), and the jury scores $\sum w_j \cdot \gamma(s_{t,i})$ (w/o jury).}
    \label{tab:cv_ablations}
\end{table}

\newpage\subsection{Jury Ablations}
Table~\ref{tab:ablations} reports results when the full jury (comprising the 4B, 8B, 14B, and 32B agents) is replaced by a single judge. Across both task types, all single-judge configurations underperform the full jury: for deep search, overall Pass@1 drops from 67.3 to 65.0–66.4, while for coding it falls from 61.1 to 59.0–60.5. Notably, no single judge is consistently superior---the 8B achieves the highest deep search accuracy (66.4), while the 14B leads on coding (60.5)---and larger judges do not reliably outperform smaller ones. Single judges also incur substantially higher costs for deep search (0.25–0.32 vs.\ 0.21 \$/Mt), suggesting the jury enables more efficient model selection. These results indicate that the jury's strength lies not in any individual member but in the diversity of perspectives: combining judges of varying capacity produces a regularizing effect that yields more robust and cost-effective decisions than any single judge alone. Crucially, this ensemble incurs negligible overhead: each judge produces only a single token (a 1–5 discrete score) and we reuse the same agents already loaded in memory as part of the pool $\mathcal{A}$. This makes the jury's benefits effectively free.

Given that individual judges underperform the full jury, we further ask whether each member is necessary by removing one judge at a time (Table~\ref{tab:judge_ablations}). Indeed, every ablation leads to a decline in pass@1, confirming that each judge contributes unique signal. For coding, removing the 4B judge causes the largest drop ($-2.9$ pass@1), despite the 4B not being the strongest individual judge, highlighting its complementary role within the ensemble. For deep search, removing the 14B has the smallest impact (66.9 vs.\ 67.3), though performance still decreases. Interestingly, removing the 4B also substantially increases deep search cost (0.28 vs.\ 0.21 \$/Mt), suggesting the smaller judge helps steer the system toward more economical selections. Overall, these results reinforce that jury diversity is not redundant: even the smallest judge provides information that improves both accuracy and efficiency.

Finally, we ablate agent self-judgment (retaining only peer-judgment) and peer-judgment (retaining only self-judgment) from \textsc{sale} (Table~\ref{tab:judgment_ablations}). Without self-judgment, the system approaches and occasionally slightly exceeds the best single agent baseline (64.1 vs.\ 63.8 for deep search; 58.8 vs.\ 58.4 for coding), though it remains below the full \textsc{sale} (67.3 and 61.1, respectively). In contrast, removing peer-judgment causes a larger drop, especially for coding, where overall pass@1 falls to 48.7. Notably, this configuration yields markedly lower costs (0.07 \$/Mt for deep search vs.\ 0.21), suggesting that smaller agents tend to be more confident self-scorers, leading the system to favor them when peer signals are unavailable. At low task complexity, relying solely on self-judgment still matches or even slightly exceeds the best single agent (e.g., 97.5 vs.\ 95.0 at $\tau \le 0.1$ for coding); yet as complexity increases, performance degrades sharply (5.6 vs.\ 12.5 at $\tau \le 60$ for deep search), indicating that self-judgment alone does not scale to harder tasks. These results underscore the complementary nature of both feedback types: peer-judgment provides the external calibration necessary for difficult problems, while self-judgment contributes efficient, low-cost signal on simpler ones.

\vspace{10pt}
\begin{table}[H]
    \centering
    \setlength{\tabcolsep}{4pt}%
    \renewcommand{\arraystretch}{1.05}%
    \scalebox{0.95}{%
    \begin{NiceTabular}{lc*{8}{c}}[cell-space-limits=2pt]
    \CodeBefore
      \rectanglecolor{metabg}{8-2}{8-10}
      \rectanglecolor{metabg}{14-2}{14-10}
    \Body
    \toprule
      \multirow{2}{*}{\textbfrm{Task type}} &
      \multirow{2}{*}{{$\tau(t)$}} &
      \multicolumn{2}{c}{\textbfrm{4B judge only}} &
      \multicolumn{2}{c}{\textbfrm{8B judge only}} &
      \multicolumn{2}{c}{\textbfrm{14B judge only}} &
      \multicolumn{2}{c}{\textbfrm{32B judge only}} \\
      \cmidrule(lr){3-4} \cmidrule(lr){5-6} \cmidrule(lr){7-8} \cmidrule(lr){9-10}
      & & Pass@1($\uparrow$) & $\$$/Mt($\downarrow$)
        & Pass@1($\uparrow$) & $\$$/Mt($\downarrow$)
        & Pass@1($\uparrow$) & $\$$/Mt($\downarrow$)
        & Pass@1($\uparrow$) & $\$$/Mt($\downarrow$) \\
    \midrule

    \multirow{6}{*}{\textit{Deep search}}
      & $\le0.1\phantom{0}$
      & 91.3$_{0.0}$ & 0.25$_{0.01}$
      & 90.0$_{0.0}$ & 0.30$_{0.00}$
      & 89.0$_{0.5}$ & 0.29$_{0.00}$
      & 90.0$_{0.0}$ & 0.28$_{0.00}$ \\
    & $\le0.5\phantom{0}$
      & 88.0$_{0.6}$ & 0.28$_{0.01}$
      & 88.8$_{0.0}$ & 0.30$_{0.01}$
      & 86.3$_{0.6}$ & 0.33$_{0.01}$
      & 88.0$_{0.6}$ & 0.32$_{0.01}$ \\
    & $\le2.5\phantom{0}$
      & 71.0$_{1.5}$ & 0.26$_{0.01}$
      & 72.5$_{0.8}$ & 0.34$_{0.01}$
      & 69.2$_{0.6}$ & 0.35$_{0.01}$
      & 69.2$_{1.0}$ & 0.32$_{0.01}$ \\
    & $\le12.5$
      & 33.4$_{0.6}$ & 0.22$_{0.01}$ 
      & 35.4$_{0.0}$ & 0.33$_{0.00}$ 
      & 36.1$_{1.0}$ & 0.31$_{0.01}$
      & 34.1$_{2.2}$ & 0.31$_{0.01}$ \\
    & $\le60\phantom{.}\phantom{0}$
      & 15.6$_{0.0}$ & 0.25$_{0.01}$
      & 15.6$_{0.0}$ & 0.31$_{0.01}$ 
      & 15.6$_{0.0}$ & 0.31$_{0.01}$
      & 14.4$_{1.5}$ & 0.29$_{0.01}$ \\
    & All
      & 65.7$_{0.4}$ & 0.25$_{0.00}$
      & 66.4$_{0.2}$ & 0.32$_{0.00}$
      & 65.0$_{0.3}$ & 0.32$_{0.00}$
      & 65.1$_{0.6}$ & 0.31$_{0.00}$ \\
    \midrule
    \multirow{6}{*}{\textit{Coding}}
      & $\le0.1\phantom{0}$
      & 96.8$_{0.6}$ & 0.18$_{0.00}$
      & 97.5$_{0.0}$ & 0.18$_{0.00}$
      & 98.5$_{0.5}$ & 0.20$_{0.01}$
      & 97.8$_{0.5}$ & 0.19$_{0.00}$ \\
    & $\le0.5\phantom{0}$
      & 80.3$_{0.6}$ & 0.29$_{0.00}$
      & 80.3$_{1.0}$ & 0.28$_{0.00}$
      & 80.5$_{0.6}$ & 0.30$_{0.00}$ 
      & 81.5$_{0.6}$ & 0.29$_{0.00}$ \\
    & $\le2.5\phantom{0}$
      & 66.8$_{1.0}$ & 0.29$_{0.00}$
      & 67.0$_{2.8}$ & 0.28$_{0.00}$
      & 68.2$_{1.0}$ & 0.29$_{0.00}$
      & 66.5$_{1.5}$ & 0.29$_{0.00}$ \\
    & $\le12.5$
      & 27.2$_{1.4}$ & 0.27$_{0.00}$ 
      & 29.1$_{1.7}$ & 0.30$_{0.01}$ 
      & 29.4$_{1.2}$ & 0.31$_{0.00}$ 
      & 28.4$_{1.6}$ & 0.30$_{0.02}$ \\
    & $\le60\phantom{.}\phantom{0}$
      & 24.3$_{0.9}$ & 0.28$_{0.00}$ 
      & 25.3$_{1.6}$ & 0.28$_{0.01}$ 
      & 26.1$_{1.7}$ & 0.29$_{0.01}$
      & 25.8$_{2.2}$ & 0.29$_{0.00}$ \\
    & All
      & 59.0$_{0.4}$ & 0.26$_{0.00}$ 
      & 59.8$_{0.8}$ & 0.26$_{0.00}$
      & 60.5$_{0.5}$ & 0.28$_{0.00}$
      & 60.0$_{0.6}$ & 0.27$_{0.00}$ \\
    \bottomrule
    \end{NiceTabular}}
    \caption{
    Deep search and coding pass@1 and price per million tokens ($\$$/Mt) across task-complexity bins for \textsc{sale} with the agent jury replaced by one single agent judge.}
    \label{tab:ablations}\vspace{-5pt}
\end{table}

\newpage
\vspace*{35pt}
\begin{table}[H]
    \centering
    \setlength{\tabcolsep}{4pt}%
    \renewcommand{\arraystretch}{1.05}%
    \scalebox{0.95}{%
    \begin{NiceTabular}{lc*{8}{c}}[cell-space-limits=2pt]
    \CodeBefore
      \rectanglecolor{metabg}{8-2}{8-10}
      \rectanglecolor{metabg}{14-2}{14-10}
    \Body
    \toprule
      \multirow{2}{*}{\textbfrm{Task type}} &
      \multirow{2}{*}{{$\tau(t)$}} &
      \multicolumn{2}{c}{\textbfrm{w/o 4B judge}} &
      \multicolumn{2}{c}{\textbfrm{w/o 8B judge}} &
      \multicolumn{2}{c}{\textbfrm{w/o 14B judge}} &
      \multicolumn{2}{c}{\textbfrm{w/o 32B judge}} \\
      \cmidrule(lr){3-4} \cmidrule(lr){5-6} \cmidrule(lr){7-8} \cmidrule(lr){9-10}
      & & Pass@1($\uparrow$) & $\$$/Mt($\downarrow$)
        & Pass@1($\uparrow$) & $\$$/Mt($\downarrow$)
        & Pass@1($\uparrow$) & $\$$/Mt($\downarrow$)
        & Pass@1($\uparrow$) & $\$$/Mt($\downarrow$) \\
    \midrule
    \multirow{6}{*}{\textit{Deep search}}
      & $\le0.1\phantom{0}$
      & 89.2$_{0.6}$ & 0.20$_{0.00}$
      & 92.3$_{0.5}$ & 0.23$_{0.00}$
      & 91.8$_{1.0}$ & 0.22$_{0.01}$
      & 90.0$_{0.0}$ & 0.22$_{0.01}$ \\
    & $\le0.5\phantom{0}$
      & 86.0$_{0.9}$ & 0.28$_{0.00}$
      & 86.0$_{0.9}$ & 0.25$_{0.01}$
      & 88.8$_{1.1}$ & 0.22$_{0.01}$
      & 87.0$_{1.0}$ & 0.24$_{0.01}$ \\
    & $\le2.5\phantom{0}$
      & 71.2$_{0.8}$ & 0.35$_{0.00}$
      & 68.2$_{1.5}$ & 0.24$_{0.01}$
      & 73.0$_{2.0}$ & 0.23$_{0.01}$
      & 72.5$_{1.1}$ & 0.23$_{0.01}$ \\
    & $\le12.5$
      & 36.1$_{1.0}$ & 0.31$_{0.01}$
      & 34.9$_{1.8}$ & 0.18$_{0.01}$
      & 36.3$_{1.8}$ & 0.18$_{0.00}$
      & 33.9$_{2.4}$ & 0.17$_{0.00}$ \\
    & $\le60\phantom{.}\phantom{0}$
      & 14.4$_{1.5}$ & 0.28$_{0.01}$
      & 16.9$_{1.5}$ & 0.23$_{0.01}$
      & 13.8$_{1.7}$ & 0.23$_{0.02}$
      & 16.3$_{1.3}$ & 0.23$_{0.01}$ \\
    & All
      & 65.3$_{0.4}$ & 0.28$_{0.00}$
      & 65.3$_{0.6}$ & 0.23$_{0.00}$
      & 66.9$_{0.7}$ & 0.21$_{0.00}$
      & 65.7$_{0.7}$ & 0.22$_{0.00}$ \\
    \midrule
    \multirow{6}{*}{\textit{Coding}}
      & $\le0.1\phantom{0}$
      & 97.0$_{0.6}$ & 0.16$_{0.00}$
      & 97.8$_{1.2}$ & 0.18$_{0.00}$
      & 97.0$_{0.6}$ & 0.18$_{0.00}$
      & 97.8$_{1.2}$ & 0.18$_{0.00}$ \\
    & $\le0.5\phantom{0}$
      & 75.7$_{2.6}$ & 0.25$_{0.01}$
      & 79.0$_{1.3}$ & 0.27$_{0.00}$
      & 79.5$_{1.5}$ & 0.27$_{0.01}$
      & 77.7$_{1.5}$ & 0.27$_{0.01}$ \\
    & $\le2.5\phantom{0}$
      & 62.8$_{1.8}$ & 0.22$_{0.01}$ 
      & 65.8$_{1.3}$ & 0.28$_{0.00}$ 
      & 65.0$_{2.2}$ & 0.26$_{0.00}$
      & 66.0$_{2.2}$ & 0.27$_{0.00}$ \\
    & $\le12.5$
      & 29.9$_{0.9}$ & 0.30$_{0.00}$ 
      & 29.4$_{1.4}$ & 0.31$_{0.01}$
      & 29.6$_{1.6}$ & 0.30$_{0.01}$ 
      & 28.9$_{1.7}$ & 0.30$_{0.00}$ \\
    & $\le60\phantom{.}\phantom{0}$
      & 25.8$_{1.5}$ & 0.28$_{0.00}$ 
      & 25.6$_{1.2}$ & 0.28$_{0.00}$
      & 25.1$_{1.2}$ & 0.28$_{0.00}$
      & 24.6$_{1.0}$ & 0.27$_{0.00}$ \\
    & All
      & 58.2$_{0.7}$ &0.24$_{0.00}$
      & 59.5$_{0.6}$ &0.26$_{0.00}$
      & 59.2$_{0.7}$ &0.26$_{0.00}$ 
      & 59.0$_{0.7}$ &0.26$_{0.00}$ \\
    \bottomrule
    \end{NiceTabular}}
    \caption{
     Deep search and coding pass@1 and price per million tokens ($\$$/Mt) across task-complexity bins for \textsc{sale} with one judge removed from the jury in turn.}
    \label{tab:judge_ablations}
\end{table}

\vspace{50pt}
\begin{table}[H]
    \centering
    \setlength{\tabcolsep}{20pt}%
    \renewcommand{\arraystretch}{1.05}%
    \scalebox{0.95}{%
    \begin{NiceTabular}{lc*{4}{c}}[cell-space-limits=2pt]
    \CodeBefore
      \rectanglecolor{metabg}{8-2}{8-6}
      \rectanglecolor{metabg}{14-2}{14-6}
    \Body
    \toprule
      \multirow{2}{*}{\textbfrm{Task type}} &
      \multirow{2}{*}{{$\tau(t)$}} &
      \multicolumn{2}{c}{\textbfrm{w/o self-feedback}} &
      \multicolumn{2}{c}{\textbfrm{w/o peer-feedback}} \\
      \cmidrule(lr){3-4} \cmidrule(lr){5-6}
      & & Pass@1($\uparrow$) & $\$$/Mt($\downarrow$)
        & Pass@1($\uparrow$) & $\$$/Mt($\downarrow$) \\
    \midrule
    \multirow{6}{*}{\textit{Deep search}}
      & $\le0.1\phantom{0}$ & 92.3$_{0.5}$ & 0.26$_{0.01}$ & 91.0$_{0.9}$ & 0.06$_{0.01}$ \\
    & $\le0.5\phantom{0}$ & 86.0$_{0.9}$ & 0.21$_{0.01}$ & 87.3$_{1.8}$ & 0.07$_{0.01}$ \\
    & $\le2.5\phantom{0}$ & 68.2$_{0.6}$ & 0.22$_{0.01}$ & 65.8$_{2.0}$ & 0.07$_{0.00}$ \\
    & $\le12.5$ & 31.5$_{2.9}$ & 0.17$_{0.01}$ & 32.7$_{1.6}$ & 0.09$_{0.01}$ \\
    & $\le60\phantom{.}\phantom{0}$ & 12.5$_{1.9}$ & 0.21$_{0.02}$ & 5.6$_{1.3}$ & 0.06$_{0.01}$ \\
    & All & 64.1$_{0.7}$ & 0.21$_{0.00}$ & 63.2$_{0.7}$ & 0.07$_{0.00}$ \\
    \midrule
    \multirow{6}{*}{\textit{Coding}}
      & $\le0.1\phantom{0}$ & 97.5$_{0.8}$ & 0.17$_{0.00}$ & 97.5$_{0.0}$ & 0.05$_{0.01}$ \\
    & $\le0.5\phantom{0}$ & 81.0$_{1.1}$ & 0.27$_{0.01}$ & 71.9$_{1.9}$ & 0.05$_{0.00}$ \\
    & $\le2.5\phantom{0}$ & 65.3$_{2.9}$ & 0.27$_{0.01}$ & 49.0$_{1.7}$ & 0.06$_{0.00}$ \\
    & $\le12.5$ & 26.8$_{0.6}$ & 0.29$_{0.01}$ & 15.1$_{1.5}$ & 0.08$_{0.00}$ \\
    & $\le60\phantom{.}\phantom{0}$ & 23.8$_{0.5}$ & 0.29$_{0.01}$ & 10.4$_{1.8}$ & 0.12$_{0.01}$ \\
    & All & 58.8$_{0.7}$ & 0.26$_{0.00}$ & 48.7$_{0.7}$ & 0.07$_{0.00}$ \\
    \bottomrule
    \end{NiceTabular}}
    \caption{
     Deep search and coding pass@1 and price per million tokens ($\$$/Mt) across task-complexity bins for \textsc{SALE} with ablated self-feedback and peer-feedback.}
    \label{tab:judgment_ablations}
\end{table}

\newpage\subsection{Memory Ablations}
\label{appendix:memory_robustness}

\vspace{10pt}
To assess the robustness of the memory-based refinement stage to aggressive pruning, we conduct two complementary experiments on a fixed pool of 100 tasks using a leave-one-out protocol. First, we vary the memory bank sample size $M \in \{0, 25, 50, 75, 99\}$, with 5 random seeds for sampling at each size. Second, we apply TTL pruning, retaining only the most recent $W$ records for $W \in \{10, 25, 50, 99\}$. Results are shown in Tables~\ref{tab:memory_size} and~\ref{tab:ttl_window}. For each configuration we report the average jury score of the initial bid (\textit{Jury score -- Original}), the average jury score of the refined bid where applicable (\textit{Jury score -- Refined}), the fraction of tasks where refinement flips the provisional winner (\textit{Allocation -- Flip rate}), and the share of tasks ultimately executed by the 32B agent (\textit{Allocation -- 32B share}).

\vspace{5pt}
The critical transition is from $M=0$ to $M>0$: enabling memory lowers 32B reliance from 35\% to roughly 17\% and raises refined-plan jury scores by approximately 0.3 points on the 0--5 scale. Beyond this point, increasing memory size yields only small further improvements, within seed-level noise ($\sigma = 1.7$--$2.6$ percentage points). TTL pruning shows the same pattern: even an aggressive 10-task window produces comparable jury scores and flip rates, with only a modestly higher 32B share (21\% vs.\ 18\% at $W=99$). Together, these results indicate that memory-based refinement is robust to substantial pruning, and that a sliding window of $\sim$10--25 recent tasks is sufficient for effective refinement in practice.

\begin{table}[h]
\centering
\vspace{30pt}
\setlength{\tabcolsep}{28pt}
\renewcommand{\arraystretch}{1.05}
\caption{Effect of memory bank size $M$ on refinement quality and agent allocation. $M=0$ corresponds to disabling memory-based refinement.}
\label{tab:memory_size}
\begin{tabular}{lcccc}
\toprule
& \multicolumn{2}{c}{\textbf{Jury score}} & \multicolumn{2}{c}{\textbf{Allocation}} \\
\cmidrule(lr){2-3} \cmidrule(lr){4-5}
Memory & Original & Refined & Flip rate & 32B share \\
\midrule
$M=0$  & 3.36 & ---  & 0\%    & 35\% \\
$M=25$ & 3.38 & 3.68 & 19.1\% & 17\% \\
$M=50$ & 3.36 & 3.67 & 17.5\% & 17\% \\
$M=75$ & 3.42 & 3.72 & 17.3\% & 16\% \\
$M=99$ & 3.41 & 3.74 & 17.2\% & 17\% \\
\bottomrule
\end{tabular}
\end{table}

\begin{table}[h]
\centering
\vspace{30pt}
\setlength{\tabcolsep}{26pt}
\renewcommand{\arraystretch}{1.05}
\caption{Effect of TTL window size $W$ on refinement quality and agent allocation. Only the most recent $W$ records are retained for retrieval.}
\label{tab:ttl_window}
\begin{tabular}{lcccc}
\toprule
& \multicolumn{2}{c}{\textbf{Jury score}} & \multicolumn{2}{c}{\textbf{Allocation}} \\
\cmidrule(lr){2-3} \cmidrule(lr){4-5}
TTL window & Original & Refined & Flip rate & 32B share \\
\midrule
$W=10$ & 3.47 & 3.68 & 16.1\% & 21\% \\
$W=25$ & 3.45 & 3.68 & 16.1\% & 21\% \\
$W=50$ & 3.45 & 3.65 & 16.7\% & 19\% \\
$W=99$ & 3.46 & 3.74 & 16.7\% & 18\% \\
\bottomrule
\end{tabular}
\end{table}

%%%%%%%%%%%%%%%%%%%%%%%%%%%%%%%%%%%%%%%%%%%%%%%%%%%%%%%%%%%%%%%%%%%%%%%%%

\newpage\section{Jury Validation}
\label{app:jury_human}

\vspace{10pt}
To validate that LLM-jury scores produce a meaningful signal, we run two complementary analyses: (i) we measure alignment between jury scores and blind expert human annotations of plan quality, and (ii) we measure the predictiveness of jury scores with respect to downstream task success.

\vspace{15pt}
\subsection{Alignment with Human Judgments}
\label{app:jury_human_alignment}

We collect blind expert annotations on a stratified subset of 100 plans from \textsc{HST-Bench}, evenly distributed across task types, complexity bins, and agent sizes. Annotators score each plan on the same 0--5 Likert scale used by the jury, without access to the jury's scores. We then compute the Spearman rank correlation between the average jury score and the human score within each complexity bin. Results are shown in Table~\ref{tab:jury_human_corr}.

Correlations are strong across all complexity bins ($\rho \in [0.74, 0.82]$) with no meaningful degradation as task complexity increases, supporting the use of jury scores as a reliable signal in the auction.

\begin{table}[h]
\centering
\vspace{20pt}
\setlength{\tabcolsep}{89pt}%
\caption{Spearman rank correlation between SALE jury scores and blind expert human annotations on 100 stratified \textsc{Hst-Bench} plans.}
\label{tab:jury_human_corr}
\begin{tabular}{cc}
\toprule
Complexity & Spearman $\rho$ \\
\midrule
$\tau(t) \le 0.1$  & 0.76 \\
$\tau(t) \le 0.5$  & 0.74 \\
$\tau(t) \le 2.5$  & 0.82 \\
$\tau(t) \le 12.5$ & 0.75 \\
$\tau(t) \le 60$   & 0.74 \\
\bottomrule
\end{tabular}
\end{table}

\vspace{20pt}
\subsection{Predictiveness of Task Success}
\label{app:jury_predictiveness}

To assess whether jury scores predict downstream success, we compute the point-biserial correlation between continuous jury scores and binary task accuracy on the full development set. We obtain a correlation of $0.36$ ($p < 0.01$), which falls within the range conventionally interpreted as good item discrimination~\citep{alma9923329510802171}. Combined with the human-alignment results above, this provides converging evidence that the jury scoring signal is both aligned with human judgments of plan quality and predictive of downstream task success.

%%%%%%%%%%%%%%%%%%%%%%%%%%%%%%%%%%%%%%%%%%%%%%%%%%%%%%%%%%%%%%%%%%%%%%%%%

\newpage\section{Agent Selection Over Time}\label{appendix:time}
\begin{figure}[H]
    \centering
    \includegraphics[width=\linewidth]{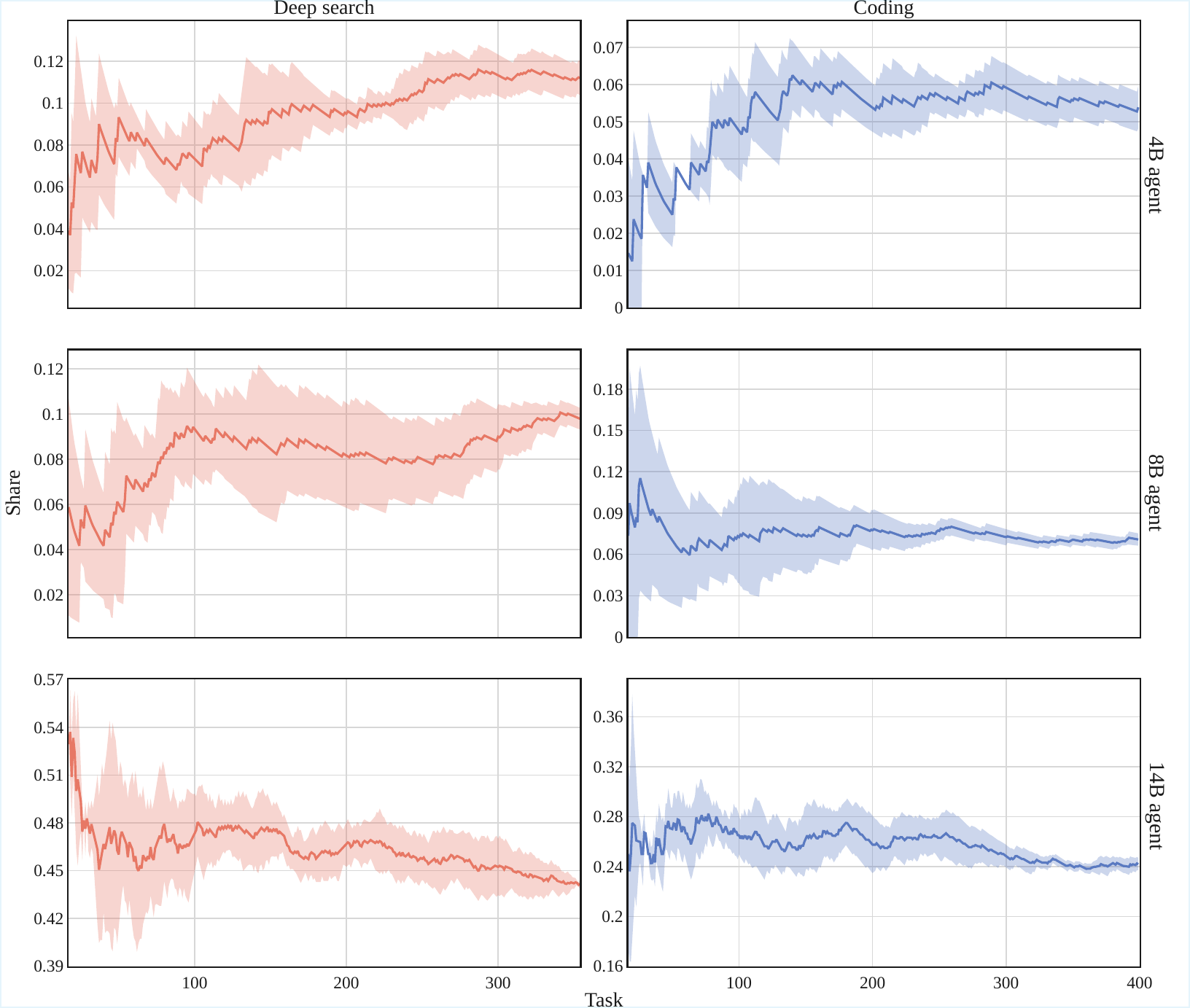}
    \caption{Cumulative selection rates over time for the 4B, 8B, and 14B agents on deep search and coding tasks, showing increasing delegation to cheaper agents while the 14B share gradually declines as auction memory accumulates.}
    \label{fig:time_all}
\end{figure}

In Figure \ref{fig:time_all}, we show the cumulative selection rate over time, defined as the running fraction of tasks ultimately delegated to each agent, for all agents that can refine their strategies during \textsc{sale}'s auction mechanism (4B, 8B, and 14B). Recall that an agent only enters the memory-based self-refinement stage when a more expensive agent is the provisional winner, so cheaper agents have more opportunities to refine their strategies than more expensive ones. As a result, the 4B agent generates a refined strategy on 92\% of samples for deep search and 93\% for coding, the 8B agent on 91\% for deep search and 92\% for coding, and the 14B agent on 62\% for deep search and 77\% for coding, leading to different patterns in how their cumulative selection rates evolve as auction memory accumulates.

Across both domains, the 4B agent's cumulative selection share (also reported in Figure~\ref{fig:time}) shows a clear upward trend, indicating that it is increasingly chosen as auction memory grows. The 8B agent's share is also broadly increasing for deep search, albeit with higher variance, but remains relatively flat for coding. In contrast, the 14B agent's cumulative share exhibits a mild downward drift in both domains, consistent with workload being gradually reallocated toward the cheaper agents. Because each curve at time step $t$ reflects the average selection rate computed over all tasks up to $t$, the trajectories are naturally more volatile at the beginning, when this average is based on few samples, and become smoother as more tasks accumulate.

%%%%%%%%%%%%%%%%%%%%%%%%%%%%%%%%%%%%%%%%%%%%%%%%%%%%%%%%%%%%%%%%%%%%%%%%%

\newpage\section{Qualitative Performance Comparison}
\label{appendix:failure_modes}
This section presents examples, spanning task complexity levels, where the largest agent (32B) fails and a smaller agent succeeds. As these examples originate from the test set, we omit the agents' final answers. Given that full traces span tens of thousands to millions of tokens, we present only the agent's solution strategy and indicate whether the final answer is correct. In many cases, the strategy alone exhibits telltale signs of failure: misunderstanding of the task, excessive complexity, or hallucinated details that ultimately lead to an incorrect answer.

\subsection{Performance Comparison on Deep Search Tasks}
For deep search tasks, we identify three recurring failure modes in the 32B agent. First, the agent occasionally bypasses tool use entirely and provides an answer from parametric memory, often incorrect. This tendency is sometimes visible in the initial strategy itself, where the eventual answer is already present despite a search plan having been outlined (Example~\hyperref[ex:1]{\ref*{ex:1}}). Second, the agent may over-engineer its approach to straightforward questions, pursuing tangential information, such as a musician's date of death (Example~\hyperref[ex:2]{\ref*{ex:2}}), or issuing excessive sequential queries that lead it further from the answer (Example~\hyperref[ex:5]{\ref*{ex:5}}). Third, the agent retrieves partial or related information from parametric knowledge that interferes with the search process: in Example~\hyperref[ex:3]{\ref*{ex:3}}, a legal case is conflated with an unrelated one, while in Example~\hyperref[ex:4]{\ref*{ex:4}}, parametric recall of an intermediate fact introduces ambiguity between a television series and its sequel.

\vspace{30pt}
\begin{examplebox}[label=ex:1]{Deep Search Example 1}
    \exquestion{%
        A patient suffers a stroke just in the tip of the paracentral artery on the left side where the artery rounds over to the lateral surface. This would likely result in\\   Answer Choices:\\
        A. More sensory loss in the right arm than the foot\\
        B. More sensory loss in the right foot than the arm\\
        C. More sensory loss in the left arm than the foot\\
        D. More sensory loss in the left foot than the arm\\
        E. More weakness of the right foot than the arm\\
        Answer with the letter of the single correct option.
    }
    \incorrectresponse{%
        \\
        \texttt{ask\_search\_agent}: What is the anatomical territory of the paracentral artery and its role in sensory/motor functions? Provide a summary of the brain regions it supplies and their functional roles.\\  
        \texttt{ask\_search\_agent}: How does a stroke in the left paracentral artery affect sensory/motor functions contralaterally? Provide a summary of the expected deficits on the right side of the body.\\  
        \texttt{ask\_search\_agent}: In the sensory homunculus, is the foot or arm more represented in the medial portion of the paracentral lobule? Provide a concise answer based on neuroanatomy resources.\\  
        \texttt{final\_answer}: B  
        }  
        {%
            % B
        }
    \correctresponse{14}{%
        \\
        \texttt{ask\_search\_agent}: What is the somatotopic representation of the arm and foot in the paracentral lobule of the somatosensory cortex? Provide a concise summary of the anatomical distribution.  
        \texttt{ask\_search\_agent}: Does a stroke in the left paracentral lobule cause contralateral sensory loss in the right arm or foot? Provide a concise summary of the clinical implications.    
    }{%
        % A
    }
\end{examplebox} % issue: uses parametric knowledge to decide the answer in advance

\newpage\vspace*{5pt}
\begin{examplebox}[label=ex:2]{Deep Search Example 2}
    \exquestion{%
        How many studio albums were published by Mercedes Sosa between 2000 and 2009 (included)? You can use the latest version of english wikipedia.
    }
    \incorrectresponse{%
        \\
        1. Use \texttt{ask\_search\_agent} to retrieve Mercedes Sosa's discography (studio albums with release years) from the English Wikipedia page. Specify the query: ``List all studio albums by Mercedes Sosa with their release years. Provide a table of album titles and years.'' Website: https://en.wikipedia.org/wiki/Mercedes\_Sosa.\\  
        2. Use \texttt{ask\_search\_agent} to confirm Mercedes Sosa's death date and check for posthumous album releases in 2009. Query: ``When did Mercedes Sosa die? Were any studio albums released posthumously in 2009? Provide dates and album names.''\\  
        3. Filter the discography data to count studio albums released between 2000–2009 (inclusive), excluding posthumous 2009 releases.\\ 
        4. Use \texttt{final\_answer} to submit the verified count.  
        }  
        {%
            % 1
        }
    \correctresponse{8}{%
        \\
        1. Use \texttt{ask\_search\_agent} to retrieve the list of studio albums by Mercedes Sosa along with their release years from the English Wikipedia page (\texttt{https://en.wikipedia.org/wiki/Mercedes\_Sosa}). Provide \texttt{output\_content} as a structured list of albums with release years.\\  
        2. Extract the release years of all studio albums from the search result.\\  
        3. Filter the albums to include only those released between 2000 and 2009 (inclusive).\\  
        4. Count the number of filtered albums.\\  
        5. Use \texttt{final\_answer} to provide the derived count.  
    }{%
        % 3
    }
\end{examplebox} % issue: overcomplicates relatively simple search (year of death and posthumous albums)

\vspace{20pt}
\begin{examplebox}[label=ex:3]{Deep Search Example 3}
    \exquestion{%
        What amount was the settlement that the character from the Son of al Qaeda got in 2017?
    }
    \incorrectresponse{%
        \\
        1. Use \texttt{ask\_search\_agent} to verify if ``Son of al Qaeda'' is a real entity or a misstatement (e.g., ``Son of Sam''). Query: ``Is `Son of al Qaeda' a real person or legal case? If not, what is the correct reference (e.g., `Son of Sam')? Provide clarification.''\\  
        2. If the search confirms a misstatement (e.g., ``Son of Sam''), use \texttt{ask\_search\_agent} to check for 2017 settlements: ``What legal settlement amount was associated with the `Son of Sam' case in 2017? Provide the numerical value.''\\ 
        3. If no 2017 data is found, cross-reference historical records for the original ``Son of Sam'' case (1979) and determine if updates or subsequent settlements occurred in 2017.\\  
        4. Use \texttt{final\_answer} to report the verified settlement amount.}  
        {%
            % 1200000
        }
    \correctresponse{8}{%
        \\
        1. Use \texttt{ask\_search\_agent} to identify what ``Son of al Qaeda'' refers to (e.g., a real person, fictional character, or legal case). Query: ``What is `Son of al Qaeda'? Provide \texttt{output\_content}''.\\  
        2. Use \texttt{ask\_search\_agent} to determine the specific settlement amount related to the identified entity in 2017. Query: ``What was the 2017 settlement amount related to [entity from Step 1]? Provide \texttt{output\_content}''.\\  
        3. Verify the settlement amount via credible sources (e.g., court records, news articles) using \texttt{ask\_search\_agent} if needed.\\  
        4. Use \texttt{final\_answer} to state the confirmed settlement amount.  
    }{%
        % 10500000
    }
\end{examplebox} % issue: parametric memory associated the legal case of the question with an unrelated legal case (Son of Sam)

\newpage\vspace*{5pt}
\begin{examplebox}[label=ex:4]{Deep Search Example 4}
    \exquestion{%
        What season was the character introduced that becomes the main antagonist in the following season, from the animated television series created by Bryan Konietzko and Michael Dante DiMartino as a sequel to ``...'', which aired from 2005 to 2008? 
    }
    \incorrectresponse{%
        \\
        1. Use \texttt{ask\_search\_agent} to confirm the name of the sequel series to ``Avatar: The Last Airbender'' created by Bryan Konietzko and Michael Dante DiMartino. Provide the name of the sequel.\\  
        2. Use \texttt{ask\_search\_agent} to identify characters introduced in each season of the sequel series who later become main antagonists in subsequent seasons. Provide a list of characters with their introduction season and antagonist season.\\  
        3. Use \texttt{ask\_search\_agent} to verify the specific season number where the character was introduced that becomes the main antagonist in the following season. Provide the season number.}
        {%
            % 1
        }
    \correctresponse{8}{%
        \\
        1. Use \texttt{ask\_search\_agent} to identify the original series (2005-2008) created by Bryan Konietzko and Michael Dante DiMartino, and confirm the sequel series name. Query: ``What is the name of the animated series created by Bryan Konietzko and Michael Dante DiMartino that aired from 2005 to 2008? Provide the name.''\\ 2. Use \texttt{ask\_search\_agent} to determine the main antagonists for each season of the sequel series. Query: ``What are the main antagonists for each season of [sequel series name]? Provide a list of seasons and corresponding antagonists.''\\
        3. Use \texttt{ask\_search\_agent} to identify which antagonist was introduced in a season and became the main antagonist in the following season. Query: ``In which season was a character introduced in [sequel series name] that became the main antagonist in the next season? Provide the season number.''\\  
        4. Use \texttt{final\_answer} to compile the result from step 3. 
    }{%
        % Season 3
    }
\end{examplebox} % issue: uses parametric memory to give a partial answer, then gets confused because of it

\vspace{25pt}
\begin{examplebox}[label=ex:5]{Deep Search Example 5}
    \exquestion{%
        D1NZ is a series based on what oversteering technique?
    }
    \incorrectresponse{%
        \\
        \texttt{ask\_search\_agent}: ``What is the D1NZ series and which oversteering technique is it based on? Provide a concise explanation of the series and the specific oversteering technique it employs.''\\ 
        \texttt{ask\_search\_agent}: ``What oversteering techniques are commonly used in drifting competitions like D1NZ? Provide a list of techniques and their association with specific drifting series or events.''\\  
        \texttt{ask\_search\_agent}: ``What vehicle specifications or rules define the D1NZ drifting series? Provide details on car configurations that indicate the required oversteering technique.''\\  
        \texttt{final\_answer}: <answer derived from cross-referencing D1NZ's vehicle specs, competition style, and associated oversteering techniques>}
        {%
            % Counter-steering
        }
    \correctresponse{4}{%
        \\
        1. Use \texttt{ask\_search\_agent} to identify the oversteering technique associated with the series D1NZ. Query: ``What oversteering technique is the series D1NZ based on? Provide the name of the technique.''\\  
        2. Use \texttt{final\_answer} to submit the identified oversteering technique as the answer.  
    }{%
        % drifting
    }
\end{examplebox} %issue: overcomplicates a simple search

\subsection{Performance Comparison on Coding Tasks}
For coding tasks, we observe analogous failure modes in the 32B agent. First, the agent over-engineers straightforward problems: in Example~\hyperref[ex:1_coding]{\ref*{ex:1_coding}}, it plans a recursive solution for what is explicitly fixed two-level nesting, whereas the 4B correctly identifies the specific structure. Such unnecessary complexity increases the surface area for errors. Second, the agent draws on parametric knowledge at the expense of task constraints: Example~\hyperref[ex:2_coding]{\ref*{ex:2_coding}} shows the 32B planning to import a library function not in the available import list, while the 14B opts for a simpler, more robust approach. This overconfidence extends to skipping the planning phase entirely: in Examples~\hyperref[ex:3_coding]{\ref*{ex:3_coding}} and~\hyperref[ex:4]{\ref*{ex:4_coding}}, the 32B produces code directly in place of a high-level strategy, without using the Python shell for verification and pursuing over-optimized solutions that introduce subtle errors, e.g., ignoring intermediate computation steps or applying mathematically incorrect shortcuts, where simpler approaches would have fewer failure points. Third, the agent neglects edge cases: in Example~\hyperref[ex:5_coding]{\ref*{ex:5_coding}}, the 32B fails to handle the ``no duplicate found'' case, resulting in an incorrect return type that breaks hidden tests.

\vspace{100pt}
\begin{examplebox}[label=ex:1_coding]{Coding Example 6}
    \exquestion{%
        You are an expert Python programmer, and here is your task: Write a function to convert a tuple of string values to a tuple of integer values. Your code should pass the following test: \texttt{assert tuple\_int\_str((('333', '33'), ('1416', '55')))==((333, 33), (1416, 55))}
    }
    \incorrectresponse{%
        \\
        1. Analyze the task: Create a function that converts nested tuples of strings to integers.\\
        2. Implement a solution using nested tuple comprehensions to recursively convert each string element to an integer.\\
        3. Validate the solution against the provided test case to ensure correctness. 
        }  
        {%
        }
    \correctresponse{4}{%
        \\
        The plan is to create a function that converts each string in the input tuple of tuples to an integer, preserving the nested structure.\\ 
        This is achieved by using a generator expression within a \texttt{tuple()} to iterate over each inner tuple, convert each string to an integer, and collect the results into a new tuple of tuples.   
    }{%
    }
\end{examplebox} % issue: 32B plan says "recursively convert" but the problem only shows fixed 2-level nesting. This over-generalization suggests the model may be solving a harder problem than what's asked, which can introduce unnecessary complexity and potential edge case failures. (in contrast, the 4B plan explicitly mentions "inner tuple", showing it correctly understood the specific structure: a tuple of tuples of strings).

\newpage
\begin{examplebox}[label=ex:2_coding]{Coding Example 7}
    \exquestion{%
        You are an expert Python programmer, and here is your task: Write a function to merge multiple sorted inputs into a single sorted iterator using heap queue algorithm. Your code should pass the following tests: \texttt{assert merge\_sorted\_list([25, 24, 15, 4, 5, 29, 110],[19, 20, 11, 56, 25, 233, 154],[24, 26, 54, 48])==[4, 5, 11, 15, 19, 20, 24, 24, 25, 25, 26, 29, 48, 54, 56, 110, 154, 233]}
    }
    \incorrectresponse{%
        \\
        1. Import the \texttt{heapq} module to utilize the merge function for efficiently combining sorted iterables.\\
        2. Define a function \texttt{merge\_sorted\_list} that accepts multiple lists as input.\\
        3. For each input list, sort it individually to ensure they are in ascending order.\\
        4. Use \texttt{heapq.merge} to merge the sorted lists into a single sorted iterator.\\
        5. Convert the merged iterator to a list and return it as the final result.
        }  
        {%
        }
    \correctresponse{14}{%
        \\
        1. Define the function \texttt{merge\_sorted\_list} to accept multiple lists as input.\\
        2. Sort each input list individually to ensure they are in ascending order.\\
        3. Initialize a min-heap to track the smallest elements from each sorted list, storing tuples of (value, list index, element index).\\
        4. While the heap is not empty, extract the smallest element, append it to the result, and push the next element from the same list into the heap if available.\\
        5. Return the merged sorted list as the final result. 
    }{%
    }
\end{examplebox} % issue: 32B plan includes importing a specific library function (heapq.merge) that is not an available import, ignoring the available import list, and without a fallback. The 14B plan doesn't even use a real heap—it uses a simpler, less optimal approach which is however more robust than attempting an elegant solution and failing.

\vspace{20pt}
\begin{examplebox}[label=ex:3_coding]{Coding Example 8}
    \exquestion{%
    You are an expert Python programmer. You will be given a question (problem specification) and will generate a correct Python program that matches the specification and passes all tests.\\
    \\
    \#\#\# Question:\\
    You are given an array of integers nums of size \texttt{n} and a positive integer threshold.\\
    There is a graph consisting of n nodes with the \texttt{i}th node having a value of \texttt{nums[i]}. Two nodes \texttt{i} and \texttt{j} in the graph are connected via an undirected edge if \texttt{lcm(nums[i], nums[j]) <= threshold}.\\
    Return the number of connected components in this graph.\\
    A connected component is a subgraph of a graph in which there exists a path between any two vertices, and no vertex of the subgraph shares an edge with a vertex outside of the subgraph.\\
    The term \texttt{lcm(a, b)} denotes the least common multiple of \texttt{a} and \texttt{b}.\\
    \\
    Example 1:\\
    \\
    Input: \texttt{nums = [2,4,8,3,9], threshold = 5}\\
    Output: \texttt{4}\\
    Explanation:\\
    The four connected components are \texttt{(2, 4), (3), (8), (9)}.\\
    \\
    Example 2:\\
    \\
    Input: \texttt{nums = [2,4,8,3,9,12], threshold = 10}\\
    Output: \texttt{2}\\
    Explanation:\\
    The two connected components are \texttt{(2, 3, 4, 8, 9)}, and \texttt{(12)}.\\
    \\
    \\
    Constraints:\\
    \\
    \texttt{1 <= nums.length <= 105}\\
    \texttt{1 <= nums[i] <= 109}\\
    All elements of nums are unique.\\
    \texttt{1 <= threshold <= 2 * 105}\\
    \\
    \\
    \#\#\# Format: You will use the following starter code to write the solution to the problem and enclose your code within delimiters.\\
    \\
    \texttt{\textasciigrave\textasciigrave\textasciigrave python final\_answer("""\\
    class Solution:\\
    \hspace*{25pt}def countComponents(self, nums: List[int], threshold: int) -> int:\\
    \\
    """)\textasciigrave\textasciigrave\textasciigrave}\\
    \\
    \#\#\# Answer: (use the provided format with backticks)
    }
    \incorrectresponse{%
    \\
    To solve this problem, we need to determine the number of connected components in a graph where nodes are connected if the least common multiple (LCM) of their corresponding values is less than or equal to a given threshold. The solution involves efficiently using Union-Find (Disjoint Set Union, DSU) to group connected nodes.\\
    \\
    \#\#\# Approach\\
    \\
    1. **Filter and Sort**: First, filter out elements from the input array that are greater than the threshold and sort the remaining elements.\\
    2. **Union-Find Data Structure**: Use a Union-Find (DSU) structure to manage the connected components.\\
    3. **Process Multiples**: For each number in the filtered list, connect it with all its multiples that are also in the list and within the threshold.\\
    4. **Process Pairs with LCM Check**: For each number, check pairs where the LCM is within the threshold and connect them if they meet the condition.\\
    5. **Count Components**: After processing all connections, count the number of unique connected components using the DSU structure and add components from elements that were filtered out due to exceeding the threshold.\\
    \\
    \#\#\# Solution Code\\
    \\
    \texttt{\textasciigrave\textasciigrave\textasciigrave python\\
    import bisect\\
    from math import gcd\\
    from typing import List\\
    \\
    class Solution:\\
    \hspace*{25pt}def countComponents(self, nums: List[int], threshold: int) -> int:\\
        \hspace*{50pt}\# Filter and sort the numbers that are <= threshold\\
        \hspace*{50pt}filtered = [x for x in nums if x <= threshold]\\
        \hspace*{50pt}filtered.sort()\\
        \hspace*{50pt}n = len(filtered)\\
        \hspace*{50pt}if n == 0:\\
            \hspace*{75pt}return len(nums)\\
        \\
        \hspace*{50pt}\# Map each value to its index in the filtered list\\
        \hspace*{50pt}value\_to\_index = \{v: i for i, v in enumerate(filtered)\}\\
        \\
        \hspace*{50pt}\# Define DSU (Disjoint Set Union) structure\\
        \hspace*{50pt}class DSU:\\
            \hspace*{75pt}def \_\_init\_\_(self, size):\\
                \hspace*{100pt}self.parent = list(range(size))\\
                \hspace*{100pt}self.rank = [0] * size\\
                \\
            \hspace*{75pt}def find(self, x):\\
                \hspace*{100pt}if self.parent[x] != x:\\
                    \hspace*{125pt}self.parent[x] = self.find(self.parent[x])\\
                \hspace*{100pt}return self.parent[x]\\
                \\
            \hspace*{75pt}def union(self, x, y):\\
                \hspace*{100pt}x\_root = self.find(x)\\
                \hspace*{100pt}y\_root = self.find(y)\\
                \hspace*{100pt}if x\_root == y\_root:\\
                    \hspace*{125pt}return\\
                \hspace*{100pt}if self.rank[x\_root] < self.rank[y\_root]:\\
                    \hspace*{125pt}self.parent[x\_root] = y\_root\\
                \hspace*{100pt}else:\\
                    \hspace*{125pt}self.parent[y\_root] = x\_root\\
                    \hspace*{125pt}if self.rank[x\_root] == self.rank[y\_root]:\\
                        \hspace*{150pt}self.rank[x\_root] += 1\\
            \\
            \hspace*{50pt}dsu = DSU(n)\\
            \\
            \hspace*{50pt}\# Process each element in the filtered list\\
            \hspace*{50pt}for i in range(n):\\
                \hspace*{75pt}x = filtered[i]\\
                \hspace*{75pt}\# Connect with multiples of x\\
                \hspace*{75pt}m = 2 * x\\
                \hspace*{75pt}while m <= threshold:\\
                    \hspace*{100pt}if m in value\_to\_index:\\
                        \hspace*{125pt}idx = value\_to\_index[m]\\
                        \hspace*{125pt}dsu.union(i, idx)\\
                    \hspace*{100pt}m += x\\
                \hspace*{75pt}\# Connect with other numbers where LCM(x, y) <= threshold\\
                \hspace*{75pt}upper = threshold // x\\
                \hspace*{75pt}if upper >= x:\\
                    \hspace*{100pt}left = i + 1\\
                    \hspace*{100pt}right\_idx = bisect.bisect\_right(filtered, upper) - 1\\
                    \hspace*{100pt}if left <= right\_idx:\\
                        \hspace*{125pt}for j in range(left, right\_idx + 1):\\
                            \hspace*{150pt}y = filtered[j]\\
                            \hspace*{150pt}if (x * y) // gcd(x, y) <= threshold:\\
                                \hspace*{175pt}dsu.union(i, j)\\
            \\
            \hspace*{50pt}\# Count the number of unique roots in DSU\\
            \hspace*{50pt}roots = set()\\
            \hspace*{50pt}for i in range(n):\\
                \hspace*{75pt}roots.add(dsu.find(i))\\
            \\
            \hspace*{50pt}\# Total components is the number of roots in DSU plus elements > threshold\\
            \hspace*{50pt}return len(roots) + (len(nums) - n)\\
    \textasciigrave\textasciigrave\textasciigrave}\\
    \\
    \#\#\# Explanation\\
    \\
    1. **Filtering and Sorting**: The input array is filtered to include only values that are less than or equal to the threshold. This helps in reducing unnecessary computations.\\
    2. **Union-Find Structure**: The DSU structure is used to efficiently manage and merge connected components.\\
    3. **Multiples Processing**: For each number, we connect it with all its multiples within the threshold. This ensures that all direct connections based on multiples are considered.\\
    4. **LCM Check for Pairs**: For each number, we check pairs where the LCM is within the threshold. This ensures that even non-multiples that meet the LCM condition are connected.\\
    5. **Component Counting**: After processing all connections, the number of unique roots in the DSU structure gives the number of connected components for the filtered values. Components from values exceeding the threshold are added separately.\\
    \\
    This approach efficiently handles the constraints and ensures that all valid connections are considered, leading to the correct count of connected components.
    }  
      {%
       }
    \correctresponse{14}{%
    \\
    1. **Sort the input array \textsc{nums}** to facilitate efficient processing of multiples.\\
    2. **Create a mapping from values to their indices** to quickly check if a number exists in the array.\\
    3. **Use a Disjoint Set Union (DSU) structure** to manage connected components.\\
    4. **For each number in \textsc{nums}**, iterate through its multiples up to the threshold and connect the number with its multiples if they exist in the array.\\
    5. **Count the number of unique parents** in the DSU to determine the number of connected components.\\
    6. **Return the count of connected components** as the result. 
    }{%
    }
\end{examplebox} %LC Hard. issue: 32B puts code in plan and the code contains a mathematical error (that is, an extra optimization where the LCM pair checking with threshold // x bound. this is mathematically incorrect). the 14B plan avoids this over complication and uses only the multiples approach (Simpler algorithms have fewer places to introduce bugs) also it does not rush into producing code and produces a high level plan as required by the instructions.

\newpage
\begin{examplebox}[label=ex:4_coding]{Coding Example 9}
    \exquestion{%
    You are an expert Python programmer. You will be given a question (problem specification) and will generate a correct Python program that matches the specification and passes all tests.\\
    \\
    \#\#\# Question:\\
    You are given a string \texttt{initialCurrency}, and you start with 1.0 of \texttt{initialCurrency}.\\
    You are also given four arrays with currency pairs (strings) and rates (real numbers):\\
    \\
    \texttt{pairs1[i] = [startCurrencyi, targetCurrencyi]} denotes that you can convert from \texttt{startCurrencyi} to \texttt{targetCurrencyi} at a rate of \texttt{rates1[i]} on day 1.\\
    \texttt{pairs2[i] = [startCurrencyi, targetCurrencyi]} denotes that you can convert from \texttt{startCurrencyi} to \texttt{targetCurrencyi} at a rate of \texttt{rates2[i]} on day 2.\\
    Also, each \texttt{targetCurrency} can be converted back to its corresponding \texttt{startCurrency} at a rate of \texttt{1 / rate}.\\
    \\
    You can perform any number of conversions, including zero, using \texttt{rates1} on day 1, followed by any number of additional conversions, including zero, using \texttt{rates2} on day 2.\\
    Return the maximum amount of \texttt{initialCurrency} you can have after performing any number of conversions on both days in order.\\
    Note: Conversion rates are valid, and there will be no contradictions in the rates for either day. The rates for the days are independent of each other.\\
    \\
    Example 1:\\
    \\
    Input: \texttt{initialCurrency = "EUR", pairs1 = [["EUR","USD"],["USD","JPY"]], rates1 = [2.0,3.0], pairs2 = [["JPY","USD"],["USD","CHF"],["CHF","EUR"]], rates2 = [4.0,5.0,6.0]}\\
    Output: \texttt{720.00000}\\
    Explanation:\\
    To get the maximum amount of EUR, starting with 1.0 EUR:\\
    \\
    On Day 1:\\
    \\	
    Convert EUR to USD to get 2.0 USD.\\
    Convert USD to JPY to get 6.0 JPY.\\
    \\
    On Day 2:\\
    \\
    Convert JPY to USD to get 24.0 USD.\\
    Convert USD to CHF to get 120.0 CHF.\\
    Finally, convert CHF to EUR to get 720.0 EUR.\\
    \\
    \\
    Example 2:\\
    \\
    Input: \texttt{initialCurrency = "NGN", pairs1 = [["NGN","EUR"]], rates1 = [9.0], pairs2 = [["NGN","EUR"]], rates2 = [6.0]}\\
    Output: \texttt{1.50000}\\
    Explanation:\\
    Converting NGN to EUR on day 1 and EUR to NGN using the inverse rate on day 2 gives the maximum amount.\\
    \\
    \\
    Example 3:\\
    \\
    Input: \texttt{initialCurrency = "USD", pairs1 = [["USD","EUR"]], rates1 = [1.0], pairs2 = [["EUR","JPY"]], rates2 = [10.0]}\\
    Output: \texttt{1.00000}\\
    Explanation:\\
    In this example, there is no need to make any conversions on either day.\\
    \\
    \\
    Constraints:\\
    \\
    \texttt{1 <= initialCurrency.length <= 3}\\
    \texttt{initialCurrency} consists only of uppercase English letters.\\
    \texttt{1 <= n == pairs1.length <= 10}\\
    \texttt{1 <= m == pairs2.length <= 10}\\
    \texttt{pairs1[i] == [startCurrencyi, targetCurrencyi]}\\
    \texttt{pairs2[i] == [startCurrencyi, targetCurrencyi]}\\
    \texttt{1 <= startCurrencyi.length, targetCurrencyi.length <= 3}\\
    \texttt{startCurrencyi} and \texttt{targetCurrencyi} consist only of uppercase English letters.\\
    \texttt{rates1.length == n}\\
    \texttt{rates2.length == m}\\
    \texttt{1.0 <= rates1[i], rates2[i] <= 10.0}\\
    The input is generated such that there are no contradictions or cycles in the conversion graphs for either day.\\
    The input is generated such that the output is at most \texttt{5 * 1010}.\\
    \\
    \\
    \#\#\# Format: You will use the following starter code to write the solution to the problem and enclose your code within delimiters.\\
    \texttt{\textasciigrave\textasciigrave\textasciigrave python final\_answer("""\\
    class Solution:\\
    \hspace*{25pt} def maxAmount(self, initialCurrency: str, pairs1: List[List[str]], rates1: List[float], pairs2: List[List[str]], rates2: List[float]) -> float:\\
    \\
""")\textasciigrave\textasciigrave\textasciigrave}
    \\
    \\
    \#\#\# Answer: (use the provided format with backticks)
    }
    \incorrectresponse{%
    \\
    \texttt{\textasciigrave\textasciigrave\textasciigrave python\\
    class Solution:\\
    \hspace*{25pt}def maxAmount(self, initialCurrency: str, pairs1: List[List[str]], rates1: List[float], pairs2: List[List[str]], rates2: List[float]) -> float:\\
        \hspace*{50pt}def bellman\_ford(start, edges, all\_currencies):\\
            \hspace*{75pt}max\_amounts = {currency: -float('inf') for currency in all\_currencies}\\
            \hspace*{75pt}max\_amounts[start] = 1.0\\
            \hspace*{75pt}V = len(all\_currencies)\\
            \hspace*{75pt}for \_ in range(V - 1):\\
                \hspace*{100pt}updated = False\\
                \hspace*{100pt}for (u, v, rate) in edges:\\
                    \hspace*{125pt}if max\_amounts[u] != -float('inf'):\\
                        \hspace*{150pt}if max\_amounts[u] * rate > max\_amounts[v]:\\
                            \hspace*{175pt}max\_amounts[v] = max\_amounts[u] * rate\\
                            \hspace*{175pt}updated = True\\
                \hspace*{100pt}if not updated:\\
                    \hspace*{125pt}break\\
            \hspace*{75pt}return max\_amounts\\
        \\
        \hspace*{50pt}\# Process day1\\
        \hspace*{50pt}edges1 = []\\
        \hspace*{50pt}for i in range(len(pairs1)):\\
            \hspace*{75pt}a, b = pairs1[i]\\
            \hspace*{75pt}rate = rates1[i]\\
            \hspace*{75pt}edges1.append((a, b, rate))\\
            \hspace*{75pt}edges1.append((b, a, 1.0 / rate))\\
        \hspace*{50pt}currencies\_day1 = set()\\
        \hspace*{50pt}for a, b in pairs1:\\
            \hspace*{75pt}currencies\_day1.add(a)\\
            \hspace*{75pt}currencies\_day1.add(b)\\
        \hspace*{50pt}currencies\_day1.add(initialCurrency)\\
        \hspace*{50pt}day1\_max = bellman\_ford(initialCurrency, edges1, currencies\_day1)\\
        \\
        \hspace*{50pt}\# Process day2\\
        \hspace*{50pt}edges2 = []\\
        \hspace*{50pt}for i in range(len(pairs2)):\\
            \hspace*{75pt}a, b = pairs2[i]\\
            \hspace*{75pt}rate = rates2[i]\\
            \hspace*{75pt}edges2.append((a, b, rate))\\
            \hspace*{75pt}edges2.append((b, a, 1.0 / rate))\\
        \hspace*{50pt}currencies\_day2 = set()\\
        \hspace*{50pt}for a, b in pairs2:\\
            \hspace*{75pt}currencies\_day2.add(a)\\
            \hspace*{75pt}currencies\_day2.add(b)\\
        \hspace*{50pt}max\_result = 1.0  \# initial amount\\
        \\
        \hspace*{50pt}for c in day1\_max:\\
            \hspace*{75pt}amount\_c = day1\_max[c]\\
            \hspace*{75pt}if amount\_c == -float('inf'):\\
                \hspace*{100pt}continue\\
            \hspace*{75pt}all\_currencies\_day2\_run = currencies\_day2.copy()\\
            \hspace*{75pt}all\_currencies\_day2\_run.add(c)\\
            \hspace*{75pt}day2\_max = bellman\_ford(c, edges2, all\_currencies\_day2\_run)\\
            \hspace*{75pt}current\_value = day2\_max.get(initialCurrency, -float('inf'))\\
            \hspace*{75pt}if current\_value > max\_result:\\
                \hspace*{100pt}max\_result = current\_value\\
        \\
        \hspace*{50pt}return max\_result\textasciigrave\textasciigrave\textasciigrave}
        }  
        {%
        }
    \correctresponse{8}{%
        \\
        To solve the problem of maximizing the amount of the initial currency after performing conversions on two days, we need to model the conversion process as a graph traversal problem. The key steps involve:\\
        \\
        1. **Building the Conversion Graphs**: For each day, construct a graph where nodes represent currencies and edges represent conversion rates (both forward and reverse).\\
        \\
        2. **Topological Sorting**: Since the input guarantees no cycles, we can perform a topological sort on the graph to process nodes in an order that ensures all dependencies are resolved.\\
        \\
        3. **Maximizing Currency Amounts**: Using dynamic programming on the topologically sorted nodes, compute the maximum amount of each currency achievable after each day's conversions.\\
        \\
        4. **Combining Results**: After processing both days, compute the maximum amount of the initial currency by considering all possible paths from the currencies obtained after day 2 back to the initial currency.\\
        \\
        ---\\
        \\
        \#\#\# Step-by-Step Plan\\
        \\
        1. **Build the Graph for Day 1**:\\
        - Construct a graph where each currency pair from \texttt{pairs1} and \texttt{rates1} is represented as bidirectional edges (forward and reverse conversions).\\
        2. **Compute Day 1 Amounts**:\\
        - Starting with 1.0 of the initial currency, use topological sorting to compute the maximum amount of each currency achievable after day 1 conversions.\\
        3. **Build the Graph for Day 2**:\\
        - Construct a similar graph using \texttt{pairs2} and \texttt{rates2}.\\
        4. **Compute Day 2 Amounts**:\\
        - Use the results from day 1 as the starting point for day 2 conversions. Again, use topological sorting to compute the maximum amount of each currency after day 2.\\
        5. **Convert Back to Initial Currency**:\\
         - For each currency obtained after day 2, compute the maximum amount of the initial currency that can be obtained by converting through day 2's graph.\\
        6. **Return the Maximum Result**:\\
        - The final answer is the maximum value among all possible conversions back to the initial currency.\\
        \\
        ---\\
        \\
        \#\#\# Final Plan\\
        \\
        1. **Build the day1 graph** using \texttt{pairs1} and \texttt{rates1}, including both forward and reverse conversions.\\
        2. **Compute day1\_amounts** by traversing the graph in topological order, starting from the initial currency.\\
        3. **Build the day2 graph** using \texttt{pairs2} and \texttt{rates2}, including both forward and reverse conversions.\\
        4. **Compute day2\_amounts** by traversing the day2 graph in topological order, starting from the day1\_amounts.\\
        5. **For each currency in day2\_amounts**, compute the maximum amount of the initial currency by traversing the day2 graph again, starting from that currency.\\
        6. **Return the maximum value** among all computed results.
    }{%
    }
\end{examplebox} % LC medium. issue: the 32B plan already shows the code, which is propagated to the final answer without using the python shell tool for tests. indeed the code is wrong, it returns only current_value (the day 2 multiplier), completely ignoring the day 1 gains. Whereas the 8B plan explicitly accounts for it ("Use the results from day 1 as the starting point for day 2 conversions." and "Compute day2_amounts by traversing the day2 graph in topological order, starting from the day1_amounts."). Note that 8B plan is better but not perfect, it mention topological sort which is inaccurate, but the model corrects this while doing the trace and still obtains the correct final answer (ie correct code).

\newpage
\begin{examplebox}[label=ex:5_coding]{Coding Example 10}
    \exquestion{%
        You are an expert Python programmer, and here is your task: Write a function to find the first duplicate element in a given array of integers. Your code should pass the following test: \texttt{assert find\_first\_duplicate(([1, 2, 3, 4, 4, 5]))==4}
    }
    \incorrectresponse{%
        \\
        1. Create a function \texttt{find\_first\_duplicate} that takes an array as input.\\
        2. Initialize an empty set to track seen elements.\\
        3. Iterate through each element in the array:\\
        a. If the element is already in the set, return it immediately.\\
        b. Otherwise, add the element to the set.\\
        4. Use the \texttt{final\_answer} tool to return the implemented function code.  
        }  
        {%
            % def find_first_duplicate(arr):
            %     seen = set()
            %     for num in arr:
            %         if num in seen:
            %             return num
            %         seen.add(num)
            %     return None
        }
    \correctresponse{14}{%
        \\
        The task is to write a Python function that identifies the first duplicate element in an array of integers. The approach involves iterating through the array while tracking seen elements using a set. When a duplicate is encountered, it is immediately returned as the result.\\
        \\
        1. Initialize an empty set to keep track of elements that have been encountered.\\
        2. Iterate through each element in the input array.\\
        3. For each element, check if it is already in the set:\\
         - If it is, return the element as it is the first duplicate.\\
         - If it is not, add the element to the set.\\
        4. If no duplicates are found during the iteration, return a default value (e.g., -1) to indicate the absence of duplicates.\\
        \\
        This method ensures that the first duplicate is identified efficiently with a time complexity of $O(n)$ and a space complexity of $O(n)$, where n is the number of elements in the array.    
    }{%
        % def find_first_duplicate(arr):
        %     seen = set()
        %     for num in arr:
        %         if num in seen:
        %             return num
        %         seen.add(num)
        %     return -1
    }
\end{examplebox} % issue: 32B plan does not deal with what to do if the loop completes without finding a duplicate. Leading to an incorrect return type for the "no duplicate found" edge case that breaks one of the (hidden) tests.

%%%%%%%%%%%%%%%%%%%%%%%%%%%%%%%%%%%%%%%%%%%%%%%%%%%%%%%%%%%%%%%%%%%%%%%%%

\newpage\section{Cross-Family Generalization}
\label{app:cross_family}

To assess whether the auction mechanism and its learned weights transfer beyond the Qwen3 family, we run two complementary experiments: (i) we re-run the auction on a different open-weight family (Gemma 3) and check whether the same mechanism produces complexity-sensitive allocations and comparable auction signals, and (ii) we introduce cross-family agents (Llama 3.1, Ministral 3) into the Qwen3 pool without retraining the MILP weights and verify that they receive sensible allocations.

\subsection{Auction Transfer to Gemma 3}
\label{app:cross_family_gemma}

We run \textsc{sale}'s auction stage on the Gemma 3 family (4B, 12B, 27B), which differs from Qwen3 in both architectural details and training recipe. We generate strategic plans for 100 \textsc{HST-Bench} tasks (evenly stratified by type and complexity), compute the cost--value proxies (with a Gemma 3 jury), and run the MILP independently on this Gemma-only pool to learn auction weights from scratch.

The auction mechanism transfers cleanly: the MILP converges and produces complexity-sensitive allocations matching those observed for Qwen3. On the hardest coding tasks ($\tau(t) \in (12.5, 60]$), the auction routes 80\% of work to Gemma 27B, mirroring the pattern observed for Qwen3 32B. On easier tasks, work is distributed more evenly across agent sizes, consistent with our main finding that smaller agents are sufficient at low complexity.

Auction signals also follow a similar trend across families. Cross-family comparisons between size-matched Gemma 3 and Qwen3 agents (4B--4B, 12B--14B, 27B--32B) show convergence across all three signals: strategy-length distributions have Wasserstein distance as low as 9 tokens on average for the largest models, jury scores correlate strongly (Spearman $\rho \in [0.45, 0.76]$, all $p < 0.001$), and entropy correlates significantly (Spearman $\rho \in [0.46, 0.60]$, all $p < 0.001$) despite different absolute scales, which the MILP compensates for via its learned weight $w_h$. The two families' juries thus converge on the same task-level quality signal.

\subsection{Heterogeneous Agent Pool}
\label{app:cross_family_heterogeneous}

To test \textsc{sale} under heterogeneous conditions, we expand the Qwen3 pool (4B, 8B, 14B, 32B) with two agents from different families: Llama 3.1 8B and Ministral 3 8B. The two cross-family agents act as bidders only; the jury remains the original Qwen3 four-agent panel, so the per-agent judge weights $w_j$ are unchanged. This isolates the question of whether the global cost and entropy weights ($w_c$, $w_h$) transfer to a heterogeneous bidder pool; assigning new judges would require either re-running the MILP or transferring weights from size-matched agents, which we leave for future work. Crucially, we freeze the MILP weights learned on Qwen3 and perform no retraining: the cross-family agents are dropped into the existing auction unchanged. On the same 100-task subset, both new agents receive non-trivial workload shares that decrease with task complexity, consistent with their 8B-class size (Table~\ref{tab:cross_family_allocation}).

We find that the cross-family agents behave as expected under the auction: their entropy and jury scores are close to Qwen3 8B (e.g., Llama jury mean 3.48 vs Qwen3 8B 3.56), and per-task cost-minus-value scores correlate with Qwen3 8B rankings (Spearman $\rho = 0.40$, $p < 0.01$). Taken together with the Gemma 3 results in Section~\ref{app:cross_family_gemma}, these findings show that \textsc{sale}'s auction signals (i.e., the strategy text, entropy, and peer jury scores) are not family-specific, and that the mechanism is able to coordinate agents across different architectures and training recipes.

\begin{table}[h]
\centering
\vspace{15pt}
\setlength{\tabcolsep}{54pt}%
\caption{Workload allocation to Llama 3.1 and Ministral 3 when added to the pool with weights frozen on Qwen3.}
\label{tab:cross_family_allocation}
\scalebox{0.91}{%
\begin{tabular}{ccc}
\toprule
Complexity & Llama 3.1 8B & Ministral 3 8B \\
\midrule
$\tau(t) \le 0.1$  & 25\% & 20\% \\
$\tau(t) \le 0.5$  & 20\% & 15\% \\
$\tau(t) \le 2.5$  & 15\% & 10\% \\
$\tau(t) \le 12.5$ & 15\% & 10\% \\
$\tau(t) \le 60$   & 5\%  & 5\%  \\
\midrule
Overall            & 16\% & 12\% \\
\bottomrule
\end{tabular}%
}
\end{table}

\end{document}